%% file: TMD-evol-sub-leadingP_v2.tex
\documentclass[a4paper]{article}
\usepackage{jheppub} % for details on the use of the package, please
\usepackage{graphicx,color}
 \usepackage{bm}% bold math
   \usepackage{amsmath}
    \usepackage{amssymb}    
     \usepackage{pifont}
      \usepackage{simplewick}
\usepackage{tikz}
\usetikzlibrary{decorations.pathmorphing}
\usetikzlibrary{decorations.markings}
\usetikzlibrary{calc}
\usetikzlibrary{math}
\usetikzlibrary{patterns}
\usepackage{standalone}
\usepackage{slashed}
\usepackage{multirow}
\usepackage{hhline}

\newcommand{\Ds}{\displaystyle}

\newcommand{\nn}{\nonumber}

\newcommand{\Tr}{\mathrm{Tr}}

\newcommand{\ot}{\leftarrow}

\renewcommand{\(}{\left(}
\renewcommand{\)}{\right)}
\renewcommand{\[}{\left[}
\renewcommand{\]}{\right]}

\title{Definition and evolution of transverse momentum dependent distribution of twist-three}

\author{Simone~Rodini}
\author{Alexey~Vladimirov}

\affiliation{Institut f\"ur Theoretische Physik, Universit\"at Regensburg, D-93040 Regensburg, Germany}

\emailAdd{simone.rodini@physik.uni-regensburg.de}
\emailAdd{alexey.vladimirov@physik.uni-regensburg.de}

\abstract{
We present an in-depth analysis of transverse momentum dependent (TMD) distributions of  twist-three. In particular, we focus on evolution equations, symmetry relations, parameterization, interpretation, small-b asymptotic behaviour and the structure of singularities. The starting point of discussion are the correlators with the definite TMD-twist. 
By considering suitable combinations of these correlators, we introduce physical TMD distribution of twist-three that can be used for practical applications. We also establish relations with generic TMD distribution of twist-three, and demonstrate that their evolution equations are autonomous in the large-$N_c$ limit.
}

\begin{document} 
\allowdisplaybreaks
\maketitle 

\section{Introduction}

The transverse-momentum dependent (TMD) distributions are universal functions that enter the TMD factorization theorems and describe various aspects of partons' nonperturbative dynamics. The TMD distributions that appear in the leading power (LP) term of the factorization theorem for a variety of processes are known very well. 
For the review of the recent theoretical and phenomenological development and further references, see refs. \cite{Angeles-Martinez:2015sea, AbdulKhalek:2021gbh, Khalek:2022bzd, Scimemi:2019mlf}. 
At next-to-leading power (NLP) accuracy, a new set of TMD distributions emerges. These TMD distributions are usually referred to as twist-three distributions, and in general are studied much less.
%The TMD distributions that appear at the next-to-leading power (NLP) factorization and are generally called TMD distributions of twist-three are studied much less. 
Despite being introduced long ago \cite{Mulders:1995dh}, and discussed in many works, twist-three TMD distributions suffer from a lack of a  systematic approach. 
This is a natural consequence of the NLP TMD factorization theory being at its earlier stage of development. The first convincing studies appeared only recently \cite{Balitsky:2017gis, Balitsky:2020jzt, Inglis-Whalen:2021bea, Vladimirov:2021hdn, Ebert:2021jhy}. 
Further progress in this direction requires the systematization of knowledge about TMD distributions of twist-three, which we present in this paper.

The term higher-twist (and particularly twist-three) distributions requires some clarification. There are several meanings of the term ``twist'', which are often mixed up in the literature. The exact definitions and pedagogical explanations can be found in refs.~\cite{Jaffe:1996zw, Braun:2003rp, Braun:2011dg}. 
One should clearly distinguish between genuine and generic higher-twist distributions. The genuine distributions are defined by the specification of quantum numbers of the corresponding operator. They are mathematically self-contained, although often too complicated for direct practical applications. On the contrary, generic distributions are usually defined by the power counting for observables, and thus have no simple theoretical structure, but are convenient practically. The best common sense analogies for the generic distributions are structure functions, which are functionals of genuine distributions.

The relations and properties of genuine and generic twist-three collinear distributions are known well (some recent development and further references can be found, e.g., in \cite{Hatta:2019csj, Braun:2021aon, Bastami:2020rxn}). For the TMD distributions, the theory is much less developed. 
So far, all twist-three TMD distributions that have been discussed were generic and, therefore, their properties remain almost unexplored. For instance, there is no complete classification, and even evolution equations, which are the central element of definition, are unknown. The reason is that a mathematically-rigorous definition of a ``genuine twist'' for TMD distributions has been introduced only recently in ref.\cite{Vladimirov:2021hdn}.

%absence of a mathematically rigorous definition of a ``genuine twist'' for TMD distributions. It has been introduced recently in ref.\cite{Vladimirov:2021hdn}.
In ref.\cite{Vladimirov:2021hdn} the TMD-twist is defined as a composition of geometric twists (``dimension-spin'') of the light-ray parts of the TMD operators. The accurate definition is presented in sec.~\ref{sec:def-twist}. The decomposition with respect to this parameter guarantees the independence of operators in the sense that their evolution is autonomous. As a result, the basis of corresponding TMD distributions is orthogonal. Each TMD distribution with a definite TMD-twist is an independent nonperturbative function. In this way, the TMD distributions with the specific TMD-twist serve as elementary blocks for constructing other twist-three TMD distributions. However, as we discuss in this paper, the proper definition of TMD requires extra details. It appears that individual terms of cross-section, computed with the na\"ive definition, are divergent (although the sum of all terms is finite). Therefore, the definition of physical TMD distributions (i.e., such that provide finite cross-section term-by-term) includes extra subtraction terms. 

The aim of this work is to find a suitable definition of physical TMD distributions of twist-three and derive their main properties (such as evolution equations, symmetries, support properties, etc.). To find a good definition, we start with the most mathematically convenient  definite-TMD-twist TMD distribution in position space. Then, step-by-step, we modify it improving its properties. This procedure allows us to derive evolution equations and other aspects of physical TMD distributions without significant effort. The resulting functions are well-defined and suitable for phenomenological studies. Additionally, we establish relations with widely-used generic distributions and compute the evolution equations for the latter for the first time. 

To make the exposition concentrated, we deal only with the quark TMD parton distribution function (TMDPDFs) in the singlet color representation. These distributions are the most important since they are the main twist-three building blocks for cross-sections of the Drell-Yan (DY) process and Semi-Inclusive Deep-Inelastic scattering (SIDIS). Other distributions, such as gluon TMDPDFs, TMD fragmentation functions (TMDFFs), or distributions with octet color representation, can be considered in the same way. Moreover, many properties are the same (e.g., evolution equations for TMDFFs can be easily obtained from those for TMDPDFs).

The paper is split into five sections, each representing a complete discussion on a particular stage of manipulations.
\begin{itemize}
\item We start in sec.~\ref{sec:TMD-in-pos} by introducing the definite-TMD-twist TMD distribution in position space, as it was proposed in ref.\cite{Vladimirov:2021hdn}. This is the defining representation, since the operator expansion that gives rise to the factorization theorem is derived in the position space. However, in such a form, distributions are useless for phenomenology. They are complex functions with indefinite T-parity and unclear interpretation.
\item In sec.~\ref{sec:TMD-in-mom}, we define definite-TMD-twist TMD distribution in momentum-fraction space, discussing their interpretation and support properties. However, the Fourier transformation to the momentum-fraction space makes evolution equations bulky.
\item In sec.~\ref{sec:TMD-Tdefinite}, we introduce particular combinations of the definite-TMD-twist TMD distributions, in such a way that they present more `natural' symmetry properties: specifically they have definite complexity (i.e., they are either real or purely imaginary) and definite T-parity. This is the most convenient stage to introduce the parameterization of particular components of TMD correlators. These distributions are not entirely autonomous but obey pairwise mixing under evolution. These distributions are not usual functions but rather generalized functions, i.e., they have indefinite values at certain points but definite integrals.
\item The expression for the cross-section involves combinations of the zeroth Mellin moments of TMD distributions, which are individually singular but their sum, relevant for the cross-section, is finite. In sec.~\ref{sec:physical}, we explicitly compute the singular part of the zeroth moment and demonstrate that it is equivalent to the rapidity divergence. Finally, we introduce the physical TMD distribution by subtracting the divergent terms. The definition of physical distributions is the final product of this work. The cross-section evaluated in terms of physical TMD distributions is finite term-by-term.
\item In sec.~\ref{sec:bi-quark}, we discuss the connection of physical TMD distributions, and generic TMD distributions introduced in works \cite{Mulders:1995dh,Goeke:2005hb,Bacchetta:2006tn}. In particular, we demonstrate that evolution equations for generic twist-three distributions are not closed. However, the evolution equation essentially simplifies and closes in the large-$N_c$ limit which allows the phenomenology of these distributions with controllable precision.
\end{itemize}

We support the work with three appendices, which contains the synopsis of vector algebra definition (appendix \ref{app:conventions}), additional details on the LO evolution kernels (appendix \ref{app:collinear}), and leading small-b asymptotic form for TMD distribution of twist-three (appendix \ref{sec:smallb}).

\section{TMD correlators in position space}
\label{sec:TMD-in-pos}

In this section, we define and discuss the TMD correlators in the position space. The general discussion about the construction of TMD operators and distributions at sub-leading powers of TMD factorization can be found in ref.\cite{Balitsky:2017gis, Inglis-Whalen:2021bea, Vladimirov:2021hdn}. The LO expressions for the evolution kernels are taken from ref.\cite{Vladimirov:2021hdn}, where one can also find details of its derivation.

\subsection{Definition}
\label{sec:def-twist}

All TMD correlators at any power of the TMD factorization theorem have the same general structure. Namely, they are a composition of two semi-compact light-cone operators $U$, separated by a transverse distance $b$
\begin{eqnarray}\label{def:O=UU}
\widetilde{\Phi}_{AB}(z_1,...,z_a,z_b,...,z_n;b)&\sim& \langle p,s|U_A(z_1,...,z_a;b)U_B(z_b,...,z_n;0)|p,s\rangle,
\end{eqnarray}
where $A$ and $B$ indicate the quantum numbers of $U$. The operator $U(\{z_i\};b)$ is a product of QCD fields positioned at points $(z_in+b)$, where $n^\mu$ is the light-cone vector and $b^\mu$ is the transverse vector. The fields are first transported to the light-cone infinity by light-cone Wilson lines, and then to transverse infinity by a transverse Wilson line\footnote{
\label{footnote:transverse-link} Transverse gauge links do not contribute to perturbative calculation in regular gauges (such as the covariant gauge), because in these gauge the gluon field vanishes at infinities. In singular gauges (such as the light-cone gauge), transverse links could produce a non-vanishing contribution \cite{Belitsky:2002sm}. However, one can select an auxiliary gauge condition such that interactions with the transverse link vanish. Explicit inclusion of transverse links essentially complicates the computation, and spoils determination of some global properties. For instance, it makes impossible the definition of twist in any form. Therefore, we omit writing transverse links along the paper, assuming that all relations are valid in regular gauges or gauges with appropriate boundary conditions. An extended discussion on inclusion of traverse links into the perturbation theory and selection of convenient gauge can be found in refs.~\cite{Vladimirov:2021hdn,Belitsky:2002sm,Idilbi:2010im}.
}%
, such that the matrix element (\ref{def:O=UU}) is explicitly gauge invariant.
The operator $U$ spans an infinite range, and for that reason it is called \textit{semi-compact}.

The properties of TMD correlators are related to the \textit{TMD-twist} of the defining operator, the notion of which was introduced in ref.\cite{Vladimirov:2021hdn}. The TMD-twist is given by a pair of integer numbers $(N,M)$, where $N$ and $M$ are geometrical twists (i.e. ``dimension-spin'') of semi-compact operators\footnote{
\label{footnote:TMD-twist}Formally, the geometrical twist is defined for a local operator, and generalized to non-local operators by means of the generating function \cite{Anikin:1978tj}. Applying this definition to semi-compact operators one needs to regularize them by setting $L$ being finite. Regularized operators can be presented as a series of local operators, for which geometrical twist are computed by the usual ``dimension-spin'-rule.  For example, at finite $L$, the operator $U_1$ can be presented as $$U_1(\{0\},0_T)=\sum_{n=0}^\infty \frac{i^n L^n}{n!}D_+^nq(0).$$
For good components of the quark field, operators $D_+^nq$ have the dimension $n+3/2$ and the spin $n+1/2$, and thus $U_1$ has geometrical twist-1. For bad components of quark field, operator has no definite geometrical twist, since it can have spin $n+1/2$ or $n-1/2$. After the twist-decomposition the limit $L\to\pm\infty$ can be taken. Such scheme correctly reproduces properties of the operator product expansion as demonstrated in ref.\cite{Moos:2020wvd}.}
$U$ in eqn.~(\ref{def:O=UU}). The LP TMD factorization theorem deals with distributions of the TMD-twist (1,1).  At NLP, a new set of TMD distributions appears with TMD-twists (1,2) and (2,1). At NNLP one has TMD-twists (1,3), (3,1) and (2,2), and so on. This structure resembles the ordinary structure of power expansion in the collinear factorizations with the twist equals to the sum $(N+M)$. Importantly, even if $(N+M)$ is the same, operators with different TMD-twists $(N,M)$ obey different evolution equations and symmetry properties as a  consequence of the Lorentz invariance. However, for simplicity, we refer to TMD-twist (1,1) as twist-two and TMD-twists (2,1) and (1,2) as \textit{twist-three} if it does not create a confusion.

At LP and NLP of the TMD factorization theorem for DY and SIDIS processes only two kinds of semi-compact operators appear. They are
\begin{eqnarray}\label{def:U}
U_{1,i}(\{z\},b)&=&[Ln+b,zn+b]q_i(zn+b),
\\\nn
U_{2,i}^\mu(\{z_1,z_2\},b)&=&g[Ln+b,z_1n+b]F^{\mu+}[z_1n+b,z_2n+b]q_i(z_2n+b),
\end{eqnarray}
where $g$ is the QCD coupling constant, $q$ is the quark field, $F^{\mu+}=F^{\mu\nu}_At^An_\nu$ is the gluon field-strength tensor with $t^A$ being the generator of $SU(N_c)$. The index $i$ is the spinor index. The gauge links are defined as usual
\begin{eqnarray}
[a,b]=P\exp\(-ig\int_a^b dz^\mu A_\mu(z)\).
\end{eqnarray}
The letter $L$ denotes the infinity and depends on the process for which the factorization is derived. So,
\begin{eqnarray}\label{def:L}
L=s\infty,\qquad \text{with } s=\left\{
\begin{array}{ll}
+1, & \text{for SIDIS,}
\\
-1, & \text{for DY,}
\end{array}\right. %}
\end{eqnarray}
where $s$ is introduced for future convenience.

For convenience, we explicitly write the conjugated of the operators in eqns.\eqref{def:U}:
\begin{eqnarray}\label{def:Ubar}
\overline{U}_{1,i}(\{z\},b)&=&\bar q_i(zn+b)[zn+b,Ln+b],
\\\nn
\overline{U}_{2,i}^\mu(\{z_1,z_2\},b)&=&g\bar q_i(z_1n+b)[z_1n+b,z_2n+b]F^{\mu+}[z_2n+b,Ln+b],
\end{eqnarray}
Semi-compact operators (\ref{def:U}) and (\ref{def:Ubar}) have triplet and anti-triplet color representations. Other semi-compact operators with same power counting have higher color representations. We do not considered neither these operators nor the pure gluon operators in this work. 

The product of semi-compact operators (\ref{def:U}) and (\ref{def:Ubar}) produces the TMD operators, whose matrix elements define TMD correlators and distributions. There is a single TMD correlator of twist-two
\begin{eqnarray}
\widetilde{\Phi}^{ij}_{11,\text{bare}}(z_1,z_2,b)&=&
\langle p,s| \overline{U}_1^j(\{z_1\},b)\times U_1^i(\{z_2\},0)|p,s\rangle,
\\\nn
&=&
\langle p,s|T\{\bar q_j(z_1n+b)[z_1n+b,Ln+b]\,\times\, [Ln,z_2n]q_i(z_2n)\}|p,s\rangle,
\end{eqnarray}
and two TMD correlators of twist-three
\begin{eqnarray}
\label{def:Phi-pos}
&&\widetilde{\Phi}_{21,\text{bare}}^{ij,\mu}(z_1,z_2,z_3,b)=
\langle p,s| \overline{U}_2^{j,\mu}(\{z_1,z_2\},b)\times U_1^i(\{z_3\},0)|p,s\rangle
\\\nn
&&\qquad\qquad=
g\langle p,s|T\{\bar q_j(z_1n+b)[z_1n+b,z_2n+b]F^{\mu+}[z_2n+b,Ln+b]\,\times\,[Ln,z_3n]q_i(z_3n)\}|p,s\rangle
\\\nn
&&\widetilde{\Phi}^{ij,\mu}_{12,\text{bare}}(z_1,z_2,z_3,b)=
\langle p,s| \overline{U}_1^j(\{z_1\},b)\times U_2^{i,\mu}(\{z_2,z_3\},0)|p,s\rangle
\\\nn &&\qquad\qquad=
g\langle p,s|T\{\bar q_j(z_1n+b)[z_1n+b,Ln+b]\,\times\,[Ln,z_2n]F^{\mu+}[z_2n,z_3n]q_i(z_3n)\}|p,s\rangle.
\end{eqnarray}
In these expressions, the symbol $\times$ indicates the position where the transverse link should be inserted (see footnote \ref{footnote:transverse-link}). The color indices are not explicitly indicated but contracted in a natural way from quark to anti-quark fields along gauge links, such that the operator is color-neutral. The subscript ``bare'' indicates that these TMD correlators still require renormalization, which is discussed in the next section. The ket $|p,s\rangle$ denotes the hadron state with the spin $s^\mu$ and the momentum $p^\mu$. The large and small components of the hadrons' momentum defines the light-cone vectors $\bar n^\mu$ and $n^\mu$, 
\begin{eqnarray}
p^\mu = p^+ \bar n^\mu+ \frac{M^2}{2p^+}n^\mu,\qquad n^2=\bar n^2=0,\qquad (n\bar n)=1,
\end{eqnarray}
where $M$ is the hadron's mass. Along the paper we neglect mass corrections, and effectively consider massless hadron. We use the standard notation for vector decomposition $a^\mu=a^+\bar n^\mu+a^- n^\mu+a_T^\mu$, where $a_T^\mu$ is the transverse component. The synopsis of vector algebra definitions is presented in app.\ref{app:conventions}.

The spinor indices $i,j$ play a crucial role for the correlator's properties. In particular, the TMD-twist of the operator is different for different components of spinor matrix, since they define the projection of the quark's spin. The genuine twist-two and twist-three operators require good components of both quark fields. 

It is convenient to adopt the standard notation
\begin{eqnarray}
\widetilde{\Phi}^{[\Gamma]}_{11}=\frac{1}{2}\Tr(\Gamma \, \widetilde{\Phi}_{11}),
\qquad
\widetilde{\Phi}^{[\Gamma]}_{\mu,12}=\frac{1}{2}\Tr(\Gamma \, \widetilde{\Phi}_{\mu,12}),
\qquad
\widetilde{\Phi}^{[\Gamma]}_{\mu,21}=\frac{1}{2}\Tr(\Gamma \, \widetilde{\Phi}_{\mu,21}),
\end{eqnarray}
where $\Gamma$ is a Dirac matrix. Let us denote the set of Dirac matrices which project both quark fields to good components as $\Gamma^+$. Such matrices form a vector space with the standard basis
\begin{eqnarray}
\Gamma^+_{\text{basis}}&=&\{\gamma^+, \gamma^+\gamma^5, i\sigma^{\alpha +}\gamma^5\},
\end{eqnarray}
where $\alpha$ is a transverse index, and $\gamma^5=i\gamma^0\gamma^1\gamma^2\gamma^3$. For $\Gamma\in\Gamma^+$, the TMD correlator $\widetilde{\Phi}_{11}^{[\Gamma]}$ has TMD-twist (1,1), whereas $\widetilde{\Phi}^{[\Gamma]}_{\mu,21}$ and $\widetilde{\Phi}^{[\Gamma]}_{\mu,12}$ have TMD-twists (2,1) and (1,2) correspondingly.

The basis of TMD correlators $\{\widetilde{\Phi}_{11}^{[\Gamma]}, \widetilde{\Phi}^{[\Gamma]}_{\mu,21}, \widetilde{\Phi}^{[\Gamma]}_{\mu,12}\}$ (with $\Gamma\in\Gamma^+$) is complete at NLP. In the sense, that no other quark TMD correlators appear in the TMD factorization theorem \cite{Vladimirov:2021hdn}.

In some applications, it is convenient to introduce TMD correlators $\widetilde{\Phi}_{11}^{[\Gamma]}$ with a Dirac matrix $\Gamma\not\in\Gamma^+$ that projects a good and a bad components of quark fields (see e.g. refs. \cite{Ebert:2021jhy,Bacchetta:2006tn}). The set of such $\Gamma$'s is denoted as $\Gamma_T$, see (\ref{app:GammaT}). The operator $U_1$ with a bad component has a mixed geometrical twist, and therefore, TMD correlator $\widetilde{\Phi}_{11}^{[\Gamma_T]}$ also do not have a definite TMD-twist. As the result, their evolution equations are not closed, and are expressed via the basis correlators. Additionally, the rapidity divergences of such correlators cannot by renomalized in the usual manner, as it is shown in sec.~\ref{sec:physical}. In the sec.~\ref{sec:bi-quark}, we discuss the relation between $\widetilde{\Phi}_{11}^{[\Gamma_T]}$ and TMD correlators with the definite twist and derive some of their properties.

\subsection{Renormalization}

Any TMD operator has two kinds of singularities that are separately renormalized. These are ultraviolet (UV) singularities and rapidity singularities. The UV singularities are specific to each semi-compact operator, and therefore, a TMD correlator is renormalized by two UV renormalization constants. The rapidity divergences appear due to the interaction of a semi-compact operator with the Wilson line of another semi-compact operator, and thus their renormalization is common for the whole TMD correlator. The renormalized TMD correlators (\ref{def:Phi-pos}) read
\begin{eqnarray}\nn
\widetilde{\Phi}^{[\Gamma]}_{11,\text{bare}}(z_1,z_2,b)&=&R\(b^2\)Z_{U1}(z_1)Z_{U1}(z_2)
\widetilde{\Phi}_{11}^{[\Gamma]}(z_1,z_2,b;\mu,\zeta),
\\\label{def:Phi-renorm}
\widetilde{\Phi}_{\mu,21,\text{bare}}^{[\Gamma]}(z_1,z_2,z_3,b)&=&R\(b^2\)Z_{U2}(z_2,z_1)Z_{U1}(z_3)\otimes
\widetilde{\Phi}^{[\Gamma]}_{\mu,21}(z_1,z_2,z_3,b;\mu,\zeta),
\\\nn
\widetilde{\Phi}_{\mu,12,\text{bare}}^{[\Gamma]}(z_1,z_2,z_3,b)&=&R\(b^2\)Z_{U1}(z_1)Z_{U2}(z_2,z_3)\otimes
\widetilde{\Phi}^{[\Gamma]}_{\mu,12}(z_1,z_2,z_3,b;\mu,\zeta),
\end{eqnarray}
where $R$ is the rapidity renormalization constant, $Z_{U1}$ and $Z_{U2}$ are the UV renormalization constants for the semi-compact operators $U_1$ and $U_2$, correspondingly. The symbol $\otimes$ denotes the integral convolution between $Z_{U2}$ and the TMD correlator. The scales $\mu$ and $\zeta$ are renormalization scales for UV and rapidity divergences, respectively.

The UV- and rapidity-renormalized TMD correlators are called subtracted \cite{Collins:2011zzd,Aybat:2011zv}. Since they are the main object of this paper, we restrain from introducing any special label. 

The expressions (\ref{def:Phi-renorm}) implicitly depend on the process in which they appear. This dependence enters through the renormalization condition for rapidity divergences. The standard renormalization condition states that the cross-section for DY and SIDIS processes is proportional to a convolution of two TMD distributions without extra nonperturbative factors. Schematically,
\begin{eqnarray}\label{def:normalizationcondition}
d\sigma \sim \int d^2b e^{-i(qb)}\widetilde{\Phi}(z^-,b;\mu,\zeta)\widetilde{\Phi}(z^+,b;\mu,\bar \zeta),
\end{eqnarray}
where $\zeta\bar \zeta=(2q^+q^-)^2$. This scheme is defined nonperturbatively. In particular, it implies that the nonperturbative LP soft factor can be written as a product
\begin{eqnarray}\label{def:S=RR}
\mathcal{S}_{\text{LP}}(\delta^+\delta^-,b^2)=R\(\frac{\delta^+}{q^+},\zeta,b^2\)R\(\frac{\delta^-}{q^-},\bar \zeta,b^2\),
\end{eqnarray}
where $\delta^\pm$ are the chosen regulators for rapidity divergences in the  $n$ and $\bar n$ directions. Therefore, the rapidity renormalization factor contains nonperturbative parts. For a detailed discussion we refer to refs.~\cite{Collins:2011zzd,Echevarria:2012js,Vladimirov:2017ksc}. 

Note that the soft factor has it own UV renormalization factor. Moreover, $Z_{U_N}$ and $R$ have collinear divergences which cancel in the product. Therefore, the explicit expressions for $Z_{U_N}$'s and $R$ depend on the order of the renormalization. This mutual dependence gives rise to the Collins-Soper equation (\ref{def:CS-equation}). Apart from it, renormalization constants are independent.  The explicit expression for $R$ up to NNLO (three-loops) can be found in ref.\cite{Vladimirov:2017ksc}. The renormalization constant $Z_{U1}$ is equal to the quark-field renormalization in the light-cone gauge, and it is know at NNLO (three-loops), see e.g. \cite{Gehrmann:2010ue}. The renormalization constant $Z_{U2}$ has been derived in ref.\cite{Vladimirov:2021hdn} at LO.

\subsection{Evolution equations}

The presence of two independent renormalization factors produces two scales, traditionally denoted as $\mu$ (for UV scaling) and $\zeta$ (for rapidity scaling). Correspondingly, each TMD correlator obeys a pair of evolution equations. 

The evolution equations with respect to $\mu$ are
\begin{eqnarray}\nn
\mu^2 \frac{d}{d\mu^2}\widetilde{\Phi}_{11}^{[\Gamma]}(z_1,z_2,b;\mu,\zeta)&=&
\(\widetilde{\gamma}_1(z_1,\mu,\zeta)+\widetilde{\gamma}_1(z_2,\mu,\zeta)\)\widetilde{\Phi}^{[\Gamma]}_{11}(z_1,z_2,b;\mu,\zeta),
\\\label{evol:UV:position}
\mu^2 \frac{d}{d\mu^2}\widetilde{\Phi}^{[\Gamma]}_{\mu,21}(z_1,z_2,z_3,b;\mu,\zeta)&=&
(\widetilde{\gamma}_2(z_2,z_1,\mu,\zeta)+\widetilde{\gamma}_1(z_3;\mu,\zeta))\widetilde{\Phi}^{[\Gamma]}_{\mu,21}(z_1,z_2,z_3,b;\mu,\zeta),
\\\nn
\mu^2 \frac{d}{d\mu^2}\widetilde{\Phi}^{[\Gamma]}_{\mu,12}(z_1,z_2,z_3,b;\mu,\zeta)&=&
(\widetilde{\gamma}_1(z_1,\mu,\zeta)+\widetilde{\gamma}_2(z_2,z_3;\mu,\zeta))\widetilde{\Phi}^{[\Gamma]}_{\mu,12}(z_1,z_2,z_3,b;\mu,\zeta),
\end{eqnarray}
where anomalous dimensions are integro-differential operators. At LO they are
\begin{eqnarray}
\widetilde{\gamma}_1(z,\mu,\zeta)&=&a_s(\mu)C_F\(\frac{3}{2}+\ln\(\frac{\mu^2}{\zeta}\)+2\ln\(\frac{q^+}{-s\partial^+_z}\)\)+\mathcal{O}(a_s^2),
\\
\widetilde{\gamma}_2(z_2,z_3,\mu,\zeta)&=&a_s(\mu)\Big\{\mathbb{H}_{z_2z_3}+
C_F\(\frac{3}{2}+\ln\(\frac{\mu^2}{\zeta}\)\)
\\\nn &&+C_A\ln\(\frac{q^+}{-s\partial^+_{z_2}}\) +2\(C_F-\frac{C_A}{2}\)\ln\(\frac{q^+}{-s\partial^+_{z_3}}\)
\Big\}+\mathcal{O}(a_s^2),\label{def:gamma_2}
\end{eqnarray}
where $a_s=g^2/(4\pi)^2$, $C_F=(N^2_c-1)/2N_c$, $C_A=N_c$ for $N_c$ colors. The parameter $q^+$ arises due to the redefinition of the rapidity renormalization scale in the boost invariant form \cite{Collins:2011zzd, Echevarria:2012js, Vladimirov:2017ksc}, and is defined together with the values of $\zeta$ and $\bar \zeta$. A convenient choice for $q^+$ is
\begin{eqnarray}\label{def:q+}
q^+=\left\{
\begin{array}{ll}
|p_{\bar q}|=|p_{q}|, & \text{for } \widetilde{\Phi}_{11}^{[\Gamma]},
\\
|p_{\bar q}|, & \text{for } \widetilde{\Phi}^{[\Gamma]}_{12},
\\
|p_{q}|, & \text{for } \widetilde{\Phi}^{[\Gamma]}_{21},
\end{array}\right.%}
\end{eqnarray}
where $p_{\bar q}$ and $p_q$ are the momenta of anti-quark and quark fields, correspondingly. Such selection naturally appears in the DY and SIDIS cross-sections due to the renormalization condition (\ref{def:normalizationcondition}), and implies $\zeta \bar \zeta=(2q^+q^-)^2=(Q^2-q_T^2)^2$ (where $Q$ and $q_T$ are the invariant mass and the transverse momentum of the hard current). In the following, we adopt this convention. The expressions for a different (positive) value of $q^+$ can be found by the rescaling $q^+\to \alpha q^+$ and $\zeta\to\alpha^{-2}\zeta$. 

The logarithms $\ln(q^+/\partial^+)$ are a short (and convenient) notation for particular integral convolutions which are understood formally by their action on the generating function: $\ln(q^+/\partial^+)e^{izp^+}=\ln(q^+/ip^+)e^{izp^+}$. Such notation allows us to operate in position space (where expressions are generally shorter), without the necessity to write the convolution explicitly.
The notation is also convenient for the subsequent Fourier transformation.

\begin{figure}[t]
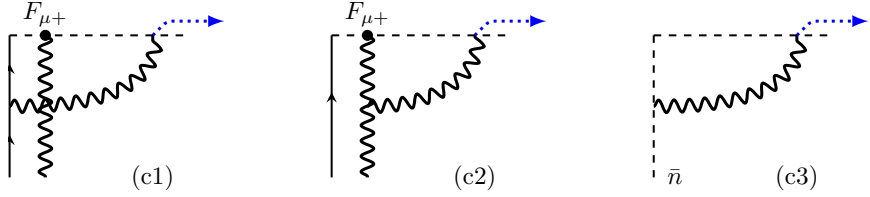

\begin{center}
\includestandalone[width=0.75\textwidth]{Figures/diagrams3}%     without .tex extension
\caption{Diagrams (c1) and (c2) produce the collinear divergence in the UV evolution kernels for operator $U_2$. The diagrams (c3) produces the corresponding divergence in the LP soft factor. The blue arrows indicate the limits in which in the loop diagram the divergences occur. Divergences cancel in the ratio between the correlator and the soft factor, leaving the remnant logarithms $\ln(q^+/(-s\partial^+))$ in egn.(\ref{evol:UV:position}). These logarithms are responsible for the complex part of the evolution kernel.}
\label{fig:collinear-diagrams}    
\end{center}
\end{figure}

Importantly, due to the convention (\ref{def:q+}), the logarithms $\ln(q^+/\partial^+)$ are in general complex-valued\footnote{\label{foot:logarithm}
To determine the complex part of expression, one should inspect the regularized expression prior to the cancellation. In the $\delta$-regularization, the integral that produces these logarithms has the form \cite{Vladimirov:2021hdn}
\begin{eqnarray}
\lim_{\delta\to0}\int_0^1 d\alpha\frac{-\partial_+}{\alpha \partial_+-s\delta}=
-\lim_{\delta\to0}\ln\(\frac{\partial_+-s\delta}{-s\delta}\)=\ln\(\frac{\delta}{-s\partial_+}\).
\end{eqnarray}
Since $\partial_+$ is pure imaginary (and $s$ is defined in eqn.~(\ref{def:L})) this expression is regular and totally determines the phase.
}. %
The complexness of evolution kernel translates into important physics effects discussed in the following sections. The origin of these logarithms are the collinear divergences that appear in the UV divergent diagrams shown in fig.\ref{fig:collinear-diagrams}. In the $\delta$-regularization, the collinear divergence of diagrams (c1) and (c2) are $\sim (C_F-C_A/2)/\epsilon\ln\left(\delta^+/(-s\partial_{z_1}^+)\right)$ and $\sim C_A/(2\epsilon)\ln\left(\delta^+/(-s\partial_{z_2}^+)\right)$, correspondingly. These divergences are cancelled by the collinear divergence of the LP soft factor, $\sim C_F/\epsilon \ln(\delta^+/q^+)$. The logarithms $\ln(q^+/\partial^+)$ are remnants of this cancellation. The  soft-factor is momentum-independent and real, while the momenta of quark and gluon can have any sign. Therefore, for any choice of $q^+$ these logarithms have complex parts. The same takes place for $U_1$ operators.

In the case of $\tilde \gamma_1(z_1)$ and $\tilde \gamma_1(z_2)$ acting to $\Phi_{11}$, the imaginary parts of logarithms have opposite signs, and thus $\Phi_{11}$ has a real-valued evolution kernel. This is the consequence of the translation invariance, which for the forward matrix element\footnote{
The evolution equations (\ref{evol:UV:position}) represent the properties of the TMD operator only, and is valid for any matrix element. In particular, they are also valid for the case of generalized TMD distributions (GTMDs), which are off-forward matrix elements of TMD operators \cite{Meissner:2009ww}.}
implies $(\partial_{z_1}+\partial_{z_2})\widetilde{\Phi}_{11}=0$.

The integral kernel $\mathbb{H}$ in eqn.\eqref{def:gamma_2} is the ordinary evolution kernels for quasi-partonic operators \cite{Bukhvostov:1985rn,Braun:2009vc}. Its action to the correlator reads 
\begin{eqnarray}
&&\mathbb{H}_{z_2z_3}\widetilde{\Phi}_{\mu,12}^{[\Gamma]}(z_1,z_2,z_3)=
\\\nn &&
\qquad
C_A\int_0^1 \frac{d\alpha}{\alpha}\(
\bar \alpha^2 \widetilde{\Phi}_{\mu,12}^{[\Gamma]}(z_1,z_{23}^\alpha,z_3)
+\bar \alpha \widetilde{\Phi}_{\mu,12}^{[\Gamma]}(z_1,z_2,z_{32}^\alpha)
-2\widetilde{\Phi}^{[\Gamma]}_{\mu,12}(z_1,z_2,z_3)
\)
\\\nn &&
\qquad 
+C_A\int_0^1 d\alpha \int_0^{\bar \alpha}d\beta\,\bar \alpha \widetilde{\Phi}_{\nu,12}^{[\Gamma\gamma_\mu\gamma^\nu]}(z_1,z_{23}^\alpha,z_{32}^\beta)
-2\!\(C_F-\frac{C_A}{2}\)\!\int_0^1 d\alpha \!\int_{\bar \alpha}^1d\beta\,\bar \alpha \widetilde{\Phi}_{\nu,12}^{[\Gamma\gamma_\mu\gamma^\nu]}(z_1,z_{23}^\alpha,z_{32}^\beta)
\\\nn &&
\qquad
+\(C_F-\frac{C_A}{2}\) \int_0^1 d\alpha \,\bar \alpha \widetilde{\Phi}_{\nu,12}^{[\Gamma\gamma^\nu\gamma_\mu]}(z_1,z_{32}^\alpha,z_2),
\end{eqnarray}
where the scaling arguments are omitted for brevity,  $\bar \alpha=1-\alpha$ and
\begin{eqnarray}\label{def:zij}
z_{ij}^\alpha=z_i(1-\alpha)+z_j \alpha.
\end{eqnarray}
The action of $\mathbb{H}$ to $\widetilde{\Phi}_{\mu,21}^{[\Gamma]}$ is analogous but the order of Dirac matrices should be reversed. In the appendix \ref{app:collinear}, we provide extra details about the structure of $\mathbb{H}$.

The evolution in the rapidity scale is same for all TMD correlators. It reads
\begin{eqnarray}\nn
\zeta \frac{d}{d\zeta}\widetilde{\Phi}_{11}^{[\Gamma]}(z_1,z_2,b;\mu,\zeta)&=&-\mathcal{D}(b,\mu)\widetilde{\Phi}_{11}^{[\Gamma]}(z_1,z_2,b;\mu,\zeta),
\\\label{evol:rap:position}
\zeta \frac{d}{d\zeta}\widetilde{\Phi}^{[\Gamma]}_{\mu,12}(z_1,z_2,z_3,b;\mu,\zeta)&=&-\mathcal{D}(b,\mu)\widetilde{\Phi}^{[\Gamma]}_{\mu,12}(z_1,z_2,z_3,b;\mu,\zeta),
\\\nn
\zeta \frac{d}{d\zeta}\widetilde{\Phi}^{[\Gamma]}_{\mu,21}(z_1,z_2,z_3,b;\mu,\zeta)&=&-\mathcal{D}(b,\mu)\widetilde{\Phi}^{[\Gamma]}_{\mu,21}(z_1,z_2,z_3,b;\mu,\zeta),
\end{eqnarray}
where $\mathcal{D}$ is the Collins-Soper kernel \cite{Collins:1981uk}, which is sometimes denoted as $\tilde K=-2\mathcal{D}$ \cite{Collins:2011zzd}. The Collins-Soper kernel is known up to NNLO (three-loops) \cite{Vladimirov:2016dll}. The LO expression in the dimensional regularization ($d=4-2\epsilon$) is
\begin{eqnarray}
\mathcal{D}(b,\mu)&=&-2a_sC_F\Big[\Gamma(-\epsilon)\(\frac{-b^2 \mu^2}{4e^{-2\gamma_E}}\)^\epsilon+\frac{1}{\epsilon}\Big]+\mathcal{O}(a_s^2)
\\\nn
&=&2a_sC_F\ln\(\frac{-b^2 \mu^2}{4e^{-2\gamma_E}}\)+\mathcal{O}(a_s^2),
\end{eqnarray}
where in the second line the limit $\epsilon\to0$ is taken. 

The anomalous dimensions $\widetilde{\gamma}$ and the Collins-Soper kernel satisfy the relation
\begin{eqnarray}\label{def:CS-equation}
-\zeta\frac{d }{d\zeta}\left(\widetilde{\gamma}_N(\mu,\zeta)+\widetilde{\gamma}_M(\mu,\zeta)\right)=\mu^2\frac{d }{d\mu^2}\mathcal{D}(b,\mu)=\frac{\Gamma_{\text{cusp}}(\mu)}{2},
\end{eqnarray}
where $N$ and $M$ are 1 or 2, and $\Gamma_{\text{cusp}}$ is the anomalous dimension for the cusp of the light-like Wilson lines. This relation represents the integrability condition of the pair of differential equations (\ref{evol:UV:position}) and (\ref{evol:rap:position}), and guaranties the existence of a common solution.

\subsection{Symmetry properties}
\label{sec:position:sym}

There are three essential symmetry relations for TMD distributions and correlators. They follow from complex conjugation, parity transformation and time-reversal transformation. Their derivation is straightforward and discussed in many articles, see e.g. refs. \cite{Boer:2003cm,Boer:2011xd,Scimemi:2018mmi}. Here, we summarize these relations without derivation. 

The complex conjugation gives
\begin{eqnarray}\nn
[\widetilde{\Phi}_{11}^{[\Gamma]}(z_1,z_2,b)]^*&=&\widetilde{\Phi}_{11}^{[\gamma^0\Gamma^\dagger\gamma^0 ]}(z_2,z_1,-b),
\\\label{sym:complex:corr}
[\widetilde{\Phi}^{[\Gamma]}_{\mu,21}(z_1,z_2,z_3,b)]^*&=&\widetilde{\Phi}_{\mu,12}^{[\gamma^0\Gamma^\dagger\gamma^0 ]}(z_3,z_2,z_1,-b),
\\\nn
[\widetilde{\Phi}^{[\Gamma]}_{\mu,12}(z_1,z_2,z_3,b)]^*&=&\widetilde{\Phi}_{\mu,21}^{[\gamma^0\Gamma^\dagger\gamma^0 ]}(z_3,z_2,z_1,-b).
\end{eqnarray}
Importantly, the complex conjugate of $\widetilde{\Phi}^{[\Gamma]}_{\mu,12}$ is expressed via $\widetilde{\Phi}^{[\Gamma]}_{\mu,21}$, and vise versa. It implies that these correlators cannot be parametrized by real distributions, but only combinations of them can.

Under parity transformation, we have
\begin{eqnarray}\nn
\mathcal{P}\widetilde{\Phi}_{11}^{[\Gamma]}(z_1,z_2,b;p,s,n)\mathcal{P}^{-1}&=&\widetilde{\Phi}_{11}^{[\gamma^0\Gamma\gamma^0 ]}(z_1,z_2,-b;\bar p,-\bar s,\bar n),
\\\label{sym:P:corr}
\mathcal{P}\widetilde{\Phi}^{[\Gamma]}_{\mu,21}(z_1,z_2,z_3,b;p,s,n)\mathcal{P}^{-1}&=&
-\widetilde{\Phi}_{\mu,21}^{[\gamma^0\Gamma\gamma^0 ]}(z_1,z_2,z_3,-b;\bar p,-\bar s,\bar n),
\\\nn
\mathcal{P}\widetilde{\Phi}^{[\Gamma]}_{\mu,12}(z_1,z_2,z_3,b;p,s,n)\mathcal{P}^{-1}&=&
-\widetilde{\Phi}_{\mu,12}^{[\gamma^0\Gamma\gamma^0 ]}(z_1,z_2,z_3,-b;\bar p,-\bar s,\bar n),
\end{eqnarray}
where %$(-1)^\mu=+1$ for $\mu=0$ and $(-1)^\mu=-1$ for $\mu=1,2,3$, 
$\bar p=(p_0,-p_1,-p_2,-p_3)$ and same for $\bar s$. 

The time-reversal transformation is conveniently combined with the parity transformation (which effectively gives charge-conjugation), producing
\begin{eqnarray}\nn
\mathcal{PT}\widetilde{\Phi}_{11}^{[\Gamma]}(z_1,z_2,b;s,L)(\mathcal{PT})^{-1}&=&\widetilde{\Phi}_{11}^{[\gamma^0T\Gamma^*T^{-1}\gamma^0 ]}(-z_2,-z_1,-b;-s,-L),
\\\label{sym:PT:corr}
\mathcal{PT}\widetilde{\Phi}^{[\Gamma]}_{\mu,21}(z_1,z_2,z_3,b;s,L)(\mathcal{PT})^{-1}&=&-\widetilde{\Phi}_{\mu,12}^{[\gamma^0T\Gamma^*T^{-1}\gamma^0 ]}(-z_3,-z_2,-z_1,-b;-s,-L),
\\\nn
\mathcal{PT}\widetilde{\Phi}^{[\Gamma]}_{\mu,12}(z_1,z_2,z_3,b;s,L)(\mathcal{PT})^{-1}&=&-\widetilde{\Phi}_{\mu,21}^{[\gamma^0T\Gamma^*T^{-1}\gamma^0 ]}(-z_3,-z_2,-z_1,-b;-s,-L),
\end{eqnarray}
where $T(\gamma^0)^*T^{-1}=\gamma^0$ and $T(\gamma^i)^*T^{-1}=-\gamma^i$. Note, that the PT transformation preserves the orientation of the light-cone vectors, but changes the position of the light-cone infinity $(L\to-L)$. It effectively connects TMD correlators defined in DY and SIDIS kinematics. Alike the complex conjugation, PT transformation relates DY and SIDIS definition of TMD distributions of the TMD twist-(1,1), but turns TMD-twist-(1,2) to (2,1) (and vise-versa). Therefore, TMD correlators $\Phi_{12}$ and $\Phi_{21}$ do not have definite T-parity, which was pointed out in ref.\cite{Boer:2003cm}.

Finally, being defined by forward matrix elements, the TMD correlators are independent on the global position of the operator, i.e.
\begin{eqnarray}\nn
\widetilde{\Phi}_{11}^{[\Gamma]}(z_1,z_2,b)&=&\widetilde{\Phi}_{11}^{[\Gamma]}(z_1+y,z_2+y,b),
\\\label{sym:shift:corr}
\widetilde{\Phi}^{[\Gamma]}_{\mu,12}(z_1,z_2,z_3,b)&=&\widetilde{\Phi}_{\mu,12}^{[\Gamma]}(z_1+y,z_2+y,z_3+y,b),
\\\nn
\widetilde{\Phi}^{[\Gamma]}_{\mu,21}(z_1,z_2,z_3,b)&=&\widetilde{\Phi}_{\mu,21}^{[\Gamma]}(z_1+y,z_2+y,z_3+y,b),
\end{eqnarray}
where $y$ is an arbitrary number.
%Note that, since $b$ really denotes the difference between the transverse position of the semi-compact operators, we do not have invariance under $b\rightarrow b+c$.

\section{TMD correlators in momentum-fraction space}
\label{sec:TMD-in-mom}

In this section we discuss the TMD correlators in the momentum-fraction space. They are related to distributions in position space by a Fourier transform, and represent distributions of partons with specified collinear momenta. The momentum-fraction distributions are certainly more useful in phenomenology. However, the distributions in position space are simpler theoretically, due to much more compact expressions. Therefore, in the following sections, we often return to the position space definition, even talking about the momentum-fraction representation.

\begin{figure}[t]
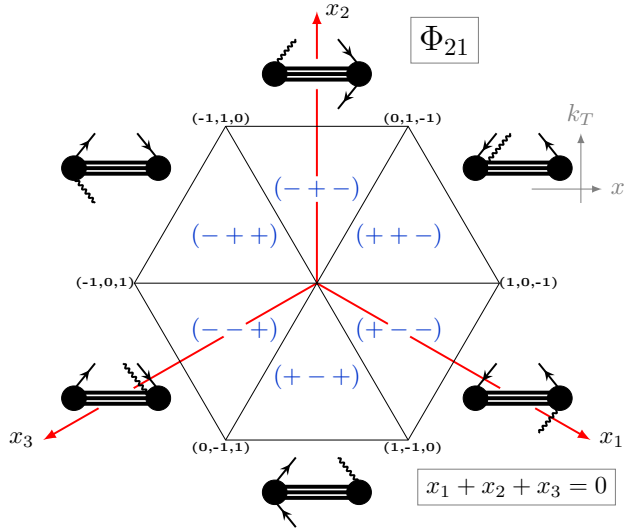

\begin{center}
\includestandalone[width=0.55\textwidth]{Figures/hexagon21}%     without .tex extension
\caption{Support of the TMD correlator $\Phi_{21}$ in the space $x_1,x_2,x_3$. The momentum fractions are constrained to $x_1+x_2+x_3=0$ and $x_i\in[-1,1]$. Each sector of the hexagon has independent interpretation as a process with radiation/absorption of partons with positive collinear momentum fractions. The difference between transverse momenta of quark and anti-quark-gluon pair is $k_T$. The interpretation process for each sector is shown by diagrams, where horizontal/vertical axis is for collinear/transverse momentum. For the correlator $\Phi_{12}$ the picture is analogous, but with the transverse momentum of the gluon reversed.}
\label{fig:hexagon}    
\end{center}
\end{figure}

\subsection{Definition and support properties}

The TMD correlators in momentum-fraction space are defined as
\begin{eqnarray}\nn
\widetilde{\Phi}_{11}^{[\Gamma]}(z_1,z_2,b)&=&p^+\int_{-1}^1 dx e^{ix(z_1-z_2)p^+}\Phi_{11}^{[\Gamma]}(x,b),
\\\label{def:z->x}
\widetilde{\Phi}^{[\Gamma]}_{\mu,21}(z_1,z_2,z_3,b)&=&(p^+)^2\int [dx] e^{-i(x_1z_1+x_2z_2+x_3z_3)p^+}\Phi_{\mu,21}^{[\Gamma]}(x_1,x_2,x_3,b),
\\\nn
\widetilde{\Phi}^{[\Gamma]}_{\mu,12}(z_1,z_2,z_3,b)&=&(p^+)^2\int [dx] e^{-i(x_1z_1+x_2z_2+x_3z_3)p^+}\Phi_{\mu,12}^{[\Gamma]}(x_1,x_2,x_3,b),
\end{eqnarray}
where the integration measure is
\begin{eqnarray}\label{def:support}
\int [dx]=\int_{-1}^1 dx_1 \int_{-1}^1 dx_2 \int_{-1}^1 dx_3 \,\delta(x_1+x_2+x_3).
\end{eqnarray}
The delta-function in the integration measure is required in order to fulfill the translation invariance (\ref{sym:shift:corr}). Note, that Fourier exponents for $\Phi^{[\Gamma]}_{11}$ and $\{\Phi^{[\Gamma]}_{12}, \Phi^{[\Gamma]}_{21}\}$ have different signs. This choice is made in agreement with the usual convention for twist-two and twist-three collinear distributions, see appendix \ref{app:small-b}, or compare to refs.~\cite{Jaffe:1996zw,Belitsky:2000vx,Braun:2009mi}. Also note, that the canonical dimension of $\Phi^{[\Gamma]}$ in the units of mass is $[\Phi^{[\Gamma]}]=0$, while $[\Phi^{[\Gamma]}_{12}]=[\Phi^{[\Gamma]}_{21}]=1$.

The TMD correlators of $\Phi^{[\Gamma]}_{12}$ and $\Phi^{[\Gamma]}_{21}$ depend on two momentum fractions. The three argument notation implicitly implies that $\Phi(x_1,x_2,x_3)$ is defined only for $x_1+x_2+x_3=0$. It is convenient to keep three $x$'s explicit, because in this form TMD correlators obey simple symmetry properties and simpler evolution equations\footnote{
In works \cite{Beneke:2017ztn,Ebert:2021jhy}, an alternative two-variable notation is used with $x$ being the (positive) momentum fraction of a single parton, and $\xi$ being imbalance of momentum fractions in the quark-gluon pair. Such as,
\begin{eqnarray}\label{example:x-xi}
\Phi^{[\Gamma]}_{\mu,12}(x,\xi)_{\text{ref.\cite{Ebert:2021jhy}}}\sim\Phi^{[\Gamma]}_{\mu,12}(-x,(1-\xi)x,\xi x),
\qquad
\Phi^{[\Gamma]}_{\mu,21}(x,\xi)_{\text{ref.\cite{Ebert:2021jhy}}}\sim\Phi^{[\Gamma]}_{\mu,21}(\xi x,(1-\xi)x,-x).
\end{eqnarray} 
In this notation some expressions are a bit shorter (see e.g. expressions for evolution kernels in appendix C of ref.\cite{Vladimirov:2021hdn}), but the support of these distributions is more cumbersome,
\begin{eqnarray}
-\frac{1-x}{x}<\xi<\frac{1}{x}~(\text{for }0<x<1),\qquad
-\frac{1}{x}<\xi<\frac{1+x}{x}~(\text{for }-1<x<0).
\end{eqnarray}
Therefore, manipulations in these variables are generally more involved.
}.
The support (\ref{def:support}) is conveniently drawn in the barycentric coordinate system, where it takes the form of the hexagon \cite{Braun:2009mi,Scimemi:2019gge} shown in fig.\ref{fig:hexagon}. 

Formally, the Fourier transformation between position and momentum-fraction spaces is made by the integration over the infinite domain. However, the correlators $\Phi$ are zero for $|x_i|>1$. This statement is derived following the same route as the one used in ref.\cite{Jaffe:1983hp} for collinear distributions. First, one observes that the T-ordering in eqn.~(\ref{def:Phi-pos}) can be omitted, because the (anti-)commutators of fields are zero due to the causality relation. Next, inserting the compete set of states in-between fields and solving the momentum conservation condition (for $p^+>0$) one finds that correlators vanish at $|x|>1$. In the TMD case (in contrast to the collinear case discussed in ref.\cite{Jaffe:1983hp}) one should also account for the transverse distance/momentum. However, it only makes momentum inequalities sharper.

The same consideration gives rise to the partonic interpretation for correlators, as the probability of emission (for $x>0$) and absorption (for $x<0$) of partons in the target's infinite momentum frame. Ordering of parton fields from positive to negative $x$ (such that energies of partons are positive) allows to identify twist-two TMD correlators in positive/negative domains of $x$ with quark/anti-quarks distributions. The relations are
\begin{eqnarray}\nn
\Phi_{11}^{[\gamma^+]}(x,b)&=&\theta(x)\Phi^{[\gamma^+]}_q(x,b)-\theta(-x)\Phi^{[\gamma^+]}_{\bar q}(-x,b),
-\\\label{eq:q<->anti-q}
\Phi_{11}^{[\gamma^+\gamma^5]}(x,b)&=&\theta(x)\Phi^{[\gamma^+\gamma^5]}_q(x,b)+\theta(-x)\Phi^{[\gamma^+\gamma^5]}_{\bar q}(-x,-b).
\\\nn
\Phi_{11}^{[i\sigma^{\alpha+}\gamma^5]}(x,b)&=&\theta(x)\Phi^{[i\sigma^{\alpha+}\gamma^5]}_q(x,b)-\theta(-x)\Phi^{[i\sigma^{\alpha+}\gamma^5]}_{\bar q}(-x,-b),
\end{eqnarray}
where $\theta(x)$ is a theta-function. The relative minus sign is due to the sign produced by the charge-conjugation \cite{Mulders:1995dh}. The distributions $\Phi_{q}$ and $\Phi_{\bar q}$ are defined for $0<x<1$. These distributions are entirely independent, and do not mix under evolution.

The interpretation of twist-three correlators is more cumbersome. They are interpreted as amplitudes of processes with radiation/absorption of a parton versus a parton pair. For example, the segment $x_1<0$ and $x_2,x_3>0$ can be interpreted as the radiation of a quark-gluon pair and absorption of a quark, whereas the segment $x_2>0$ and $x_1,x_3<0$ can be interpreted as the radiation of a gluon and absorption of a quark-anti-quark pair. This is analogous to collinear distributions \cite{Jaffe:1983hp}.
The TMD correlators additionally store the information about the transverse momentum of emitted/absorbed partons. The transverse momentum $k_T$ (Fourier conjugated to $b$) is the relative transverse momentum of the quark and the quark-gluon pair, irrespectively their $x$'s signs. Therefore, distributions of transverse momentum is different for $\Phi^{[\Gamma]}_{12}$ and $\Phi^{[\Gamma]}_{21}$ in any segment of $x$'s. For example, in the segment $x_1<0$ and $x_2,x_3>0$, the transverse momentum $k_T$ is the difference between momenta the quark-gluon pair and the quark for the correlator $\Phi_{12}^{[\Gamma]}$, while it is the difference between momenta of the anti-quark-gluon pair and the anti-quark for the correlator $\Phi_{21}^{[\Gamma]}$. In total, there are six interpretation regions for each correlator $\Phi^{[\Gamma]}_{12}$ and $\Phi^{[\Gamma]}_{21}$, which are schematically shown in fig.\ref{fig:hexagon}. These twelve combinations of $x$'s and $k_T$ shows all possible quark-gluon-quark-transition processes with a single measured $k_T$.

The involved interpretation makes impossible to identify quark and anti-quark twist-three distributions in a similar way to \eqref{eq:q<->anti-q}. However, one can introduce a similar notation, which also helps conveying similar properties. We define
\begin{eqnarray}\label{eq:q<->anti-q-tw-3}
\Phi_{21}^{[\Gamma]}(x_1,x_2,x_3,b)&=&\theta(x_3)\Phi^{[\Gamma]}_{q,21}(x_1,x_2,x_3,b)\pm\theta(-x_3)\Phi^{[\Gamma]}_{\bar q,21}(x_1,x_2,-x_3,-b),
\\\nn
\Phi_{12}^{[\Gamma]}(x_1,x_2,x_3,b)&=&\theta(x_1)\Phi^{[\Gamma]}_{q,12}(x_1,x_2,x_3,b)\pm\theta(-x_1)\Phi^{[\Gamma]}_{\bar q,12}(-x_1,x_2,x_3,-b),
\end{eqnarray}
where the sign $\pm$ is selected in accordance to $\Gamma$ in the same way as in eqn.\eqref{eq:q<->anti-q}. The distributions $\Phi_{q}$ and $\Phi_{\bar q}$ do not mix with each other under evolution (this statement is proved in the following section), and therefore, they are entirely independent. However, the signs of momenta of any parton pair are not fixed, and they mixes in the evolution. Therefore, any further decomposition is not practical. The splitting (\ref{eq:q<->anti-q-tw-3}) preserves the general picture of physical process and naturally applied to many practical cases (see e.g. sec.~\ref{sec:bi-quark})). For the sake of compactness of results, without loss of generality we will not employ the decompositon \eqref{eq:q<->anti-q-tw-3}.

The inverse transformation (from position to momentum-fraction spaces) requires fixation of the global position of the correlators. For example, fixing the position of the quark-field at the origin, one gets
\begin{eqnarray}\nn
\Phi^{[\Gamma]}_{11}(x,b)&=&\int_{-\infty}^\infty \frac{dz}{2\pi} e^{-ix zp^+}\widetilde{\Phi}_{11}^{[\Gamma]}(z,0,b),
\\\label{def:x->z}
\Phi_{\mu,21}^{[\Gamma]}(x_1,x_2,x_3,b)&=&\int_{-\infty}^\infty \frac{dz_1 dz_2}{(2\pi)^2} e^{i(x_1z_1+x_2z_2)p^+}\widetilde{\Phi}^{[\Gamma]}_{\mu,21}(z_1,z_2,0,b),
\\\nn
\Phi_{\mu,12}^{[\Gamma]}(x_1,x_2,x_3,b)&=&\int_{-\infty}^\infty \frac{dz_1 dz_2}{(2\pi)^2} e^{i(x_1z_1+x_2z_2)p^+}\widetilde{\Phi}^{[\Gamma]}_{\mu,12}(z_1,z_2,0,b),
\end{eqnarray}
where $x_3=-x_1-x_2$. If a distribution with a different global position appears, it is always possible to shift it using  eqn.~(\ref{sym:shift:corr}).

We emphasize that negative values of momentum fractions are physical and contribute to the factorized expression. We also stress that the evolution equation (see the next section) mixes contributions of different sectors. This effect is well-known in the case of collinear distributions of twist-three and higher, see e.g. discussion in ref.\cite{Belitsky:2005qn,Braun:2009mi}. However, in contrast to collinear twist-three distributions, the TMD distributions are not definite at points $x_i=0$. This point is elaborated on in details in the following sections. 

\subsection{Evolution equations}

\begin{figure}[t]
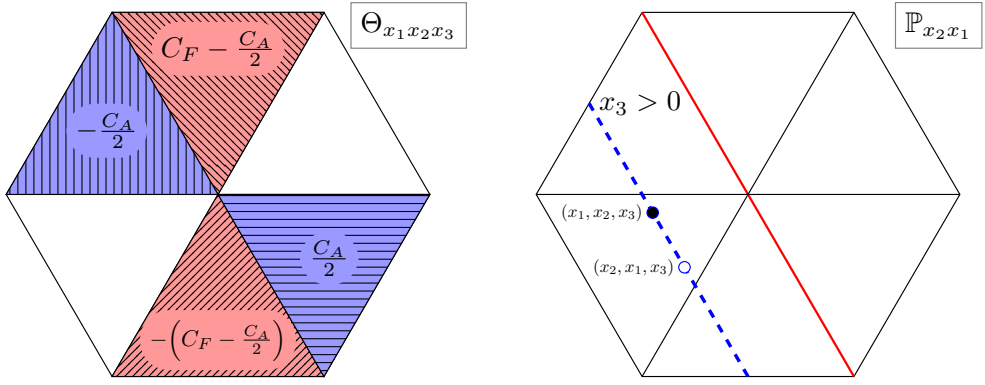

\begin{center}
\includestandalone[width=0.75\textwidth]{Figures/theta_kernel}%     without .tex extension
\caption{Illustration for the elements of the evolution kernel. Left panel: the function $\Theta_{x_1x_2x_3}$ defined in eqn.~(\ref{def:Theta123}). The same shading of the segment corresponds to the same coefficient. The LO value for the coefficient is written inside each sector. Right panel: example of the integration paths for the evolution kernel $\mathbb{P}_{x_2x_1}$ in the barycentric coordinates. The black and white dot represent the coordinates of the special points of the integral kernels ($v=0$). The black one for the `natural' ordered coordinates. The white one for the exchanged quark and gluon momenta, which take place in the last lines of eqns. (\ref{def:PA}, \ref{def:PB}).}
\label{fig:thetas}    
\end{center}
\end{figure}

The pair of evolution equations for the LP TMD correlator is well known (see \cite{Aybat:2011zv,Echevarria:2011epo}):
\begin{eqnarray}\label{evol:corr:11}
\mu^2 \frac{d}{d\mu^2}\Phi^{[\Gamma]}_{11}(x,b;\mu,\zeta)&=&\(\frac{\Gamma_{\text{cusp}}}{2}\ln\(\frac{\mu^2}{\zeta}\)-\frac{\gamma_V}{2}\)\Phi^{[\Gamma]}_{11}(x,b;\mu,\zeta),
\\\nn
\zeta \frac{d}{d\zeta}\Phi^{[\Gamma]}_{11}(x,b;\mu,\zeta)&=&-\mathcal{D}(b,\mu)\Phi^{[\Gamma]}_{11}(x,b;\mu,\zeta),
\end{eqnarray}
where 
%$\Gamma_{\text{cusp}}$ is the cusp anomalous dimension for the light-like Wilson lines, and 
$\gamma_V$ is the anomalous dimension of the quark vector form factor. At LO one has
\begin{eqnarray}\label{eqn:gammaCusp-LO}
\Gamma_{\text{cusp}}&=&4C_F\,a_s+\mathcal{O}(a_s^2),\qquad \gamma_V=-6C_F\,a_s+\mathcal{O}(a_s^2).
\end{eqnarray}
The higher perturbative orders can be found, e.g. in ref.\cite{Echevarria:2016scs}.

The evolution equation for the TMD correlators of twist-three are more complicated. The expression for the UV evolution kernels is computed by
\begin{eqnarray}\label{eqn:tranform-UV}
&&\delta(x_1+x_2+x_3)\gamma_N\otimes \Phi(x_1,x_2,x_3,b)
\\\nn&&\qquad =(p^+)^3\int_{-\infty}^\infty \frac{dz_1dz_2dz_3}{(2\pi)^3}e^{i(x_1z_1+x_2z_2+x_3z_3)p^+}\int [dy]\widetilde{\gamma}_N\otimes e^{-i(y_1z_1+y_2z_2+y_3z_3)p^+}\Phi(y_1,y_2,y_3,b),
\end{eqnarray}
where $\otimes$ are integral convolutions. The resulting equations are conveniently written in the form
\begin{eqnarray}\nn
\mu^2 \frac{d}{d\mu^2}\Phi_{\mu,21}^{[\Gamma]}&=&
\(\frac{\Gamma_{\text{cusp}}}{2}\ln\(\frac{\mu^2}{\zeta}\)+\Upsilon_{x_1x_2x_3}+2\pi i \, s \, \Theta_{x_1x_2x_3}\) \Phi_{\mu,21}^{[\Gamma]}
\\ &&\qquad
+\mathbb{P}^{A}_{x_2x_1}\otimes \Phi_{\nu,21}^{[\gamma^{\nu}\gamma^{\mu}\Gamma]}
+\mathbb{P}^{B}_{x_2x_1}\otimes \Phi_{\nu,21}^{[\gamma^\mu\gamma^\nu\Gamma]}
,
\\\nn
\mu^2 \frac{d}{d\mu^2}\Phi_{\mu,12}^{[\Gamma]}&=&
\(\frac{\Gamma_{\text{cusp}}}{2}\ln\(\frac{\mu^2}{\zeta}\)+\Upsilon_{x_3x_2x_1}+2\pi i \, s \, \Theta_{x_3x_2x_1}\) \Phi_{\mu,12}^{[\Gamma]}
\\\nn &&\qquad
+\mathbb{P}^{A}_{x_2x_3}\otimes \Phi_{\nu,12}^{[\Gamma\gamma^{\mu}\gamma^{\nu}]}
+\mathbb{P}^{B}_{x_2x_3}\otimes \Phi_{\nu,12}^{[\Gamma\gamma^\nu\gamma^\mu]}
,
\end{eqnarray}
where we suppressed the argument $(x_1,x_2,x_3,b;\mu,\zeta)$ of the TMD correlators for brevity. 
%$\Gamma_{\text{cusp}}$ is the cusp anomalous dimension for the light-like Wilson lines, and $\otimes$ is the integral convolution.
The elements $\mathbb{P}$, $\Upsilon$ and $\Theta$ are defined below. The coefficient in front of the $\ln\zeta$ (i.e. $\Gamma_{\text{cusp}}$) is fixed by the integrability condition (\ref{def:CS-equation}) and at LO is given in eqn.~(\ref{eqn:gammaCusp-LO}). Although this equation is derived at LO, it is clear that the same pattern of kernels is preserved at all orders, since it is the most general pattern that can be written.

The function $\Upsilon$ incorporates the terms that multiply TMD correlator without convolution. The LO expression is
\begin{eqnarray}\label{def:Upsilon123}
\Upsilon_{x_1x_2x_3}&=&a_s
\Big[3 C_F+C_A\ln\(\frac{|x_3|}{|x_2|}\)+2\(C_F-\frac{C_A}{2}\)\ln\(\frac{|x_3|}{|x_1|}\)\Big]+\mathcal{O}(a_s^2).
\end{eqnarray}
The expression for $\Upsilon_{x_3x_2x_1}$ is obtained from $\Upsilon_{x_1x_2x_3}$ replacing $x_1\leftrightarrow x_3$. To obtain this function we used the convention (\ref{def:q+}). For arbitrary $q^+$ the function $\Upsilon$ reads
\begin{eqnarray}\label{def:Upsilon123:q+}
\Upsilon_{x_1x_2x_3}&=&\Gamma_{\text{cusp}}\ln\(\frac{q^+}{p^+}\)
%+a_s\Big[3 C_F-2C_F \ln|x_3|-C_A\ln|x_2|-2\(C_F-\frac{C_A}{2}\)\ln|x_1|\Big]+\mathcal{O}(a_s^2),
+a_s\Big[3 C_F-2C_F \ln|x_1x_3|-C_A\ln\(\frac{|x_2|}{|x_1|}\)\Big]+\mathcal{O}(a_s^2),
\end{eqnarray}
where the dependence on $q^+/p^+$ is accumulated in the first term.

The function $\Theta$ results from the complex parts of the logarithms $\ln(q^+/\partial_+)$, whose real parts are collected into $\Upsilon$'s. The values of $\Theta$ depend on the segment of the support domain. At LO, one has
\begin{eqnarray}\label{def:Theta123}
\Theta_{x_1x_2x_3}&=&a_s\times \left\{
\begin{array}{lcc}
\frac{C_A}{2} & x_{1,2,3}\in & (+,-,-),
\\
-\(C_F-\frac{C_A}{2}\) & x_{1,2,3}\in & (+,-,+),
\\
0 & x_{1,2,3}\in & (-,-,+),
\\
-\frac{C_A}{2} & x_{1,2,3}\in & (-,+,+),
\\
C_F-\frac{C_A}{2} & x_{1,2,3}\in & (-,+,-),
\\
0 & x_{1,2,3}\in & (+,+,-),
\end{array}
\right.%}
+\mathcal{O}(a_s^2),
\end{eqnarray}
where $x_{1,2,3}\in (+,-,-)$ indicates the sector $\{x_1>0,x_2<0,x_3<0\}$ (see also fig.\ref{fig:hexagon}) and similar for other cases. The function $\Theta_{x_3x_2x_1}$ is obtained from $\Theta_{x_1x_2x_3}$ with replacement $x_1\leftrightarrow x_3$. The visual representation of $\Theta_{x_1x_2x_3}$ is given in fig.\ref{fig:thetas}. 

The fact that the evolution kernel has a complex part is worrisome. However, the evolved function is complex-valued and thus it only indicates that the real and complex parts of TMD correlators evolve by different evolution kernels. The  physical observables, that are real, are expressed via the real-valued combinations of TMD distributions (see the next section). Such combinations are evolved by real kernels. We emphasize that the complex part of the evolution kernel depends on the process, via the sign-variable $s$. To our best knowledge, it is the first observation of the process-dependence in the evolution equation.

The evolution kernels $\mathbb{P}$ in momentum-fraction space are obtained from eqn.~(\ref{evol:UV:position}) via the transformation (\ref{eqn:tranform-UV}). They read
\begin{eqnarray}\label{def:PA}
&&\mathbb{P}^{A}_{x_2x_1}\otimes \Phi(x_1,x_2,x_3)=-\frac{a_s}{2}\bigg\{
\delta_{x_20}\,C_A\Phi(x_1,0,x_3)
\\\nn &&\qquad
+C_A\int_{-\infty}^\infty dv \Big[\frac{x_2}{v}[(v+x_2)\Phi(x_1,x_2,x_3)-x_2\Phi(x_1-v,x_2+v,x_3)]\frac{\theta(v,x_2)-\theta(-v,-x_2)}{(v+x_2)^2}
\\\nn && \qquad\qquad-
\frac{x_1}{v}(\Phi(x_1,x_2,x_3)-\Phi(x_1-v,x_2+v,x_3))\frac{\theta(v,-x_1)-\theta(-v,x_1)}{v-x_1}\Big]
\\\nn &&
\qquad
-C_A\int_{-\infty}^{\infty}dv\Big[\frac{x_2^2(v+2x_2+x_1)}{(x_1+x_2)^2}\frac{\theta(v,x_2)-\theta(-v,-x_2)}{(v+x_2)^2}
\\\nn &&\qquad\qquad+\frac{x_1(2x_2+x_1)}{(x_1+x_2)^2}\frac{\theta(v,-x_1)-\theta(-v,x_1)}{v-x_1}\Big]\Phi(x_1-v,x_2+v,x_3)
\\\nn &&
\qquad
+2\(C_F-\frac{C_A}{2}\)\int_{-\infty}^{\infty}dv
\Big[\frac{x_2^2}{(x_1+x_2)^2}\frac{\theta(v,x_2)-\theta(-v,-x_2)}{v+x_2}
\\\nn &&
\qquad\qquad
-\frac{x_1(x_1x_2-2vx_2-vx_1)}{(x_1+x_2)^2}
\frac{\theta(v,-x_1)-\theta(-v,x_1)}{(v-x_1)^2}\Big]\Phi(x_2+v,x_1-v,x_3)
\bigg\}
\\\nn &&\qquad+\mathcal{O}(a_s^2),
\end{eqnarray}
\begin{eqnarray}\label{def:PB}
&&\mathbb{P}^{B}_{x_2x_1}\otimes \Phi(x_1,x_2,x_3)=-\frac{a_s}{2}\bigg\{
\delta_{x_20}\,2(C_A-C_F)\Phi(x_1,0,x_3)
\\\nn &&\qquad
+C_A\int_{-\infty}^\infty dv \Big[\frac{x_2}{v}[(v+x_2)\Phi(x_1,x_2,x_3)-x_2\Phi(x_1-v,x_2+v,x_3)]\frac{\theta(v,x_2)-\theta(-v,-x_2)}{(v+x_2)^2}
\\\nn && \qquad\qquad+
\frac{x_1}{v}(\Phi(x_1,x_2,x_3)-\Phi(x_1-v,x_2+v,x_3))\frac{\theta(v,-x_1)-\theta(-v,x_1)}{v-x_1}\Big]
\\\nn &&
\qquad
+2\(C_F-\frac{C_A}{2}\)\int_{-\infty}^{\infty}dv
\,x_1\Phi(x_2+v,x_1-v,x_3)\frac{\theta(v,-x_1)-\theta(-v,x_1)}{(v-x_1)^2}
\bigg\}+\mathcal{O}(a_s^2).
\end{eqnarray}
In these expressions
\begin{eqnarray}
\theta(v,x)=\left\{
\begin{array}{ll}
1, & v>0 \text{ and } x>0\\
0, & v\leqslant 0 \text{ or }x\leqslant 0,
\end{array}\right.%}
\qquad
\delta_{x0}=\left\{
\begin{array}{ll}
1, & x=0\\
0, & x\neq 0.
\end{array}\right.%}
\end{eqnarray}
The kernels $\mathbb{P}_{x_2x_3}$ are defined analogously with exchange $x_1\leftrightarrow x_3$ and $\Phi(x_1,x_2,x_3)\to \Phi(x_3,x_2,x_1)$. The integration paths in the barycentric coordinates are shown in fig. \ref{fig:thetas}. The kernels $\mathbb{P}_{x_2x_1}$ are \textit{regular and continuous} at $x_1=0$ or $x_2=0$. Also they preserve the value of  $x_1+x_2=-x_3$ at each point of the integral, and therefore, the values of function at $x_1+x_2>0$ ($x_3<0$) does not mix with the values at $x_1+x_2<0$ ($x_3>0$). This allows identification of quark- and anti-quark components in eqn.~(\ref{eq:q<->anti-q-tw-3}). 
However, the signs of $x_1$ and $x_2$ are not preserved individually.

The evolution equations with respect to rapidity scales are the same as in the position space (\ref{evol:rap:position}):
\begin{eqnarray}
\label{evol:rap:momentum}
\zeta \frac{d}{d\zeta}\Phi^{[\Gamma]}_{\mu,12}(x_1,x_2,x_3,b;\mu,\zeta)&=&-\mathcal{D}(b,\mu)\Phi^{[\Gamma]}_{\mu,12}(x_1,x_2,x_3,b;\mu,\zeta),
\\\nn
\zeta \frac{d}{d\zeta}\Phi^{[\Gamma]}_{\mu,21}(x_1,x_2,x_3,b;\mu,\zeta)&=&-\mathcal{D}(b,\mu)\Phi^{[\Gamma]}_{\mu,21}(x_1,x_2,x_3,b;\mu,\zeta).
\end{eqnarray}

\subsection{Symmetry properties}

The symmetry relations for TMD correlators in momentum-fraction space are obtained by substituting definition (\ref{def:z->x}) into relations listed in sec.~\ref{sec:position:sym}. The complex conjugation relations (\ref{sym:complex:corr}) give
\begin{eqnarray}\nn
[\Phi^{[\Gamma]}_{11}(x,b)]^*&=&\Phi^{[\gamma^0\Gamma^\dagger\gamma^0 ]}_{11}(x,-b),
\\\label{sym:complex:corr-momentum}
[\Phi^{[\Gamma]}_{\mu,12}(x_1,x_2,x_3,b)]^*&=&\Phi_{\mu,21}^{[\gamma^0\Gamma^\dagger\gamma^0 ]}(-x_3,-x_2,-x_1,-b),
\\\nn
[\Phi^{[\Gamma]}_{\mu,21}(x_1,x_2,x_3,b)]^*&=&\Phi_{\mu,12}^{[\gamma^0\Gamma^\dagger\gamma^0 ]}(-x_3,-x_2,-x_1,-b).
\end{eqnarray}
The parity transformation relations (\ref{sym:P:corr}) give
\begin{eqnarray}\nn
\mathcal{P}\Phi^{[\Gamma]}_{11}(x,b;p,s,n)\mathcal{P}^{-1}&=&\Phi_{11}^{[\gamma^0\Gamma\gamma^0 ]}(x,-b;\bar p,-\bar s,\bar n),
\\\label{sym:P:corr-momentum}
\mathcal{P}\Phi^{[\Gamma]}_{\mu,12}(x_1,x_2,x_3,b;p,s,n)\mathcal{P}^{-1}&=&
-\Phi_{\mu,12}^{[\gamma^0\Gamma\gamma^0 ]}(x_1,x_2,x_3,-b;\bar p,-\bar s,\bar n),
\\\nn
\mathcal{P}\Phi^{[\Gamma]}_{\mu,21}(x_1,x_2,x_3,b;p,s,n)\mathcal{P}^{-1}&=&
-\Phi_{\mu,21}^{[\gamma^0\Gamma\gamma^0 ]}(x_1,x_2,x_3,-b;\bar p,-\bar s,\bar n).
\end{eqnarray}
The PT-transformation relations (\ref{sym:PT:corr}) give
\begin{eqnarray}\nn
\mathcal{PT}\Phi_{11}^{[\Gamma]}(x,b;s,L)(\mathcal{PT})^{-1}&=&\Phi_{11}^{[\gamma^0T\Gamma^*T^{-1}\gamma^0 ]}(x,-b;-s,-L),
\\\label{sym:PT:corr-momentum}
\mathcal{PT}\Phi^{[\Gamma]}_{\mu,12}(x_1,x_2,x_3,b;s,L)(\mathcal{PT})^{-1}&=&-\Phi_{\mu,21}^{[\gamma^0T\Gamma^*T^{-1}\gamma^0 ]}(-x_3,-x_2,-x_1,-b;-s,-L),
\\\nn
\mathcal{PT}\Phi^{[\Gamma]}_{\mu,21}(x_1,x_2,x_3,b;s,L)(\mathcal{PT})^{-1}&=&-\Phi_{\mu,12}^{[\gamma^0T\Gamma^*T^{-1}\gamma^0 ]}(-x_3,-x_2,-x_1,-b;-s,-L).
\end{eqnarray}
The translation invariance (\ref{sym:shift:corr}) is already accounted in the definition of TMD correlator in momentum-fraction space since it is implicitly assumed that $x_1+x_2+x_3=0$.

\section{TMD distributions with definite T-parity}
\label{sec:TMD-Tdefinite}

The main drawback of the definite-TMD-twist TMD distributions is their complex-valued evolution kernels. Also, the TMD correlators $\Phi_{\mu,12}^{[\Gamma]}$ and $\Phi_{\mu,21}^{[\Gamma]}$ do not satisfy simple rules of complex conjugation (\ref{sym:complex:corr-momentum}) and have indefinite T-parity (\ref{sym:PT:corr-momentum}). For that reason they are impractical. In this section, we introduce combinations more suited for practical applications. The new combinations remove the issues above at the price of more involved evolution equations. We refer to these combinations as TMD distributions with definite T-parity.

\subsection{TMD correlators with definite T-parity}

We define two independent combinations of the definite-TMD-twist TMD distributions,
\begin{eqnarray}\label{def:T-definite}
\Phi^{[\Gamma]}_{\mu,\oplus}(x_1,x_2,x_3,b)&=&\frac{\Phi^{[\Gamma]}_{\mu,21}(x_1,x_2,x_3,b)+\Phi^{[\Gamma]}_{\mu,12}(-x_3,-x_2,-x_1,b)}{2},
\\\nn
\Phi^{[\Gamma]}_{\mu,\ominus}(x_1,x_2,x_3,b)&=&i\frac{\Phi^{[\Gamma]}_{\mu,21}(x_1,x_2,x_3,b)-\Phi^{[\Gamma]}_{\mu,12}(-x_3,-x_2,-x_1,b)}{2},
\end{eqnarray}
where we omit arguments $(\mu,\zeta)$ for brevity. Note that the combinations involves momentum-fractions with opposite signs, but they have the same transverse separation.
Such correlators lack a simple partonic interpretation. Since the renormalization of $\Phi_{12}$ and $\Phi_{21}$ operators is independent, the renormalization of $\Phi_{\oplus}$ and $\Phi_{\ominus}$ is convoluted. For this reason, the correlators (\ref{def:T-definite}) cannot be presented as a matrix element of a single bare operator. Nonetheless, these combinations are the ones that appear in the practical applications \cite{Boer:2003cm,Bacchetta:2006tn,Vladimirov:2021hdn}.

The symmetry relations for TMD distributions with definite T-parity resemble the relations for the leading-twist TMD correlator. The complex conjugation gives
\begin{eqnarray}\label{sym:complex:corr-Tdef}
[\Phi^{[\Gamma]}_{\mu,\oplus}(x_1,x_2,x_3,b)]^*&=&\Phi_{\mu,\oplus}^{[\gamma^0\Gamma^\dagger \gamma^0]}(x_1,x_2,x_3,-b),
\\\nn
[\Phi^{[\Gamma]}_{\mu,\ominus}(x_1,x_2,x_3,b)]^*&=&\Phi_{\mu,\ominus}^{[\gamma^0\Gamma^\dagger \gamma^0]}(x_1,x_2,x_3,-b).
\end{eqnarray}
The parity transformation gives
\begin{eqnarray}\nn
\label{sym:P:corr-Tdef}
\mathcal{P}\Phi^{[\Gamma]}_{\mu,\oplus}(x_1,x_2,x_3,b;p,s,n)\mathcal{P}^{-1}&=&
-\Phi_{\mu,\oplus}^{[\gamma^0\Gamma\gamma^0 ]}(x_1,x_2,x_3,-b;\bar p,-\bar s,\bar n),
\\\nn
\mathcal{P}\Phi^{[\Gamma]}_{\mu,\ominus}(x_1,x_2,x_3,b;p,s,n)\mathcal{P}^{-1}&=&
-\Phi_{\mu,\ominus}^{[\gamma^0\Gamma\gamma^0 ]}(x_1,x_2,x_3,-b;\bar p,-\bar s,\bar n).
\end{eqnarray}
The PT-transformation gives
\begin{eqnarray}\nn
\\\label{sym:PT:corr-T-def}
\mathcal{PT}\Phi^{[\Gamma]}_{\mu,\oplus}(x_1,x_2,x_3,b;s,L)(\mathcal{PT})^{-1}&=&-\Phi_{\mu,\oplus}^{[\gamma^0T\Gamma^*T^{-1}\gamma^0 ]}(x_1,x_2,x_3,-b;-s,-L),
\\\nn
\mathcal{PT}\Phi^{[\Gamma]}_{\mu,\ominus}(x_1,x_2,x_3,b;s,L)(\mathcal{PT})^{-1}&=&+\Phi_{\mu,\ominus}^{[\gamma^0T\Gamma^*T^{-1}\gamma^0 ]}(x_1,x_2,x_3,-b;-s,-L).
\end{eqnarray}
Note that, due to the composition of momentum fraction $x$'s in the definition (\ref{def:T-definite}), the form of the split into functions with quark and anti-quark labels (\ref{eq:q<->anti-q-tw-3}) is preserved.

The evolution equations for $\Phi_{\oplus}$ and $\Phi_{\ominus}$ are
\begin{eqnarray}\nn
\mu^2 \frac{d}{d\mu^2}\Phi^{\mu [\Gamma]}_{\oplus}&=&
\(\frac{\Gamma_{\text{cusp}}}{2}\ln\(\frac{\mu^2}{\zeta}\)+\Upsilon_{x_1x_2x_3}\)\Phi^{\mu [\Gamma]}_{\oplus}
+2\pi s\,\Theta_{x_1x_2x_3}\Phi^{\mu [\Gamma]}_{\ominus}
\\ &&\label{eqn:evol-correlator+}
+\mathbb{P}^{A}_{x_2x_1}\otimes \(
\Phi_{\nu,\oplus}^{\[\frac{1}{2}(\Gamma\gamma^\mu\gamma^\nu+\gamma^{\nu}\gamma^{\mu}\Gamma)\]}
+\Phi_{\nu,\ominus}^{\[\frac{i}{2}(\Gamma\gamma^\mu\gamma^\nu-\gamma^{\nu}\gamma^{\mu}\Gamma)\]}
\)
\\\nn && 
+\mathbb{P}^{B}_{x_2x_1}\otimes \(
\Phi_{\nu,\oplus}^{\[\frac{1}{2}(\Gamma\gamma^\nu\gamma^\mu+\gamma^{\mu}\gamma^{\nu}\Gamma)\]}
+\Phi_{\nu,\ominus}^{\[\frac{i}{2}(\Gamma\gamma^\nu\gamma^\mu-\gamma^{\mu}\gamma^{\nu}\Gamma)\]}
\)
\\\nn
\mu^2 \frac{d}{d\mu^2}\Phi^{\mu [\Gamma]}_{\ominus}&=&
\(\frac{\Gamma_{\text{cusp}}}{2}\ln\(\frac{\mu^2}{\zeta}\)+\Upsilon_{x_1x_2x_3}\)\Phi^{\mu [\Gamma]}_{\ominus}
-2\pi s\,\Theta_{x_1x_2x_3}\Phi^{\mu [\Gamma]}_{\oplus}
\\ &&\label{eqn:evol-correlator-}
+\mathbb{P}^{A}_{x_2x_1}\otimes \(
\Phi_{\nu,\ominus}^{\[\frac{1}{2}(\Gamma\gamma^\mu\gamma^\nu+\gamma^{\nu}\gamma^{\mu}\Gamma)\]}
-\Phi_{\nu,\oplus}^{\[\frac{i}{2}(\Gamma\gamma^\mu\gamma^\nu-\gamma^{\nu}\gamma^{\mu}\Gamma)\]}
\)
\\\nn && 
+\mathbb{P}^{B}_{x_2x_1}\otimes \(
\Phi_{\nu,\ominus}^{\[\frac{1}{2}(\Gamma\gamma^\nu\gamma^\mu+\gamma^{\mu}\gamma^{\nu}\Gamma)\]}
-\Phi_{\nu,\oplus}^{\[\frac{i}{2}(\Gamma\gamma^\nu\gamma^\mu-\gamma^{\mu}\gamma^{\nu}\Gamma)\]}
\),
\end{eqnarray}
where it was used that $\Upsilon_{-x_1-x_2-x_3}=\Upsilon_{x_1x_2x_3}$, $\Theta_{-x_1-x_2-x_3}=-\Theta_{x_1x_2x_3}$, 
$[\mathbb{P}_{x_2x_3}\otimes\Phi](-x_3,-x_2,-x_1)$ $=[\mathbb{P}_{x_2x_1}\otimes \Phi(-x_3,-x_2,-x_1)]$. 

In this representation the evolution kernels are real. An amazing feature of these evolution equations is that they mix the functions with different T-parity. However, the mixing terms are proportional to $s$, which changes sign under $L\to-L$. Thus the T-parity of each correlator is preserved, and it remains process independent (apart of trivial sign-change for T-odd functions). To our best knowledge, it is the first example of such behaviour.

\subsection{Parameterization}
\label{sec:TMD-parameterization}

\begin{table}[tb]
\begin{center}
\begin{tabular}{|c||c|c|c|c|c|}
\hline
& U & L & T$_{J=0}$ & T$_{J=1}$ & T$_{J=2}$
\\\hline
U & $f^\perp_\bullet$ & $g^\perp_\bullet$ & & $h_\bullet$ & $h^\perp_\bullet$
\\\hline
L & $f_{\bullet L}^\perp$ & $g_{\bullet L}^\perp$ & $h_{\bullet L}$ & & $h_{\bullet L}^\perp$
\\\hline
T & $f_{\bullet T},\quad f_{\bullet T}^\perp$ & $g_{\bullet T},\quad g_{\bullet T}^\perp$ & $h_{\bullet T}^{D\perp}$ & $h_{\bullet T}^{A\perp}$ & $h_{\bullet T}^{S\perp},\quad h_{\bullet T}^{T\perp}$
\\\hline
\end{tabular}
\caption{\label{tab:tw3} Quark TMD distributions of twist-three sorted with respect to polarization properties of both the operator (columns) and the hadron (rows). The labels U, H, and T are for the unpolarized, longitudinal and transverse polarizations. The subscript $J$ differentiates different angular momentum for the transversely-polarized case. The bullet $\bullet$  stands for the $\oplus$, $\ominus$ labels.}
\end{center}
\end{table}

The discussion is matured enough to allow us to introduce the parameterization for the TMD correlators in the terms of TMD distributions. For each TMD correlator we write all possible spin and tensor structures in accordance to its parity and dimension.  As an elements of contraction one can use the vectors $b^\mu$ and $s^\mu$, and the tensors $g_T^{\mu\nu}$ and $\epsilon^{\mu\nu}_T$. The spin vector is spit into the longitudinal and transverse projections
\begin{eqnarray}
s^\mu=\lambda\frac{p^- n^\mu-p^+ \bar n^\mu}{M}+s_T^\mu,
\end{eqnarray}
where $M$ is the mass of the hadron. It implies $\lambda=M s^+/p^+$.

The standard parameterization of the leading twist TMD correlators has been carried out in ref.~\cite{Mulders:1995dh}. We present this parameterization here for completeness
\begin{eqnarray}\label{def:TMDs:1:g+}
\Phi^{[\gamma^+]}(x,b)&=&f_1(x,b)+i\epsilon^{\mu\nu}_T b_\mu s_{T\nu}M f_{1T}^\perp(x,b),
\\\label{def:TMDs:1:g+5}
\Phi^{[\gamma^+\gamma^5]}(x,b)&=&\lambda g_{1}(x,b)+i(b \cdot s_T)M g_{1T}(x,b),
\\\label{def:TMDs:1:s+}
\Phi^{[i\sigma^{\alpha+}\gamma^5]}(x,b)&=&s_T^\alpha h_{1}(x,b)-i\lambda b^\alpha M h_{1L}^\perp(x,b)
\\\nn && +i\epsilon^{\alpha\mu}b_\mu M h_1^\perp(x,b)-\frac{M^2 b^2}{2}\(\frac{g_T^{\alpha\mu}}{2}-\frac{b^\alpha b^\mu}{b^2}\)s_{T\mu}h_{1T}^\perp(x,b),
\end{eqnarray}
where $b^2<0$. All TMD distributions are dimensionsless real function that depend on $b^2$ (the argument $b$ is used for shortness).

The parameterization of the sub-leading correlators is
\begin{eqnarray}\label{def:TMDs:2:g+}
\Phi_{\bullet}^{\mu[\gamma^+]}(x_{1,2,3},b)&=&
\epsilon^{\mu\nu}s_{T\nu} M f_{\bullet T}(x_{1,2,3},b)
+ ib^\mu M^2 f^\perp_\bullet(x_{1,2,3},b)
\\\nn &&
+i\lambda \epsilon^{\mu\nu}b_\nu M^2 f^\perp_{\bullet L}(x_{1,2,3},b)
+b^2M^3\epsilon_T^{\mu\nu}\(\frac{g_{T,\nu\rho}}{2}-\frac{b_\nu b_\rho}{b^2}\)s^{\rho}_Tf_{\bullet T}^\perp(x_{1,2,3},b),
\\\label{def:TMDs:2:g+5}
\Phi_{\bullet}^{\mu[\gamma^+\gamma^5]}(x_{1,2,3},b)&=&
s_T^\mu M g_{\bullet T}(x_{1,2,3},b)
-i\epsilon^{\mu\nu}_Tb_\nu M^2 g^\perp_\bullet(x_{1,2,3},b)
\\\nn &&
+i\lambda b^\mu M^2 g_{\bullet L}^\perp(x_{1,2,3},b)
+b^2M^3\(\frac{g_T^{\mu\nu}}{2}-\frac{b^\mu b^\nu}{b^2}\)s_{T\nu}g_{\bullet T}^\perp(x_{1,2,3},b),
\\\nn
\Phi_{\bullet}^{\mu[i\sigma^{\alpha+}\gamma^5]}(x_{1,2,3},b)&=&
\lambda g_{T}^{\mu\alpha} M h_{\bullet L}(x_{1,2,3},b) 
+\epsilon^{\mu\alpha}_T M h_\bullet(x_{1,2,3},b)
+ig_T^{\mu\alpha}(b\cdot s_T)M^2 h_{\bullet T}^{D\perp}(x_{1,2,3},b)
\\\nn &&
+i(b^\mu s^\alpha_T-s_T^\mu b^\alpha)M^2h_{\bullet T}^{A\perp}(x_{1,2,3},b)
+(b^\mu \epsilon^{\alpha\beta}_Tb_\beta+\epsilon_T^{\mu\beta}b_\beta b^\alpha)M^3h_{\bullet}^\perp(x_{1,2,3},b)
\\\label{def:TMDs:2:s+} &&
+\lambda M^3 b^2\(\frac{g^{\mu\alpha}_T}{2}-\frac{b^\mu b^\alpha}{b^2}\)h_{\bullet L}^\perp(x_{1,2,3},b)
\\\nn &&
+i(b\cdot s_T) M^2\(\frac{g^{\mu\alpha}_T}{2}-\frac{b^\mu b^\alpha}{b^2}\)h_{\bullet T}^{T\perp}(x_{1,2,3},b)
\\\nn &&
+iM^2\(\frac{b^\mu s^\alpha_T+s_T^\mu b^\alpha}{2}-\frac{b^\mu b^\alpha}{b^2}(b\cdot s_T)\)h_{\bullet T}^{S\perp}(x_{1,2,3},b)
,
\end{eqnarray}
where $\bullet$ is $\oplus$ or $\ominus$, and $(x_{1,2,3},b)$ is a shorten notation for $(x_1,x_2,x_3,b;\mu,\zeta)$. We emphasize that the parameterization is written for distributions with the upper index $\mu$, i.e. $\Phi_{\bullet}^{\mu[\Gamma]}=g^{\mu\nu}_T\Phi_{\nu\bullet}^{[i\sigma^{\alpha+}\gamma^5]}$. This is important because all indices are transverse and thus change sign upon rising and lowering, i.e.  $g_T^{11}=g_T^{22}=-1$, $b^1=-b_1$, etc.

The distributions defined in (\ref{def:TMDs:2:g+}, \ref{def:TMDs:2:g+5}, \ref{def:TMDs:2:s+}) are dimensionless and real functions.  The elements of parameterization (signs and tensors) are adjusted such that the evolution equations have simpler structure and minimal mixing between distributions\footnote{
Let us mention the different tensor structures for $f_{\bullet T}^\perp$ (\ref{def:TMDs:2:g+}) and $f_T^\perp$ defined in eqn.~(\ref{def:biq:V}). The tensor structure traditionally used for $f_T^\perp$ generates extra mixing terms.}.
The notation for the TMD distributions follows the traditional pattern used in the parameterization of leading TMD distributions (\ref{def:TMDs:1:g+}, \ref{def:TMDs:1:g+5}, \ref{def:TMDs:1:s+}) (known as the Amsterdam notation \cite{Mulders:1995dh}). Namely, the proportionality to $b$ is marked by the superscript $\perp$, and the polarization by subscript $L$ (for longitudinal) or $T$ (for transverse). In the tensor case, one faces four structures $\sim b^\mu s^\alpha_T$, which are denoted as $h_T^{A\perp}$, $h_T^{D\perp}$, $h_T^{S\perp}$, $h_T^{T\perp}$ for antisymmetric, diagonal, symmetric, and traceless components. In total there are 32 TMD distributions of twist-three.

The Dirac matrices project particular components of the quark polarizations, which provide an extra layer of interpretation for TMD distributions, as densities of unpolarized (for $\gamma^+$), helicity (for $\gamma^+\gamma^5$) and transversity (for $i\sigma^{\alpha+}\gamma^5$) quark compositions. Twist-three distributions must also account for gluon polarization vector. The cases of unpolarized and helicity operators correspond to different combinations with opposite quark helicities, such as $\uparrow \Uparrow \downarrow$, $\downarrow \Uparrow \uparrow$, etc (here the single (double) arrow indicates the spin component of the quark (gluon), respectively). In the transversity case, one can split operator into three components with angular momenta 0, 1, and 2. They correspond to the trace part $\sim \sigma^{\alpha+}F_{\alpha +}$, the anti-symmetric part $\sim \epsilon_T^{\alpha\beta}\sigma_{\alpha+}F_{\beta +}$, and the symmetric-traceless combination. In the terms of helicity states, these operators are build from the combinations with the same quark helicities, such as $\uparrow \Uparrow \uparrow$, $\uparrow \Downarrow \uparrow$, etc. Note, that the exact interpretation of these distributions is not possible since the operators for (\ref{def:T-definite}) are not simple. Nonetheless, it allows to sort TMD distributions of twist-three with respect to their spin-content, see table \ref{tab:tw3}.

Among the 32 TMD distributions, 16 distributions change the sign under T-parity transformation, and 16 do not. It means that 16 distributions are na\"ively T-odd. They have different sign but same shape once measured in the Drell-Yan and SIDIS processes, similarly to the Sivers and Boer-Mulders functions \cite{Collins:2002kn, Boer:2003cm}. The T-parity of all TMD distributions discussed in this paper is summarized in the table \ref{tab:T}.

\renewcommand{\arraystretch}{1.3}
\begin{table}
\begin{center}
\begin{tabular}{c||c||c}
Correlator & T-even & T-odd
\\\hhline{=||=||=}
$\Phi^{[\Gamma]}_{11}$ & 
$f_1,~g_1,~g_{1T},~h_1,~h_{1L}^\perp,~h_{1T}^\perp$
&
$f_{1T}^\perp,~h_{1}^\perp$
\\\hhline{=||=||=}
$\Phi^{[\gamma^+]}_{\mu \bullet}$ &
$f_{\oplus T},~f_{\oplus L}^\perp,~ f_{\oplus T}^\perp,~f_\ominus^\perp$
&
$f_\oplus^\perp,~f_{\ominus T},~f_{\ominus L}^\perp,~ f_{\ominus T}^\perp$
\\\hline
$\Phi^{[\gamma^+\gamma^5]}_{\mu \bullet}$ &
$g_\oplus^\perp,~g_{\ominus T},~g_{\ominus L}^\perp,~ g_{\ominus T}^\perp$
& $g_{\oplus T},~g_{\oplus L}^\perp,~ g_{\oplus T}^\perp,~g_\ominus^\perp$
\\\hline
\multirow{2}{*}{$\Phi^{[i\sigma^{\alpha+}\gamma^5]}_{\mu \bullet}$} &
$h_\oplus, ~ h_{\oplus}^\perp,~h_{\ominus L},~ h_{\ominus T}^{D\perp},$
&
$h_{\oplus L},~ h_{\oplus T}^{D\perp},~ h_{\oplus T}^{S\perp},~ h_{\oplus T}^{A\perp},$
\\
 & $ h_{\ominus T}^{S\perp},~ h_{\ominus T}^{A\perp},~h_{\ominus T}^{T\perp},~ h_{\ominus L}^\perp$ &
$h_{\oplus T}^{T\perp},~ h_{\oplus L}^\perp,~h_\ominus, ~ h_{\ominus}^\perp$ 
\\\hhline{=||=||=}
\multirow{2}{*}{$\Phi_{\bar qq}^{[\Gamma_T]}$}
& 
$e,~f^\perp,~g_T,~g_L^\perp$
&
$e_L,~e_T,~e_T^\perp,~h$
\\
&
$g_T^\perp,~h_T^\perp,~h_L^\perp,~h_T$
&
$f_T,~f_T^\perp,~f_L^\perp,~g^\perp$
\end{tabular}
\end{center}
\caption{\label{tab:T} The T-parity of TMD distributions of the twist-two, twist-three and bi-quark distributions.}
\end{table}
\renewcommand{\arraystretch}{1}

\subsection{Evolution equations for TMD distributions}
\label{sec:evolution-T-definite}

The evolution equations for the twsit-two TMD distributions are the same as for TMD correlators, i.e.
\begin{eqnarray}
\mu^2 \frac{d}{d\mu^2}F_1(x,b;\mu,\zeta)&=&\(\frac{\Gamma_{\text{cusp}}}{2}\ln\(\frac{\mu^2}{\zeta}\)-\frac{\gamma_V}{2}\)F_1(x,b;\mu,\zeta),
\\
\zeta \frac{d F_1(x,b;\mu,\zeta)}{d\zeta}&=&-\mathcal{D}(b,\mu)F_1(x,b;\mu,\zeta),
\end{eqnarray}
where $F_1\in\{f_1,f_{1T}^\perp, g_1, g_{1T},h_1,h_{1L}^\perp,h_1^\perp,h_{1T}^\perp\}$, and anomalous dimensions are defined after (\ref{evol:corr:11}).

As already alluded, the evolution equations for TMD distributions of  twist-three take on a non-trivial matrix form. To find it we substitute the parameterizations (\ref{def:TMDs:2:g+}-\ref{def:TMDs:2:s+}) into the evolution equations for the correlators (\ref{eqn:evol-correlator+}, \ref{eqn:evol-correlator-}), and extract independent tensor combinations. We found that chiral-even and chiral-odd sectors obey different evolution equations.

For the chiral-even sector we find
\begin{eqnarray}\label{evolEQN:even}
&&\mu^2 \frac{d}{d\mu^2}\(\begin{array}{c}
F_\oplus \\ G_\ominus \\ G_\oplus \\ F_\ominus
\end{array}\)
=
\(\frac{\Gamma_{\text{cusp}}}{2}\ln\(\frac{\mu^2}{\zeta}\)+\Upsilon_{x_1x_2x_3}\)
\(\begin{array}{c}
F_\oplus \\ G_\ominus \\ G_\oplus \\ F_\ominus
\end{array}\)
\\\nn &&\qquad\qquad +
\(
\begin{array}{cccc}
\mathbb{P}^A_{x_2x_1}+\mathbb{P}^B_{x_2x_1} & 
\mathbb{P}^A_{x_2x_1}-\mathbb{P}^B_{x_2x_1} & 
0 & 2\pi s\Theta_{x_1x_2x_3}
\\
\mathbb{P}^A_{x_2x_1}-\mathbb{P}^B_{x_2x_1} & 
\mathbb{P}^A_{x_2x_1}+\mathbb{P}^B_{x_2x_1} & 
-2\pi s\Theta_{x_1x_2x_3} & 0
\\
0 & 2\pi s\Theta_{x_1x_2x_3} &
\mathbb{P}^A_{x_2x_1}+\mathbb{P}^B_{x_2x_1} & 
-\mathbb{P}^A_{x_2x_1}+\mathbb{P}^B_{x_2x_1} 
\\
-2\pi s\Theta_{x_1x_2x_3} & 0 &
-\mathbb{P}^A_{x_2x_1}+\mathbb{P}^B_{x_2x_1} &
\mathbb{P}^A_{x_2x_1}+\mathbb{P}^B_{x_2x_1} 
\end{array}
\)
\(\begin{array}{c}
F_\oplus \\ G_\ominus \\ G_\oplus \\ F_\ominus
\end{array}\),
\end{eqnarray}
where the argument of distributions $(x_1,x_2,x_3,b;\mu,\zeta)$ is omitted. The pair of distributions $\{F, G\}$ is any of the pairs out of $\{f_T,g_T\}$, $\{f^\perp, g^\perp\}$, $\{f_L^\perp, g_L^\perp\}$, $\{f_T^\perp,g_T^\perp\}$, with $\oplus$ and $\ominus$. The definitions of functions $\Upsilon$, $\Theta$ and kernels $\mathbb{P}$ are given in eqns.~(\ref{def:Upsilon123}), (\ref{def:Theta123}) and (\ref{def:PA}, \ref{def:PB}), respectively. 

The evolution equations for chiral-odd distributions split into two subsets with equations
\begin{eqnarray}\label{evolEQN:oddA}
\mu^2 \frac{d}{d\mu^2}\(\begin{array}{c}
H^A_\oplus \\ H^A_\ominus
\end{array}\)
&=&
\(\frac{\Gamma_{\text{cusp}}}{2}\ln\(\frac{\mu^2}{\zeta}\)+\Upsilon_{x_1x_2x_3}\)
\(\begin{array}{c}
H^A_\oplus \\ H^A_\ominus
\end{array}\)
\\\nn && \qquad\qquad +
\(
\begin{array}{cccc}
2\mathbb{P}^A_{x_2x_1} & 
2\pi s\Theta_{x_1x_2x_3}
\\
-2\pi s\Theta_{x_1x_2x_3} &
2\mathbb{P}^A_{x_2x_1} 
\end{array}
\)
\(\begin{array}{c}
H^A_\oplus \\ H^A_\ominus
\end{array}\),
\end{eqnarray}
\begin{eqnarray}\label{evolEQN:oddB}
\mu^2 \frac{d}{d\mu^2}\(\begin{array}{c}
H^B_\oplus \\ H^B_\ominus
\end{array}\)
&=&
\(\frac{\Gamma_{\text{cusp}}}{2}\ln\(\frac{\mu^2}{\zeta}\)+\Upsilon_{x_1x_2x_3}\)
\(\begin{array}{c}
H^B_\oplus \\ H^B_\ominus
\end{array}\)
\\\nn && \qquad\qquad +
\(
\begin{array}{cccc}
2\mathbb{P}^B_{x_2x_1} & 
2\pi s\Theta_{x_1x_2x_3}
\\
-2\pi s\Theta_{x_1x_2x_3} &
2\mathbb{P}^B_{x_2x_1} 
\end{array}
\)
\(\begin{array}{c}
H^B_\oplus \\ H^B_\ominus
\end{array}\),
\end{eqnarray}
where again the argument of distributions $(x_1,x_2,x_3,b;\mu,\zeta)$ is omitted. The two set of distributions are $H^A \in \{h, h_L, h_T^{D\perp}, h_T^{A\perp}\} $ and $H^B \in \{h^\perp, h_L^\perp, h_T^{S\perp}, h_T^{T\perp}\} $.

The evolution equations for chiral-even sector (\ref{evolEQN:even}) can be rewritten in the same form as (\ref{evolEQN:oddA}, \ref{evolEQN:oddB}). We write
\begin{eqnarray}\label{evolEQN:evenA}
\mu^2 \frac{d}{d\mu^2}\(\begin{array}{c}
F_\oplus+G_\ominus \\ F_\ominus-G_\oplus
\end{array}\)
&=&
\(\frac{\Gamma_{\text{cusp}}}{2}\ln\(\frac{\mu^2}{\zeta}\)+\Upsilon_{x_1x_2x_3}\)
\(\begin{array}{c}
F_\oplus+G_\ominus \\ F_\ominus-G_\oplus
\end{array}\)
\\\nn && \qquad\qquad +
\(
\begin{array}{cccc}
2\mathbb{P}^A_{x_2x_1} & 
2\pi s\Theta_{x_1x_2x_3}
\\
-2\pi s\Theta_{x_1x_2x_3} &
2\mathbb{P}^A_{x_2x_1} 
\end{array}
\)
\(\begin{array}{c}
F_\oplus+G_\ominus \\ F_\ominus-G_\oplus
\end{array}\),
\end{eqnarray}
\begin{eqnarray}\label{evolEQN:evenB}
\mu^2 \frac{d}{d\mu^2}\(\begin{array}{c}
F_\oplus-G_\ominus \\ F_\ominus+G_\oplus
\end{array}\)
&=&
\(\frac{\Gamma_{\text{cusp}}}{2}\ln\(\frac{\mu^2}{\zeta}\)+\Upsilon_{x_1x_2x_3}\)
\(\begin{array}{c}
F_\oplus-G_\ominus \\ F_\ominus+G_\oplus
\end{array}\)
\\\nn && \qquad\qquad +
\(
\begin{array}{cccc}
2\mathbb{P}^B_{x_2x_1} & 
2\pi s\Theta_{x_1x_2x_3}
\\
-2\pi s\Theta_{x_1x_2x_3} &
2\mathbb{P}^B_{x_2x_1} 
\end{array}
\)
\(\begin{array}{c}
F_\oplus-G_\ominus \\ F_\ominus+G_\oplus
\end{array}\),
\end{eqnarray}
where the notation is the same as in eqn.~(\ref{evolEQN:even}). These combinations of $F$ and $G$ functions naturally appear in the applications, as it is shown in sec.~\ref{sec:bi-quark}.

The evolution equations are explicitly real, and explicitly preserve the T-parity, despite mixing the distributions of different parity. The full set of 32 TMD distributions of twist-three is split into two subsets -- the one evolving with the kernel $\mathbb{P}_A$ (\ref{evolEQN:evenA}, \ref{evolEQN:oddA}), and the one evolving with the kernel $\mathbb{P}_B$  (\ref{evolEQN:evenA}, \ref{evolEQN:oddA}). The two subsets of equations correspond to the different spin content of a quark-gluon pair. So, the pair with helicity structures $\Uparrow\uparrow$ and $\Downarrow\downarrow$ evolves with $\mathbb{P}_A$, while the pair with $\Uparrow\downarrow$ evolves with $\mathbb{P}_B$. In sec.~\ref{sec:large-NC} we show that kernels $\mathbb{P}_A$ and $\mathbb{P}_B$ (and the associated distributions) have different properties in the large-$N_c$ limit.

\section{Physical TMD distributions}
\label{sec:physical}

The TMD correlators and \textit{distributions of twist-three are indefinite at} $x_i=0$. It can be seen in several ways. First, the evolution kernels are discontinuous at $x_i=0$ due to the $\Theta$-term, and also has logarithmic singularity due to $\Upsilon$-term.  As a consequence any continuous (or even vanishing) function at $x_i=0$ will turn to a discontinuous and singular function after evolution to a different scale. Second, the discontinuity is explicitly revealed in the small-b limit which can be computed explicitly. This computation is presented in appendix \ref{sec:smallb}. Already, at one-loop the expressions at $x_i=0$ are indefinite, see eqn.~(\ref{smallb:SLO-pos1}). Nonetheless, \textit{TMD distributions of twist-three are integrable} at $x_i=0$.

In a sense, TMD distributions of twist-three (and higher twists as well) are generalized functions. They do not have definite values at each point of the support, but have definite integrals. This situation is unusual, and it is not yet clear how to incorporate such functions into the phenomenology. The positive point is that all known observables are expressed via integrals, and thus can be defined.

Here, however, one faces another complication. Many physical observables, such as cross-sections of Drell-Yan and SIDIS processes \cite{Balitsky:2020jzt,Vladimirov:2021hdn,Ebert:2021jhy}, quasi-TMD distributions \cite{Rodini:2022inprep}, etc. contain integrals with an additional singularity at $x_2=0$. A typical expression contributing into the hadronic tensor incorporates zeroth Mellin moment of twist-three distributions:
\begin{eqnarray}\label{def:0moment}
\Phi^{(0)[\Gamma]}_{\mu,21}(x,b)&=&\int [dx] \delta(x-x_3)\frac{\Phi_{\mu,21}^{[\Gamma]}(x_1,x_2,x_3,b)}{x_2-is0},
\\\nn
\Phi^{(0)[\Gamma]}_{\mu,12}(x,b)&=&\int [dx] \delta(x+x_1)\frac{\Phi_{\mu,12}^{[\Gamma]}(x_1,x_2,x_3,b)}{x_2-is0}.
\end{eqnarray}
These integrals are divergent, since $\Phi$'s are non-analytical at $x_2=0$ and $\lim_{x_2\to+0}\Phi\neq \lim_{x_2\to-0}\Phi$.

In this section, we demonstrate that divergence of integrals (\ref{def:0moment}) is the rapidity divergence. It can be explicitly computed and subtracted. It gives rise to a new layer of definition of TMD distributions of twist-three, which we refer as \textit{physical distributions}. Physical distributions has finite zeroth-momentum and the observables written in their terms are finite term-by-term. 

\subsection{\texorpdfstring{ Divergence at $x_2=0$  at LO}{TEXT}}
\label{sec:x2=0}

\begin{figure}[t]
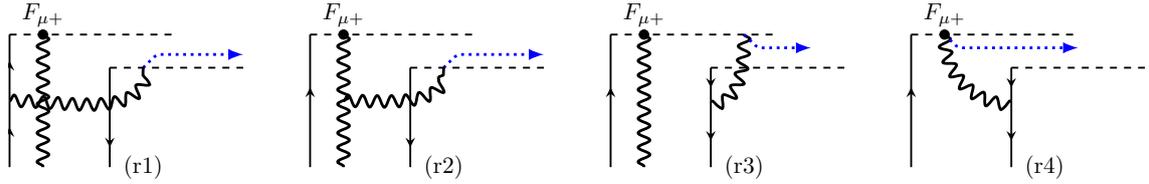

\begin{center}
\includestandalone[width=0.99\textwidth]{Figures/diagrams2}%     without .tex extension
\caption{\label{fig:rapidity-diag} The diagrams contributing to the rapidity divergence of TMD-twist-(1,2) operator. The blue arrow shows the rapidity divergent limit.}
\end{center}
\end{figure}

We start with the explicit computation of the divergent part of zeroth moments (\ref{def:0moment}) at LO. As usual, it is more intuitive to perform computation in the position space, where the zeroth moments read\footnote{
Explicitly, these integrals can be written in terms of operators as follow:
\begin{eqnarray}
\widetilde{\Phi}^{(0)[\Gamma]}_{\mu,21}(z_1,z_2,b)&=&-\langle p,s|T\{\bar q(z_1n+b)\Big(D_\mu[z_1n+b,Ln+b]\Big)\frac{\Gamma}{2}[Ln,z_2n]q(z_2n)\}|p,s\rangle,
\\\nn
\widetilde{\Phi}^{(0)[\Gamma]}_{\mu,12}(z_1,z_2,b)&=&+\langle p,s|T\{\bar q(z_1n+b)[z_1n+b,Ln+b]\frac{\Gamma}{2}\Big([Ln,z_2n]\overleftarrow{D_\mu}\Big) q(z_2n)\}|p,s\rangle,
\end{eqnarray}
where $D_\mu$ is the covariant derivative. In the terms of SCET fields, these correlators are
\begin{eqnarray}
\widetilde{\Phi}^{(0)[\Gamma]}_{\mu,21}(z_1,z_2,b)&=&i\langle p,s|T\{\bar\chi_{\bar n}(z_1n)\mathcal{B}_{\bar n,\mu}(z_1n)\frac{\Gamma}{2}\chi_{\bar n}(z_2n)\}|p,s\rangle,
\\\nn
\widetilde{\Phi}^{(0)[\Gamma]}_{\mu,12}(z_1,z_2,b)&=&i\langle p,s|T\{\bar\chi_{\bar n}(z_1n)\frac{\Gamma}{2}\mathcal{B}_{\bar n,\mu}^{\dagger}(z_2n)\chi_{\bar n}(z_2n)\}|p,s\rangle,
\end{eqnarray}
where we use the convention by ref.\cite{Inglis-Whalen:2021bea}.
}
\begin{eqnarray}\label{def:Psi}
\widetilde{\Phi}^{(0)[\Gamma]}_{\mu,21}(z_1,z_2,b)&=&-i\int_L^{z_1}d\sigma\widetilde{\Phi}_{\mu,21}^{[\Gamma]}(z_1,\sigma,z_2,b),
\\\nn
\widetilde{\Phi}^{(0)[\Gamma]}_{\mu,12}(z_1,z_2,b)&=&-i\int_L^{z_2}d\sigma\widetilde{\Phi}_{\mu,12}^{[\Gamma]}(z_1,\sigma,z_2,b).
\end{eqnarray}
Here, the transformation between position and momentum spaces is done according to ``single momentum-fraction rule'' (\ref{def:z->x}), i.e.
\begin{eqnarray}
\widetilde{\Phi}^{(0)[\Gamma]}_{\mu,21}(z_1,z_2,b)=p^+\int_{-1}^1 dx e^{ix(z_1-z_2)p^+}\Phi^{(0)[\Gamma]}_{\mu,21}(x,b),
\end{eqnarray}
and similar for $\widetilde{\Phi}^{(0)[\Gamma]}_{\mu,12}$.

We would like to compute the rapidity divergence for these operators at one-loop. The rapidity divergence appears in the loop diagrams with the gluon attached to the far-end of the light-cone Wilson line. There are two possibilities to receive divergence for discussed operators.
\begin{itemize}
\item The first case (standard) is to have a diagram with a gluon coupled to the Wilson line. Such diagrams are shown in figs.\ref{fig:rapidity-diag} (r1), (r2) and (r3). 
\item The second case (special) is to couple $F^{\mu+}$ to the rest of the operator. In the limit \\$\lim_{z_2\to L}F^{\mu+}(z_2)$ this diagram (diagram (r4) in fig.\ref{fig:rapidity-diag}) is also rapidity divergent. Importantly, the special case does not appear for the operators $\Phi$, where the position of $F^{\mu+}$ is fixed. This special case appears only for operators $\Phi^{(0)}$, and represents the divergent part of zeroth moments (\ref{def:0moment}) at $x_2=0$. This divergence can be also classified as the end-point divergence.
\end{itemize}

We regularize the rapidity divergences by the $\delta$-regularization \cite{Echevarria:2015byo, Echevarria:2016scs}. It consists in the multiplication of gluon fields on the Wilson line, including $F^{\mu+}$, by the suppressing factor $e^{-s\delta}$, with $\delta>0$.  The technique of computation with $\delta$-regulator can be found in refs.~\cite{Vladimirov:2017ksc, Scimemi:2019gge, Vladimirov:2021hdn}. In particular, the detailed computation of diagrams (r1), (r2) and (r3) is provided in sec.8.1 of ref.\cite{Vladimirov:2021hdn}. The result of computation is
\begin{eqnarray}\label{eqn:r1+r2+r3}
\widetilde{\Phi}^{(0)[\Gamma](r1+r2+r3)}_{\mu,21}(z_1,z_2,b)&=&-4a_sC_F\Gamma(-\epsilon)\(\frac{-b^2}{4}\)^{\epsilon}\ln\(\frac{\delta^+}{q^+}\)\widetilde{\Phi}^{(0)[\Gamma]}_{\mu,21}(z_1,z_2,b)+\text{fin.terms},
\\\nn
\widetilde{\Phi}^{(0)[\Gamma](r1+r2+r3)}_{\mu,12}(z_1,z_2,b)&=&-4a_sC_F\Gamma(-\epsilon)\(\frac{-b^2}{4}\)^{\epsilon}\ln\(\frac{\delta^+}{q^+}\)\widetilde{\Phi}^{(0)[\Gamma]}_{\mu,12}(z_1,z_2,b)+\text{fin.terms},
\end{eqnarray}
where ``$\text{fin.terms}$'' are terms finite at $\delta^+\to0$. The value of $q^+$ is $|p_{\bar q}|$ and $|p_q|$ for $\widetilde{\Phi}^{(0)[\Gamma]}_{\mu,12}$ and $\widetilde{\Phi}^{(0)[\Gamma]}_{\mu,21}$, correspondingly. The expressions (\ref{eqn:r1+r2+r3}) provide the bare LO expression for the rapidity renormalization factor $R$ introduced in eqn.~(\ref{def:Phi-renorm}). The complete expression for $R$ at $\mathcal{O}(a_s)$ reads \cite{Echevarria:2015byo,Vladimirov:2017ksc}
\begin{eqnarray}\label{rap:RNLO}
R(b^2)&=&1-4a_sC_F\ln\(\frac{\delta^+}{q^+}\)\(\Gamma(-\epsilon)\(\frac{-b^2}{4}\)^{\epsilon}+\frac{1}{\epsilon}\)+\mathcal{O}(a_s^2),
\end{eqnarray}
where we include $1/\epsilon$ term from the renormalization.

The diagram (r4) is calculated similarly. In this diagram, the gluon field is totally quantum and thus the final operator contains only quark fields. The difference in dimensions of operators is compensated by an inverse power of $b$. We obtain\footnote{
%%%%
The equation (\ref{rap:1}) can be also derived from know results. Using the relation
\begin{eqnarray}\nn
\partial_\mu[z_1n,z_2n]=
ig\Big(A_\mu[z_1n,z_2n]-
[z_1n,z_2n]A_\mu
+\int_{z_2}^{z_1}d\tau 
[z_1n,\tau n]F_{\mu+}[\tau n,z_2n]\Big),
\end{eqnarray}
we can relate the rapidity divergent diagrams for twist-two and twist-three operators (at least at the one-loop order). We find
\begin{eqnarray}\label{eqn:from_literature}
\widetilde{\Phi}^{(0)[\Gamma]}_{\mu,12}\Big|_{\text{rap.div.}}
=
\widetilde{\Phi}^{(0)[\Gamma]}_{\mu,21}\Big|_{\text{rap.div.}}
=-\frac{1}{2}\frac{\partial}{\partial b^\mu}\widetilde{\Phi}^{[\Gamma]}_{11}\Big|_{\text{rap.div.}},
\end{eqnarray}
where derivative does not act on the quark field, and factor $1/2$ is to compensate the absence of the symmetric contribution. The LO rapidity divergence for $\widetilde{\Phi}^{[\Gamma]}_{11}$ has been computed in many papers, see e.g. \cite{Collins:2011zzd,Scimemi:2019gge,Echevarria:2016scs,Gutierrez-Reyes:2017glx}. Explicitly, it is given in, e.g., equation (5.35) in ref. \cite{Scimemi:2019gge}). Substituting  $\widetilde{\Phi}^{[\Gamma]}_{11}\Big|_{\text{rap.div.}}$   into (\ref{eqn:from_literature}), we receive eqn.~(\ref{rap:1}).} expressions that are finite at $\epsilon\to0$,
%%%
\begin{eqnarray}\label{rap:1}
\widetilde{\Phi}^{(0)[\Gamma](r4)}_{\mu,21}(z_1,z_2,b)&=&
%-4a_sC_F\Gamma(1-\epsilon)\frac{b_\mu}{b^2}\(\frac{-b^2}{4}\)^{\epsilon}\ln\(\frac{\delta^+}{q^+}\)\widetilde{\Phi}^{[\Gamma]}_{11}(z_1,z_2,b)+\text{fin.terms},
-4a_sC_F\frac{b_\mu}{b^2}\ln\(\frac{\delta^+}{q^+}\)\widetilde{\Phi}^{[\Gamma]}_{11}(z_1,z_2,b)+\text{fin.terms},
\\\nn
\widetilde{\Phi}^{(0)[\Gamma](r4)}_{\mu,12}(z_1,z_2,b)&=&
%-4a_sC_F\Gamma(1-\epsilon)\frac{b_\mu}{b^2}\(\frac{-b^2}{4}\)^{\epsilon}\ln\(\frac{\delta^+}{q^+}\)\widetilde{\Phi}^{[\Gamma]}_{11}(z_1,z_2,b)+\text{fin.terms}.
-4a_sC_F\frac{b_\mu}{b^2}\ln\(\frac{\delta^+}{q^+}\)\widetilde{\Phi}^{[\Gamma]}_{11}(z_1,z_2,b)+\text{fin.terms}.
\end{eqnarray}
Note that the coefficient in front of the correlator $\tilde{\Phi}_{11}$ in eqn.~(\ref{rap:1}) exactly reproduces the derivative of $R$ (\ref{rap:RNLO}). 

The expressions (\ref{rap:1}) represent only the perturbative part of the rapidity renormalization factor. At larger values of $b$ the nonperturbative corrections appears. Altogether they must combine into the derivative of (nonperturbative) factor $R=\exp(-2\mathcal{D}\ln\delta +B)$, where $B$ is some finite terms \cite{Vladimirov:2017ksc}. Therefore, the complete LO expressions for rapidity divergent part of zeroth moments are
\begin{eqnarray}\label{rap:2}
\widetilde{\Phi}^{(0)[\Gamma](r4)}_{\mu,21}(z_1,z_2,b)&=&
\ln\(\frac{\delta^+}{q^+}\)\partial_{\mu}\mathcal{D}(b)\widetilde{\Phi}^{[\Gamma]}_{11}(z_1,z_2,b)+\text{fin.terms},
\\\nn
\widetilde{\Phi}^{(0)[\Gamma](r4)}_{\mu,12}(z_1,z_2,b)&=&
\ln\(\frac{\delta^+}{q^+}\)\partial_{\mu}\mathcal{D}(b)\widetilde{\Phi}^{[\Gamma]}_{11}(z_1,z_2,b)+\text{fin.terms},
\end{eqnarray}
where
\begin{eqnarray}
\partial_{\mu}\mathcal{D}(b)=\frac{\partial}{\partial b^\mu}\mathcal{D}(b)=2b^\mu \frac{\partial}{\partial b^2}\mathcal{D}(b),
\end{eqnarray}
is the derivative of the nonperturbative Collins-Soper kernel.

One could expect a contribution to the ``special'' rapidity divergence from the operators with two emitted gluons (see diagrams C and D in fig.\ref{fig:singular-diag}). The explicit computation presented in appendix \ref{sec:smallb} shows that such diagrams are finite.

We combine the expressions for rapidity divergent parts of $\Phi^{(0)}$,
\begin{eqnarray}\label{eqn:rap-div-for0}
\begin{aligned}
\widetilde{\Phi}^{(0)[\Gamma]}_{\mu,12}(z_1,z_2,b)\Big|_{\text{rap.div.}}&=
R(b^2)\widetilde{\Phi}^{(0)[\Gamma]}_{\mu,12}(z_1,z_2,b)
+\(\partial_{\mu}\mathcal{D}(b)\ln\(\frac{\delta^+}{q^+}\)+\mathcal{O}(a_s^2)\)\widetilde{\Phi}_{11}^{[\Gamma]}(z_1,z_2,b),
\\
\widetilde{\Phi}^{(0)[\Gamma]}_{\mu,21}(z_1,z_2,b)\Big|_{\text{rap.div.}}&=
R(b^2)\widetilde{\Phi}^{(0)[\Gamma]}_{\mu,21}(z_1,z_2,b)
+\(\partial_{\mu}\mathcal{D}(b)\ln\(\frac{\delta^+}{q^+}\)+\mathcal{O}(a_s^2)\)\widetilde{\Phi}_{11}^{[\Gamma]}(z_1,z_2,b).
\end{aligned}
\end{eqnarray}
The terms proportional to twist-three correlators are exact at all orders. The terms proportional to twist-two correlators at higher perturbative orders are different from derivative of $R$, and can contain integral convolutions. The term in brackets is same for both correlators at all perturbative orders as a consequence of PT-invariance (\ref{sym:PT:corr}).

\subsection{Definition of physical TMD distributions}

The physical cross-section is a sum of terms with various combinations of TMD correlators. Some of them have the following schematic form
\begin{eqnarray}
d\sigma \sim \int d^2b e^{-i(qb)} [H\otimes \Phi_{\text{tw3}}](x,b) \Phi_{11}(\tilde x,b),
\end{eqnarray}
where $H$ is a hard coefficient function, $\Phi_{\text{tw3}}$ is a twist-three TMD distribution, and $\otimes$ is a convolution with respect to momentum fractions. For example, the LO of $H\otimes \Phi_{\text{tw3}}$ is the zeroth moment (\ref{def:0moment}). After the renormalization procedure (\ref{def:Phi-renorm}) explicit divergences of $H$ and $\Phi$'s cancel. Nonetheless, the convolution $H\otimes \Phi_{\text{tw3}}$ still contains the (special) rapidity divergences.  These divergences, however, do not imply the breaking of the factorization theorem at NLP, because they cancel in-between various terms of the factorization theorem. Therefore, even if the factorization theorem holds and the cross-sections are finite, it makes impossible to straightforward compute the convolutions term-by-term, using the TMD distributions defined earlier. This is, of course, a problem for any phenomenological study involving twist-three TMD distributions.

In order to make physical quantities well-defined term-by-term, we further modify the definition of twist-three TMD distributions by subtracting the (special) divergent part. We define
\begin{eqnarray}\label{def:physical}
\mathbf{\widetilde{\Phi}}_{\mu,12}^{[\Gamma]}(z_1,z_2,z_3)&=&
\widetilde{\Phi}_{\mu,12}^{[\Gamma]}(z_1,z_2,z_3)-[\mathcal{R}_{12}\otimes \widetilde{\Phi}_{11}]^{[\Gamma]}_{\mu}(z_1,z_2,z_3,b),
\\\nn
\mathbf{\widetilde{\Phi}}_{\mu,21}^{[\Gamma]}(z_1,z_2,z_3)&=&
\widetilde{\Phi}_{\mu,21}^{[\Gamma]}(z_1,z_2,z_3)-[\mathcal{R}_{21}\otimes \widetilde{\Phi}_{11}]^{[\Gamma]}_{\mu}(z_1,z_2,z_3,b),
\end{eqnarray}
where $\otimes$ is some integral convolution. The kernel $\mathcal{R}$ is defined with respect to the hard-coefficient function $H$, such that the integral convolution $H\otimes \mathbf{\Phi}_{\text{tw3}}$ is finite. The physical observables expressed in terms of $\mathbf{\Phi}$ are finite term-by-term. For that reason, we call correlators defined in eqn.~(\ref{def:physical}), and corresponding distributions, \textit{physical}.

Physical TMD correlators satisfy the same symmetry properties as subtracted TMD correlators discussed in sections \ref{sec:TMD-in-pos} and \ref{sec:TMD-in-mom}. Thus one can define TMD correlators with definite T-parity analogously to eqn.~(\ref{def:T-definite}), and define TMD distribution as in sec.~\ref{sec:TMD-parameterization}. The physical analogs of corresponding distributions we denote by the bold font.

Note that not each physical TMD distribution has subtraction term. In some cases, the subtraction term $[\mathcal{R}\otimes \Phi]^{[\Gamma]}$ has a zero projection to corresponding tensor structure. 

The functions $\mathcal{R}$ are to be constructed order-by-order in the perturbation theory. Using the computation made in the previous section, we can construct $\mathcal{R}$ at LO. We find 
\begin{eqnarray}\label{def:R}
[\mathcal{R}_{21}\otimes \widetilde{\Phi}_{11}]^{[\Gamma]}_{\mu}(z_1,z_2,z_3,b)&=&
\partial_{\mu}\mathcal{D}(b)\int_0^1 d\alpha
\frac{\partial}{\partial z_1}
\Phi^{[\Gamma]}_{11}(z_1,z_{23}^\alpha,b)+\mathcal{O}(a_s^2),
\\\nn
[\mathcal{R}_{12}\otimes \widetilde{\Phi}_{11}]^{[\Gamma]}_{\mu}(z_1,z_2,z_3,b)&=&
\partial_{\mu}\mathcal{D}(b)\int_0^1 d\alpha \frac{\partial}{\partial z_3}
\Phi^{[\Gamma]}_{11}(z_{21}^\alpha,z_3,b)+\mathcal{O}(a_s^2).
\end{eqnarray}
It is straightforward to check that integrals (\ref{def:Psi}) computed with (\ref{def:R}) reproduces (\ref{rap:1}). Therefore the zeroth moment of the physical distribution and corresponding term in the cross-section are finite.

The subtraction term removes the ``special'' rapidity divergence of the zeroth moment. Therefore, the physical TMD distributions satisfy ordinary evolution equation with respect to the scale $\zeta$
\begin{eqnarray}\label{eqn:evol-0mom}
\zeta \frac{d}{d\zeta} \widetilde{\mathbf{\Phi}}^{(0)[\Gamma]}_{\mu,\bullet}(z_1,z_2,b;\mu,\zeta)
=-\mathcal{D}(b,\mu)\widetilde{\mathbf{\Phi}}^{(0)[\Gamma]}_{\mu,\bullet}(z_1,z_2,b;\mu,\zeta),
\end{eqnarray}
where $\bullet$ is $(12)$, $(21)$, $\oplus$ or $\ominus$. Importantly, the physical distributions also satisfy ordinary rapidity-evolution equation (\ref{evol:rap:position}) without modifications
\begin{eqnarray}
\zeta \frac{d}{d\zeta} \widetilde{\mathbf{\Phi}}^{[\Gamma]}_{\mu,\bullet}(z_1,z_2,z_3,b;\mu,\zeta)
=-\mathcal{D}(b,\mu)\widetilde{\mathbf{\Phi}}^{[\Gamma]}_{\mu,\bullet}(z_1,z_2,z_3,b;\mu,\zeta),
\end{eqnarray}
where $\bullet$ is $(12)$, $(21)$, $\oplus$ or $\ominus$. 
The evolution equations with respect to $\mu$ are identical to the corresponding equations for twist-three TMD correlators, because $d[R\otimes \Phi]/d\mu\sim \mathcal{O}(a_s^2)$ at LO. At the moment, it is not clear how the transformation (\ref{def:physical}) effects the evolution equations with respect to $\mu$ at higher perturbative orders. It is also not clear how to explicitly guarantee the finiteness of integrals in the numerical form.

In the momentum-fraction space, the expressions (\ref{def:R}) become
\begin{eqnarray}\label{eqn:R-in-x}
\begin{aligned}
\,
[\mathcal{R}_{21}\otimes \Phi_{11}]^{[\Gamma]}_{\mu}(x_1,x_2,x_3,b)=&
i\partial_{\mu}\mathcal{D}(b)\int_{-1}^1 dy\int_0^1 d\alpha \delta(x_3-\alpha y)\delta(x_2-\bar \alpha y) \,y\Phi^{[\Gamma]}_{11}(y,b)
+\mathcal{O}(a_s^2),
\\
\,
[\mathcal{R}_{12}\otimes \Phi_{11}]^{[\Gamma]}_{\mu}(x_1,x_2,x_3,b)=&
-i\partial_{\mu}\mathcal{D}(b)\int_{-1}^1 dy\int_0^1 d\alpha \delta(x_1+\alpha y)\delta(x_2+\bar \alpha y) \,y\Phi^{[\Gamma]}_{11}(y,b)
+\mathcal{O}(a_s^2),
\end{aligned}
\end{eqnarray}
At $x_2=0$ the values of expressions (\ref{eqn:R-in-x}) are not defined. In this case, the integral depends on the order in which one evaluate the Dirac-delta functions. However, the integral over $x_2$ is well-defined. Therefore, eqns.~(\ref{eqn:R-in-x}) represent generalized functions. The zeroth moments (\ref{def:0moment}) of eqns.~(\ref{eqn:R-in-x}) are rapidity divergent at $x_2=0$. Thus, the expression for the physical TMD correlator has finite zeroth momentum, and evolves according to eqn.~(\ref{eqn:evol-0mom}).

For $x_2\neq 0$, one has
\begin{eqnarray}
[\mathcal{R}_{21}\otimes \Phi_{11}]^{[\Gamma]}_{\mu}(x_1,x_2,x_3,b)&=&
i\partial_{\mu}\mathcal{D}(b) \,\Phi^{[\Gamma]}_{11}(-x_1,b)(\theta(x_2,x_3)-\theta(-x_2,-x_3))+\mathcal{O}(a_s^2),
\\\nn
[\mathcal{R}_{12}\otimes \Phi_{11}]^{[\Gamma]}_{\mu}(x_1,x_2,x_3,b)&=&
i\partial_{\mu}\mathcal{D}(b)\, \Phi^{[\Gamma]}_{11}(x_3,b)(\theta(x_1,x_2)-\theta(-x_1,-x_2))+\mathcal{O}(a_s^2).
\end{eqnarray}
The value of discontinuity at $x_2=0$ is proportional to the difference between quark and anti-quark distributions. Indeed,
\begin{eqnarray}\label{eqn:discontinuity}
\(\lim_{x_2\to +0}-\lim_{x_2\to -0}\)[\mathcal{R}_{21}\otimes \Phi_{11}]^{[\Gamma]}_{\mu}(x_1,x_2,x_3,b)
&=&
i\partial_{\mu}\mathcal{D}(b)\,\(\Phi^{[\Gamma]}_{11}(x,b)+\Phi^{[\Gamma]}_{11}(-x,b)\),
\end{eqnarray}
where $x=|x_3|=|x_1|$, and similar for the $(12)$-case. Since the physical TMD correlator has a finite zeroth moment, one can expect that it is smooth at $x_2=0$, and thus the discontinuity of $\Phi_{\mu,21}^{[\Gamma]}$ is given by eqn.~(\ref{eqn:discontinuity}).

The integrals (\ref{def:0moment}) appear in the factorized cross-section at the tree-level. At higher perturbative orders one faces more singular combinations, such as $\ln x_2/x_2$ \cite{Vladimirov:2021hdn}. To study these NLO terms one has to perform two-loop computation, which goes beyond the present work. However, we expect that the general strategy presented in this section holds at all orders of perturbation theory. It is also evident that the higher powers of TMD factorization theorem will suffer similar problems of singular integrals. We expect that the receipt suggested here is of general validity, i.e. the higher twist TMD distributions can be turned to physical ones, that provide term-by-term finite cross-section, by subtracting lower twist parts.

Since the physical distributions are defined with respect to kernel $H$, they generally are not universal, and could depend on the process. However, for SIDIS, Drell-Yan and semi-inclusive annihilation processes the hard coefficient functions are the same. It is also the same in the factorization theorem for quasi-TMD correlators \cite{Rodini:2022inprep}. Therefore, the physical TMD distributions defined here have at least a large spectrum of validity, and can be used for description of all these cases.

The present definition of physical TMD distributions requires only the finiteness of integral convolution $[H\otimes \Phi]$. It leaves a large freedom in the definition of the subtraction terms. The suggested terms (\ref{def:R}) represent the ``minimal subtraction scheme'', where only the divergent part is removed. Clearly, one can also incorporate finite terms, and use this freedom to improve the properties of physical distributions. For example, one can prepare physical distributions of twist-three regular in the small-b limit (see appendix \ref{sec:smallb}). This is a very important point which requires a detailed study in the future. 

\section{Bi-quark TMD correlators}
\label{sec:bi-quark}

The distributions introduced in the previous sections are fundamental. They have closed evolution equations, and the factorization theorem is expressed in their terms at any order of the perturbative expansion. However, they are also very complicated from the practical point of view. To start with, they are functions of two independent momentum fractions, which complicates all formulas. For practical applications, it may be convenient to also have a simpler approximation.

Until recently, the TMD factorization theorem at NLP was known only at the three order, which can be expressed in the terms of TMD correlators $\Phi_{11}^{[\Gamma_T]}$, where $\Gamma_T$ projects one good and one bad components of quark fields. Such \textit{bi-quark distributions} are visually simpler, albeit having very convoluted properties. In this section we derive the relations between bi-quark distributions and genuine twist-three TMD distributions. We also demonstrate that the evolution equations for bi-quark distributions are not closed. Finally, we show that the large-$N_c$ limit ensures a simpler system of evolution equations and closure of the evolution of bi-quark distributions.

\subsection{Definition}

The bi-quark TMD correlators were introduced in analogy to the leading-twist TMD correlators in ref.\cite{Mulders:1995dh}. They are defined as 
\begin{eqnarray}\label{def:biq}
\Phi^{[\Gamma]}_{\bar qq}(x,b)&=&\int_{-\infty}^\infty \frac{dz}{2\pi}e^{-ixzp^+}
\langle p,s|T\{\bar q(zn+b)[zn+b,Ln+b]\frac{\Gamma}{2}[Ln,0]q(0)\}|p,s\rangle,
\end{eqnarray}
where $\Gamma$ is any Dirac matrix. For $\Gamma\in\Gamma^+$ the bi-quark correlator has TMD-twist-(1,1) and coincides with $\Phi_{11}^{[\Gamma]}$. For other $\Gamma$'s bi-quark correlators do not have definite twist. 

We consider the correlators of twist-three with $\Gamma\in \Gamma_T$, where $\Gamma_T$ projects one good and one bad component of the quark fields (\ref{app:G+GTG-}). The standard parameterization for bi-quark twist-three TMD distributions reads\footnote{
The standard parameterization is defined in refs.~\cite{Mulders:1995dh,Bacchetta:2006tn}, in the momentum space. The parameterization in the transverse distance space is obtained by Fourier transformation, see e.g. refs.~\cite{Scimemi:2018mmi,Boer:2011xd}, which basically consists in the replacement $p_T^\mu\to -i b^\mu M^2$.}
%\cite{Mulders:1995dh,Bacchetta:2006tn}
\begin{eqnarray}\label{def:biq:1}
\Phi_{\bar q q}^{[1]}(x,b)&=&\frac{M}{p^+}\Big[
e(x,b)
+i\epsilon^{\mu\nu}_T b_\mu s_{T\nu} M\, e_T^\perp(x,b)
\Big],
\\\label{def:biq:5}
\Phi_{\bar q q}^{[i\gamma^5]}(x,b)&=&\frac{M}{p^+}\Big[
\lambda e_L(x,b)
+i(b\cdot s_{T}) M\, e_T(x,b)
\Big],
\\\label{def:biq:V}
\Phi_{\bar q q}^{[\gamma^\alpha]}(x,b)&=&\frac{M}{p^+}\Big[
-\epsilon^{\alpha\mu}_Ts_{T\mu} f_T(x,b)
+i\lambda \epsilon^{\alpha\mu}b_\mu M\, f_L^\perp(x,b)
-ib^\alpha M f^\perp(x,b)
\\\nn &&
-b^2M^2\(\frac{g_T^{\alpha\mu}}{2}-\frac{b^\alpha b^\mu}{b^2}\)\epsilon_{T\mu\nu}s_T^\nu f_T^\perp(x,b)
\Big],
\\\label{def:biq:A}
\Phi_{\bar q q}^{[\gamma^\alpha\gamma^5]}(x,b)&=&\frac{M}{p^+}\Big[
s_{T}^\alpha g_T(x,b)
-i\lambda b^\alpha M\, g_L^\perp(x,b)
+i\epsilon^{\alpha \mu}b_\mu M g^\perp(x,b)
\\\nn &&
-b^2M^2\(\frac{g_T^{\alpha\mu}}{2}-\frac{b^\alpha b^\mu}{b^2}\)s_{T\nu} g_T^\perp(x,b)
\Big],
\\\label{def:biq:T}
\Phi_{\bar q q}^{[i\sigma^{\alpha\beta}\gamma^5]}(x,b)&=&\frac{M}{p^+}\Big[
i(b^\alpha s_T^\beta-s_T^\alpha b^\beta)M h_T^\perp(x,b)
-\epsilon^{\alpha\beta} h(x,b)
\Big],
\\\label{def:biq:+-}
\Phi_{\bar q q}^{[i\sigma^{+-}\gamma^5]}(x,b)&=&\frac{M}{p^+}\Big[
\lambda h_L^\perp(x,b)
+i(b\cdot s_T) h_T(x,b)
\Big],
\end{eqnarray}
where all indices are transverse. There are 16 TMD bi-quark distribution of twist-3. Eight of them are T-even and eight are T-odd, see table \ref{tab:T}. These distributions are often used in the phenomenology. Among these distributions, one of the most studied is the $e$ distribution, see, e.g., ref.~\cite{Hatta:2020iin,Pasquini:2018oyz} and references therein. The $e$ distribution has one of the simplest structures in terms of twist-three distributions and, unlike others, contains no leading-twist contributions. 

\subsection{Relation between bi-quark and definite-twist distributions}

On the theory side, the bi-quark TMD-correlators for $\Gamma \not\in \Gamma^+$ are not well-defined objects, in the sense that their renormalization is troublesome. To explicitly reveal the problems, we use the equations of motion for the bad components of the quark fields to obtain \cite{Boer:2003cm,Bacchetta:2006tn}
\begin{eqnarray}\label{biQ:EOMs1}
&& \widetilde{\Phi}^{[\Gamma_T]}_{\bar qq}(z_1,z_2,b)=
-\frac{1}{2}\int_L^0d\tau \frac{\partial}{\partial b^\mu} 
\(
\widetilde{\Phi}_{11}^{[\gamma^\mu \gamma^+\Gamma_T]}(z_1+\tau,z_2,b)
-
\widetilde{\Phi}_{11}^{[\Gamma_T \gamma^+\gamma^\mu]}(z_1,z_2+\tau,b)\)
\\\nn &&\qquad
+\frac{i}{2}\int_L^0d\tau \int_L^\tau d\sigma
\(
\widetilde{\Phi}_{\mu,21}^{[\gamma^\mu \gamma^+\Gamma_T]}(z_1+\tau,z_1+\sigma,z_2,b)
-
\widetilde{\Phi}_{\mu,12}^{[\Gamma_T \gamma^+\gamma^\mu]}(z_1,z_2+\sigma,z_2+\tau,b)\).
\end{eqnarray}
Here we omit the quark mass terms for simplicity. These terms are proportional to $\tilde\Phi_{11}$, and can be simply restored if needed. The matrices $\{\gamma^\mu\gamma^+\Gamma_T,\Gamma_T \gamma^+\gamma^\mu\}\in \Gamma_+$ for any $\Gamma_T$. Therefore, the distributions in the first line of eqn.~(\ref{biQ:EOMs1}) have TMD-twist-(1,1), while in the second line TMD-twists (2,1) and (1,2). In the momentum fraction representation the relation is 
\begin{eqnarray}
\Phi_{\bar qq}^{[\Gamma_T]}(x,b)&=&
\frac{i}{2xp^+}\frac{\partial}{\partial b^\mu}\(
\Phi^{[\gamma^\mu \gamma^+\Gamma_T]}_{11}(x,b)
+
\Phi^{[\Gamma_T \gamma^+\gamma^\mu]}_{11}(x,b)
\)
\\\nn &&
+\frac{i}{2xp^+}\(\Phi_{\mu,21}^{(0)[\gamma^\mu \gamma^+\Gamma_T]}(x,b)-
\Phi_{\mu,12}^{(0)[\Gamma_T \gamma^+\gamma^\mu]}(x,b)\),
\end{eqnarray}
where the zeroth moments are defined in eqns.~(\ref{def:0moment}). This \textit{bare} relation clearly demonstrates the issues of the definition:
\begin{itemize}
\item The UV renormalization of the zeroth moments $\Phi_{\mu,\bullet}^{(0)}$ does not express via itself. Therefore, the evolution equation for $\Phi_{\bar qq}^{[\Gamma_T]}$ is not closed, but mixes with integrals of genuine twist-three TMD distributions;
\item The rapidity renormalization of the zeroth moments $\Phi_{\mu,\bullet}^{(0)}$ is not multiplicative. Namely, in addition to ordinary rapidity divergences (that are renormalized by $R$), there are special rapidity divergences alike ones considered in sec.~\ref{sec:x2=0}.
\end{itemize}
Thus, the bi-quark TMD distributions are not well-defined. 

Let us also point that the derivative of bare twist-two distributions does not commute with the rapidity renomalization factor $R(b)$. It produces extra terms $\sim \partial_\mu R$ during the renormalization procedure. These terms resemble special rapidity divergences in zeroth moments terms (\ref{eqn:rap-div-for0}). However, they do not cancel each other due to extra factor 1/2 in eqn.~(\ref{eqn:rap-div-for0}).

All mentioned problems arise at NLO of perturbative expression. In fact, bi-quark correlators were introduced just as convenient combinations which appeared at the tree-order of factorization theorem \cite{Boer:2003cm,Bacchetta:2006tn}. Already at NLO of factorization theorem, this notation is inconvenient, because genuine twist-three and twist-two parts receives different hard coefficient functions \cite{Vladimirov:2021hdn}, which break down the combinations. The situation here is similar to the bi-quark twist-three collinear distributions $g_T$, $h_L$, $e$ (see e.g. \cite{Jaffe:1996zw}), which appears in the polarized DIS. They are not more than a convenient combinations of genuine parton distributions, and their convenience does not hold beyond LO approximation. 

For applications beyond LO, the bi-quark distributions are defined as the corresponding combinations of \textit{renormalized} distributions. Similarly, we define bi-quark TMD distributions as combinations of \textit{physical} TMD distributions,
\begin{eqnarray}\label{biQ:EOMSs2}
\Phi_{\bar qq}^{[\Gamma_T]}(x,b;\mu,\zeta)&=&
\frac{i}{2xp^+}\frac{\partial}{\partial b^\mu}\(
\Phi^{[\gamma^\mu \gamma^+\Gamma_T]}_{11}(x,b;\mu,\zeta)
+
\Phi^{[\Gamma_T \gamma^+\gamma^\mu]}_{11}(x,b;\mu,\zeta)
\)
\\\nn &&
+\frac{i}{2xp^+}\(\mathbf{\Phi}_{\mu,21}^{(0)[\gamma^\mu \gamma^+\Gamma_T]}(x,b;\mu,\zeta)-
\mathbf{\Phi}_{\mu,12}^{(0)[\Gamma_T \gamma^+\gamma^\mu]}(x,b;\mu,\zeta)\).
\end{eqnarray}
This expression is well-defined at all orders in perturbation theory, and  reduces to eqn.~(\ref{def:biq}) at LO.

Comparing the parameterizations for bi-quark correlators and physical TMD distributions, we obtain the following set of relations
\begin{eqnarray}\label{def:e}
x\, e&=&
%0+
2\mathbf{h}^{(0)}_{\oplus}~,
\\
x\, e_T^\perp&=&
%0+
2\mathbf{h}^{(0)A\perp}_{\oplus,T}~,
\\
x\, e_L&=&
%0+
2\mathbf{h}^{(0)}_{\oplus,L}~,
\\
x\, e_T&=&
%0+
2\mathbf{h}^{(0)D\perp}_{\oplus,T}~,
\\
x\, f_T&=&
\phantom{+}\mathbf{f}^{(0)}_{\ominus,T}-\mathbf{g}^{(0)}_{\oplus,T}-f_{1T}^\perp-\frac{b^2M^2 }{2}\mathring{f}_{1T}^\perp~,
\\
x\, f_L^\perp&=&
%0+
-\mathbf{f}^{(0)\perp}_{\ominus,L}+\mathbf{g}^{(0)\perp}_{\oplus,L}~,
\\
x\, f^\perp&=&
\phantom{+}\mathbf{f}^{(0)\perp}_{\ominus}-\mathbf{g}^{(0)\perp}_{\oplus}+\mathring{f}_1~,
\\
x\, f_T^\perp&=&
-\mathbf{f}^{(0)\perp}_{\ominus,T}+\mathbf{g}^{(0)\perp}_{\oplus,T}+\mathring{f}_{1T}^\perp~,
\\
x\, g_T&=&
-\mathbf{f}^{(0)}_{\oplus,T}-\mathbf{g}^{(0)}_{\ominus,T}+
g_{1T}+\frac{b^2M^2}{2}\mathring{g}_{1T}~,
\\
x\, g_L^\perp&=&
\phantom{+}\mathbf{f}^{(0)\perp}_{\oplus,L}+\mathbf{g}^{(0)\perp}_{\ominus,L}+\mathring{g}_1~,
\\
x\, g^\perp&=&
%0+
\phantom{+}\mathbf{f}^{(0)\perp}_{\oplus}+\mathbf{g}^{(0)\perp}_{\ominus}~,
\\
x\, g_T^\perp&=&
\phantom{+}\mathbf{f}^{(0)\perp}_{\oplus,T}+\mathbf{g}^{(0)\perp}_{\ominus,T}+\mathring{g}_{1T}~,
\\
x\, h_T^\perp&=&
\phantom{+}2\mathbf{h}^{(0)A\perp}_{\ominus,T}
+\mathring{h}_1-h_{1T}^\perp-\frac{b^2M^2}{4}\mathring{h}_{1T}^\perp
~,
\\
x\, h&=&
-2\mathbf{h}^{(0)}_{\ominus}
-2\(h_1^\perp+\frac{b^2M^2}{2}\mathring{h}_1^\perp\)
~,
\\
x\, h_L^\perp&=&
-2\mathbf{h}^{(0)}_{\ominus,L}
-2\(h_{1L}^\perp+\frac{b^2M^2}{2}\mathring{h}_{1L}^\perp\)
~,
\\\label{def:hT}
x\, h_T&=&
-2\mathbf{h}^{(0)D\perp}_{\ominus,T}
-\mathring{h}_1-h_{1T}^\perp-\frac{b^2M^2}{4}\mathring{h}_{1T}^\perp
~,
\end{eqnarray}
where we omit the argument $(x,b;\mu,\zeta)$ for all functions, and
\begin{eqnarray}
\mathring{F}=\frac{2}{M^2}\frac{\partial}{\partial b^2}F.
\end{eqnarray}
In the space of transverse momentum (Fourier conjugated to $b$) these relations were derived in ref.\cite{Bacchetta:2006tn}.

The relations (\ref{def:e} - \ref{def:hT}) involves only a half of the twist-three quark-gluon-quark distributions, namely
\begin{eqnarray}
\label{eqn:pairs}
\(\begin{array}{c}
f_{\oplus}^\perp+g_{\ominus}^\perp\\f_{\ominus}^\perp-g_{\oplus}^\perp,
\end{array}\),
\quad
\(\begin{array}{c}
f_{\oplus,L}^\perp+g_{\ominus,L}^\perp\\f_{\ominus,L}^\perp-g_{\oplus,L}^\perp
\end{array}\),
&\quad&
\(\begin{array}{c}
f_{\oplus,T}+g_{\ominus,T}\\f_{\ominus,T}-g_{\oplus,T}
\end{array}\),
\quad
\(\begin{array}{c}
f_{\oplus,T}^\perp+g_{\ominus,T}^\perp\\ f_{\ominus,T}^\perp-g_{\oplus,T}^\perp
\end{array}\),
\\\nn
\(\begin{array}{c}
h_{\oplus}\\h_{\ominus}
\end{array}\),
\quad
\(\begin{array}{c}
h_{\oplus,L}\\h_{\ominus,L}
\end{array}\),
&\quad&
\(\begin{array}{c}
h_{\oplus,T}^{A\perp}\\h_{\ominus,T}^{A\perp}
\end{array}\),
\quad
\(\begin{array}{c}
h_{\oplus,T}^{D\perp}\\h_{\ominus,T}^{D\perp}
\end{array}\).
\end{eqnarray}
These are the distributions with the quark-gluon pair having opposite helicities only, i.e. only $\Uparrow\downarrow$ or $\Downarrow\uparrow$. The pairs (\ref{eqn:pairs}) evolve autonomously with the kernel $\mathbb{P}_A$ (\ref{def:PA}) (see sec.~\ref{sec:evolution-T-definite}), and do not mix with other distributions. Elements of a pair have opposite T-parity, and mix only due to non-zero $\Theta$. This mixture implies that pairs of twist-three bi-quark distributions $\{f,g\}$, $\{h,e\}$ (with the same labels) are intrinsically connected.

The rest distributions,
\begin{eqnarray}
&&\nn
f_{\ominus,T}+g_{\oplus,T},
\quad
f_{\ominus,L}^\perp+g_{\oplus,L}^\perp,
\quad
f_{\ominus,T}^\perp+g_{\oplus,T}^\perp,
\quad
f_{\ominus}^\perp+g_{\oplus}^\perp,
\\\label{eqn:listB}
&&
f_{\oplus,T}-g_{\ominus,T},
\quad
f_{\oplus,L}^\perp-g_{\ominus,L}^\perp,
\quad
f_{\oplus,T}^\perp-g_{\ominus,T}^\perp,
\quad
f_{\oplus}^\perp-g_{\ominus}^\perp,
\\\nn
&&
h^\perp_{\oplus},
\quad 
h_{\oplus,L}^{\perp},
\quad
h_{\oplus,T}^{T\perp},
\quad
h_{\oplus,T}^{S\perp},
\quad
h^\perp_{\ominus},
\quad 
h_{\ominus,L}^{\perp},
\quad
h_{\ominus,T}^{T\perp},
\quad
h_{\ominus,T}^{S\perp},
\end{eqnarray}
do not contribute to the NLP factorization theorem, and evolve with the kernel $\mathbb{P}_B$ (\ref{def:PB}). However, they could appear at NNLP factorization.

\subsection{Evolution equations}
\label{sec:large-NC}

To find the evolution of the bi-quark distributions one needs to compute the zeroth moment of the evolution kernels (\ref{eqn:evol-correlator+}, \ref{eqn:evol-correlator-}). This computation is presented in the appendix \ref{app:0moment}. The resulting expression does not reproduces the zeroth momentum and has also terms with convolutions of twist-three distributions. Therefore, the evolution equations with respect to $\mu$ for the bi-quark distributions has the following schematic form
\begin{eqnarray}\label{eqn:biq:evol}
\mu^2 \frac{d}{d\mu^2}
\(\begin{array}{c}F_+ \\ F_- \end{array}\)
=
\(\frac{\Gamma_{\text{cusp}}}{2}\ln\(\frac{\mu^2}{\zeta}\)+\gamma_1\)\(\begin{array}{c}F_+ \\ F_- \end{array}\)
-\(\gamma_1+\frac{\gamma_V}{2}\)\frac{1}{x}
\(\begin{array}{c} f_+ \\ f_- \end{array}\)
\\\nn+
\frac{1}{x}\int \frac{[dx]}{x_2}
\(
\begin{array}{cc}
2\mathbb{P} & 2\pi s\Theta
\\
-2\pi s\Theta & 2\mathbb{P}
\end{array}\)
\(\begin{array}{c}\Phi_+ \\ \Phi_- \end{array}\),
\end{eqnarray}
where $\{F_+,F_-\}$ is a pair bi-quark distributions ($\{f,g\}$ or $\{h,e\}$ with the same labels), $\{f_+,f_-\}$ are the corresponding twist-two parts, and $\{\Phi_+,\Phi_-\}$ are twist-three distributions (\ref{eqn:pairs}). The cusp anomalous dimension $\Gamma_{\text{cusp}}$ and the quark anomalous dimension $\gamma_V$ are defined in eqn.~(\ref{eqn:gammaCusp-LO}). The anomalous dimension $\gamma_1$ at LO is
\begin{eqnarray}\label{eqn:gammaN}
\gamma_1=a_s(\mu)C_F+\mathcal{O}(a_s^2).
\end{eqnarray}
The mixing with twist-two terms depend on the specific distribution and are given in eqns.~(\ref{def:e-evol} -- \ref{def:hT-evol}). The last ineqn~\eqref{eqn:biq:evol} term mixes distributions with different T-parity. The kernels $\mathbb{P}$ and $\Theta$ are combinations of $\mathbb{P}_A$ (\ref{def:PA}), $\Upsilon$ (\ref{def:Upsilon123}) and $\Theta$ (\ref{def:Theta123}). They are given in eqn.~(\ref{app:P}). 

Since equations (\ref{eqn:biq:evol}) do not have practical importance we do not write anomalous dimensions and kernels explicitly. If necessary they could be found combining definitions (\ref{def:e} -- \ref{def:hT}), with equations (\ref{evolEQN:evenA}, \ref{evolEQN:oddA}, \ref{app:P}). Importantly, these equations significantly simplify in the large-$N_c$ limit, where
%$\Theta=0$ and $\mathbb{P}_A=0$. I.e.
\begin{eqnarray}
\mathbb{P}=\mathcal{O}\(\frac{a_s}{N_c}\),\qquad \Theta=\mathcal{O}\(\frac{a_s}{N_c}\),
\end{eqnarray}
which is derived in appendix \ref{app:0moment}. This structure is also evident in the NLO coefficient function for TMD factorization (see eqn.~(6.15) in ref.\cite{Vladimirov:2021hdn}). 

In the large-$N_c$ limit, the zeroth moments of twist-three TMD distributions from the list (\ref{eqn:pairs}) have autonomous evolution. Alike twist-two distributions, they satisfy the pair of equations
\begin{eqnarray}\label{eqn:FA-large-Nc}
\mu^2 \frac{d}{d\mu^2}F_A^{(0)}(x,b)&=&\(\frac{\Gamma_{\text{cusp}}}{2}\ln\(\frac{\mu^2}{\zeta}\)+\gamma_1\)F_A^{(0)}(x,b)+\mathcal{O}\(\frac{a_s}{N_c}\),
\\\nn
\zeta \frac{d}{d\zeta}F_A^{(0)}(x,b)&=&-\mathcal{D}(b,\mu)F_A^{(0)}(x,b),
\end{eqnarray}
where $F_A^{(0)}$ is the zeroth moment (\ref{def:0moment}) of any twist-three distribution listed in eqn.~(\ref{eqn:pairs}), $\Gamma_{\text{cusp}}$ is the cusp anomalous dimension, and $\gamma_1$ is given in (\ref{eqn:gammaN}). The evolution for the zeroth moments of other twist-three distributions, i.e. those that are listed in eqn.~(\ref{eqn:listB}), does not simplify in the large-$N_c$ limit. It is given in equation (\ref{app:P}).

Consequently, the evolution equations for bi-quark distributions in the large-$N_c$ limit are also closed. This observation is analogous to the situation with the twist-three structure functions for DIS and baryon distribution amplitudes, for which the evolution equations at LO and large-$N_c$ are also closed \cite{Belitsky:1999bf,Braun:1998id}. In the collinear case, this effect happens due to the conformal properties of the operator \cite{Braun:2003rp}. In the TMD case, we do not have any general explanation yet.

For completeness we provide the complete list of evolution equations in the large-$N_c$ limit explicitly
\begin{align}
\label{def:e-evol}
\mu^2 \frac{d}{d\mu^2}e&=
\(\frac{\Gamma_{\text{cusp}}}{2}\ln\(\frac{\mu^2}{\zeta}\)+a_sC_F\)e~,
\\
\mu^2 \frac{d}{d\mu^2}e_T^\perp&=
\(\frac{\Gamma_{\text{cusp}}}{2}\ln\(\frac{\mu^2}{\zeta}\)+a_sC_F\)e_T^\perp~,
\\
\mu^2 \frac{d}{d\mu^2}e_L&=
\(\frac{\Gamma_{\text{cusp}}}{2}\ln\(\frac{\mu^2}{\zeta}\)+a_sC_F\)e_L~,
\\
\mu^2 \frac{d}{d\mu^2}e_T&=
\(\frac{\Gamma_{\text{cusp}}}{2}\ln\(\frac{\mu^2}{\zeta}\)+a_sC_F\)e_T~,
\\
\mu^2 \frac{d}{d\mu^2}f_T&=
\(\frac{\Gamma_{\text{cusp}}}{2}\ln\(\frac{\mu^2}{\zeta}\)+a_sC_F\)f_T
-\frac{2a_sC_F}{x}\(
f_{1T}^\perp+\frac{b^2M^2 }{2}\mathring{f}_{1T}^\perp\)~,
\\
\mu^2 \frac{d}{d\mu^2}f_L^\perp&=
\(\frac{\Gamma_{\text{cusp}}}{2}\ln\(\frac{\mu^2}{\zeta}\)+a_sC_F\)f_L^\perp~,
\\
\mu^2 \frac{d}{d\mu^2}f^\perp&=
\(\frac{\Gamma_{\text{cusp}}}{2}\ln\(\frac{\mu^2}{\zeta}\)+a_sC_F\)f^\perp
+\frac{2a_sC_F}{x}\mathring{f}_1
~,
\\
\mu^2 \frac{d}{d\mu^2}f_T^\perp&=
\(\frac{\Gamma_{\text{cusp}}}{2}\ln\(\frac{\mu^2}{\zeta}\)+a_sC_F\)f_T^\perp
+\frac{2a_sC_F}{x}\mathring{f}_{1T}^\perp
~,
\\
\mu^2 \frac{d}{d\mu^2}g_T&=
\(\frac{\Gamma_{\text{cusp}}}{2}\ln\(\frac{\mu^2}{\zeta}\)+a_sC_F\)g_T
+\frac{2a_sC_F}{x}\(
g_{1T}+\frac{b^2M^2}{2}\mathring{g}_{1T}\)~,
\\
\mu^2 \frac{d}{d\mu^2}g_L^\perp&=
\(\frac{\Gamma_{\text{cusp}}}{2}\ln\(\frac{\mu^2}{\zeta}\)+a_sC_F\)g_L^\perp
+\frac{2a_sC_F}{x}\mathring{g}_1~,
\\
\mu^2 \frac{d}{d\mu^2}g^\perp&=
\(\frac{\Gamma_{\text{cusp}}}{2}\ln\(\frac{\mu^2}{\zeta}\)+a_sC_F\)g^\perp~,
\\
\mu^2 \frac{d}{d\mu^2}g_T^\perp&=
\(\frac{\Gamma_{\text{cusp}}}{2}\ln\(\frac{\mu^2}{\zeta}\)+a_sC_F\)g_T^\perp
+\frac{2a_sC_F}{x} \mathring{g}_{1T}~,
\\
\mu^2 \frac{d}{d\mu^2}h_T^\perp&=
\(\frac{\Gamma_{\text{cusp}}}{2}\ln\(\frac{\mu^2}{\zeta}\)+a_sC_F\)h_T^\perp
+\frac{2a_sC_F}{x}\(
\mathring{h}_1-h_{1T}^\perp-\frac{b^2M^2}{4}\mathring{h}_{1T}^\perp\)~,
\\
\mu^2 \frac{d}{d\mu^2}h&=
\(\frac{\Gamma_{\text{cusp}}}{2}\ln\(\frac{\mu^2}{\zeta}\)+a_sC_F\)h
-\frac{4a_sC_F}{x}\(h_1^\perp+\frac{b^2M^2}{2}\mathring{h}_1^\perp\)~,
\\
\mu^2 \frac{d}{d\mu^2}h_L^\perp&=
\(\frac{\Gamma_{\text{cusp}}}{2}\ln\(\frac{\mu^2}{\zeta}\)+a_sC_F\)h_L^\perp
-\frac{2a_sC_F}{x}\(h_{1L}^\perp+\frac{b^2M^2}{4}\mathring{h}_{1L}^\perp\)~,
\\\label{def:hT-evol}
\mu^2 \frac{d}{d\mu^2}h_T&=
\(\frac{\Gamma_{\text{cusp}}}{2}\ln\(\frac{\mu^2}{\zeta}\)+a_sC_F\)h_T
-\frac{2a_sC_F}{x}\(\mathring{h}_1+h_{1T}^\perp+\frac{b^2M^2}{4}\mathring{h}_{1T}^\perp\),
\end{align}
where we omit the arguments $(x,b;\mu,\zeta)$ for each function, and the term $\mathcal{O}(a_s/N_c)$ for brevity. Note, that the twist-two and twist-three parts of bi-quark distributions can be evolved separately, and added together afterwards.

The evolution with respect to scale $\zeta$ is much simpler, thanks to the physical distributions $\mathbf{\Phi}$ in the definition (\ref{biQ:EOMSs2}). Since twist-two and twist-three distributions satisfies the same evolution with $\zeta$, the only deviation from ordinary evolution equations happen due derivatives of twist-two terms. For completeness we provide the list of evolution equations with respect to $\zeta$
\begin{align}
\label{def:e-evol-rap}
\zeta \frac{d}{d\zeta}e&=
-\mathcal{D} \,e~,
&
\zeta \frac{d}{d\zeta}e_T^\perp&=
-\mathcal{D}\,e_T^\perp~,
\\
\zeta \frac{d}{d\zeta}e_L&=
-\mathcal{D}\,e_L~,
&
\zeta \frac{d}{d\zeta}e_T&=
-\mathcal{D}\,e_T~,
\\
\zeta \frac{d}{d\zeta}f_T&=
-\mathcal{D}\,f_T
+\frac{b^2M^2 }{2 x}\mathring{\mathcal{D}}\,f_{1T}^\perp~,
&
\zeta \frac{d}{d\zeta}f_L^\perp&=
-\mathcal{D}f_L^\perp~,
\\
\zeta \frac{d}{d\zeta}f^\perp&=
-\mathcal{D}\,f^\perp
-\mathring{\mathcal{D}}\,\frac{f_1}{x}
~,
&
\zeta \frac{d}{d\zeta}f_T^\perp&=
-\mathcal{D}\,f_T^\perp
-\mathring{\mathcal{D}}\,\frac{f_{1T}^\perp}{x}
~,
\\
\zeta \frac{d}{d\zeta}g_T&=
-\mathcal{D}\,g_T
-\frac{b^2M^2}{2x}\mathring{\mathcal{D}}\,g_{1T}~,
&
\zeta \frac{d}{d\zeta}g_L^\perp&=
-\mathcal{D}\,g_L^\perp
-\mathring{\mathcal{D}}\frac{g_1}{x}~,
\\
\zeta \frac{d}{d\zeta}g^\perp&=
-\mathcal{D}\,g^\perp~,
&
\zeta \frac{d}{d\zeta}g_T^\perp&=
-\mathcal{D}\,g_T^\perp
-\mathring{\mathcal{D}}\frac{g_{1T}}{x}~,
\\
\zeta \frac{d}{d\zeta}h_T^\perp&=
-\mathcal{D}\,h_T^\perp
-\mathring{\mathcal{D}}\frac{h_1}{x}+\frac{b^2M^2}{4x}\mathring{\mathcal{D}}\,h_{1T}^\perp~,
&
\zeta \frac{d}{d\zeta}h&=
-\mathcal{D}\,h
+\frac{b^2M^2}{x}\mathring{\mathcal{D}}\,h_1^\perp~,
\\
\zeta \frac{d}{d\zeta}h_L^\perp&=
-\mathcal{D}\,h_L^\perp
+\frac{b^2M^2}{4x}\mathring{\mathcal{D}}\,h_{1L}^\perp~,
&
\label{def:hT-evol-rap}
\zeta \frac{d}{d\zeta}h_T&=
-\mathcal{D}\,h_T
+\mathring{\mathcal{D}}\frac{h_1}{x}+\frac{b^2M^2}{4x}\mathring{\mathcal{D}}\,h_{1T}^\perp,
\end{align}
where we omit the arguments $(x,b;\mu,\zeta)$ for each distribution and arguments $(b,\mu)$ for the Collins-Soper kernels for brevity. Note that these equations are exact at all orders of perturbation theory.

The derivative of the Collis-Soper kernel is singular at $b\to0$,
\begin{eqnarray}
\mathring{\mathcal{D}}(b,\mu)=\frac{\Gamma_{\text{cusp}}}{2 b^2M^2}+2C_F\beta_0 \frac{a_s}{b^2M^2}+\mathcal{O}\(\frac{a_s^2}{b^2M^2}\)+\text{NP-terms},
\end{eqnarray}
where $\beta_0$ is the LO coefficient of the QCD beta-function. The ``NP-terms'' are non-perturbative terms which are $\mathcal{O}(1)$ \cite{Vladimirov:2020umg}. Additionally, the terms with $\mathring{\mathcal{D}}$ are multiplied by $x^{-1}$.  Therefore, the contribution of $\mathring{\mathcal{D}}$ to the evolution is numerically large.
From this, we expect that the functions $\{f^\perp, f_{T}^\perp, g_L^\perp,  h_T^\perp, h_T\}$ are larger than other bi-quark TMD distributions. 
Although this effect can be isolated to the twist-two part, we observe that the genuine twist-three part of these functions have a singular small-$b$ behavior (see appendix \ref{app:small-b}). This implies that the functions $\{f^\perp, f_{T}^\perp, g_L^\perp,  h_T^\perp, h_T\}$ contribute to the cross-section (at large $p_T$) at one power higher in comparison to na\"ive counting. This observation could resolve the old-stated problem of mismatches in power counting between fixed-order and TMD factorization for some observable \cite{Bacchetta:2008xw}. The possibility of such a scenario was discussed in ref.\cite{Bacchetta:2019qkv}. The detailed study of this mechanism is left for future publications.

\section{Conclusion}

This paper presents a study of twist-three TMD distributions starting from the operator definition obtained within the NLP factorization theorem \cite{Vladimirov:2021hdn}. We discuss and demonstrate issues of this definition, such as complex-valued functions, discontinuities at $x_i=0$, and others. We revise these issues step-by-step, modifying the definition of twist-three TMD distributions in order to solve them. In such a way, we introduce physical TMD distributions of twist-three, which are well-defined in all aspects and can be used for practical applications. The results of this work are of principal importance for the description of available and future data on spin asymmetries in Drell-Yan and SIDIS processes. The recent measurements by CLAS \cite{CLAS:2021opg} allow testing of twist-three distributions' predicted scaling and asymptotic values. 

The bottom-top strategy allows us to keep track of the properties of TMD distributions and derive them with minimal efforts. So, we equip the definition with the evolution equations, LO evolution kernels, symmetry relations, discussions on interpretation, support properties, and small-b limit. We also link the genuine distributions discussed here with the generic distributions. It provides the evolution equations for the latter. In this work, we consider only quark TMDPDFs. The study of TMDFFs and/or gluon distributions can be made in the same way. 

To our best knowledge, it is the first systematic first-principles study of genuine TMD distributions of twist-three. Let us emphasize several points that we consider to be the most important.
\begin{itemize}
\item There are 32 genuine TMDPDFs of twist-three. Only half of them contribute to the NLP cross-section of the SIDIS or Drell-Yan process. Out of these TMDPDFs, 16 are T-odd, i.e., they change signs between DY and SIDIS definitions.
\item The evolution equations for twist-three TMD distributions mix T-even and T-odd TMD distributions. This mixture is proportional to a process-dependent sign factor, such that each function preserves its T-parity and in-between-process universality.
\item The genuine twist-three TMD distributions are generalized functions. Namely, their value is not defined for some points of support domain (most importantly for the vanishing gluon momentum), albeit they have definite integrals.
\item The convolutions that contribute to the factorized cross-section are singular due to discontinuities of TMD distributions, albeit producing a finite cross-section thanks to the cancellation of divergences between different terms. The value of discontinuity can be computed and subtracted. It gives rise to the definition of the physical TMD distribution of twist-three, which provides a term-by-term finite cross-section.
\item Some of the genuine twist-three TMD distributions have singular behavior at small-$b$, namely as $b^{-1}$. This behavior changes the na\"ive counting of terms in cross-section at large transverse momenta, such as the one studied in ref.\cite{Bacchetta:2008xw}.
\item The evolution of bi-quark twist-three TMD distributions is autonomous in the large-$N_c$ limit.
\end{itemize}
Each of these points is important for the further development of the theory and practice of twist-three TMD distributions.

\acknowledgments Authors are thankful to V.Braun, I.Scimemi and H.Avakian for discussions, helpful comments and suggestions. This study was supported by Deutsche Forschungsgemeinschaft (DFG) through the research Unit FOR 2926, “Next Generation pQCD for Hadron Structure: Preparing for the EIC”, project number 30824754. 

\appendix

\section{Convention for vector algebra}
\label{app:conventions}

Along the paper, we use the following convention for light-cone decomposition with vectors $n^\mu$ and $\bar n^\mu$ ($n^2=\bar n^2=0$ and $(n\cdot \bar n)=1$). The components of vectors are 
\begin{eqnarray}
v^\mu=v^+\bar n^\mu+v^-n^\mu+v_T^\mu,
\end{eqnarray}
where $v_T$ is the transverse component. The symmetric and anti-symmetric tensors acting in the transverse space are
\begin{eqnarray}\label{def:gT,eT}
g_T^{\mu\nu}=g^{\mu\nu}-n^\mu\bar n^\nu-\bar n^\mu n^\nu,\qquad \epsilon_T^{\mu\nu}=\bar n_\alpha n_\beta \epsilon^{\alpha\beta \mu\nu}=\epsilon^{-+\mu\nu},
\end{eqnarray}
with $\epsilon^{0123}=+1$. 

The (massless) hadron moves along $\bar n^\mu$, i.e. its momentum is $p^\mu=p^+ \bar n^\mu$. It defines the good component ($\xi$) and  the bad component ($\eta$) of the quark field as
\begin{eqnarray}
\xi=\frac{\gamma^-\gamma^+}{2}q,\qquad 
\eta=\frac{\gamma^+\gamma^-}{2}q.
\end{eqnarray}
These projectors split the space of Dirac matrices $\Gamma$ into three subspaces $\{\Gamma^+,\Gamma_T,\Gamma^-\}$, such that any quark-bilinear has decomposition
\begin{eqnarray}\label{app:qGq=...}
\bar q\Gamma q=\bar \xi\Gamma^+\xi+\bar \xi \Gamma_T\eta+\bar \eta \Gamma_T\xi+\bar \eta \Gamma^- \eta.
\end{eqnarray}
The only non-vanishing projections of Dirac matrices for each set are
\begin{eqnarray}\label{app:G+GTG-}
\frac{\gamma^+\gamma^-}{2}\Gamma^+\frac{\gamma^-\gamma^+}{2}\neq 0,
\qquad
\frac{\gamma^+\gamma^-}{2}\Gamma_T\frac{\gamma^+\gamma^-}{2}\neq 0,
\qquad
\frac{\gamma^-\gamma^+}{2}\Gamma^-\frac{\gamma^+\gamma^-}{2}\neq 0.
\end{eqnarray}
The standard decomposition bases for $\Gamma^+$  and $\Gamma_T$ are
\begin{eqnarray}
\Gamma^+_{\text{basis}}&=&\{\gamma^+, \gamma^+\gamma^5, i\sigma^{\alpha +}\gamma^5\},
\\\label{app:GammaT}
\Gamma_{T,\text{basis}}&=&\{1,i\gamma^5,\gamma^\alpha,\gamma^\alpha\gamma^5,i\sigma^{\alpha\beta}\gamma^5,i\sigma^{+-}\gamma^5\},
\end{eqnarray}
where $\alpha$ and $\beta$ are transverse indices. The $\sigma$ and $\gamma^5$ matrices defined as
\begin{eqnarray}
\sigma^{\mu\nu}=\frac{i(\gamma^\mu\gamma^\nu-\gamma^\nu\gamma^\mu)}{2},\qquad
\gamma^5=i\gamma^0\gamma^1\gamma^2\gamma^3=\frac{-i}{4!}\epsilon^{\mu\nu\rho\sigma}\gamma_\mu \gamma_\nu \gamma_\rho \gamma_\sigma=\frac{-i}{2}\epsilon_T^{\mu\nu}\gamma^+\gamma^-\gamma^\mu\gamma^\nu.
\end{eqnarray}
\begin{eqnarray}
\gamma^\alpha \gamma^5=i \epsilon^{\alpha\beta}_T\gamma^+\gamma^-\gamma_{\beta},
\end{eqnarray}
where $\alpha$ is the transverse index.

\section{Evolution kernels}
\label{app:collinear}

In this appendix we provide additional information about evolution kernels for twist-three distributions.

\subsection{Elementary kernels}

The evolution kernel $\mathbb{H}$ has an convoluted form. For practical manipulations, it is convenient to consider it as a sum of elementary kernels acting to the parton fields. We have
\begin{eqnarray}\label{app:H}
\mathbb{H}_{z_2z_3}U_{\mu}(z_2,z_3)&=&-C_A \widehat{\mathcal{H}}U_{\mu}(z_2,z_3)+C_A\mathcal{H}^+ \gamma_\mu\gamma^\nu U_{\nu}(z_2,z_3)
\\\nn &&-2\(C_F-\frac{C_A}{2}\)\mathcal{H}^- \gamma_\mu\gamma^\nu U_{\nu}(z_2,z_3)+\(C_F-\frac{C_A}{2}\)P_{12}\mathcal{H}^{e,(1)} \gamma_\nu\gamma^\mu U_{\nu}(z_2,z_3),
\end{eqnarray}
where
\begin{eqnarray}
U_\mu(z_2,z_3)&=&F_{\mu+}(z_2n)q(z_3n).
\end{eqnarray}
The elementary evolution kernels are \cite{Braun:2009vc},
\begin{eqnarray}\nn
\widehat{\mathcal{H}}f(z_2,z_3)&=&\int_0^1 \frac{d\alpha}{\alpha}\(2f(z_2,z_3)-\bar \alpha^2 f(z_{23}^\alpha,z_3)-\bar \alpha f(z_2,z_{32}^\alpha)\),
\\\label{app:Hs}
\mathcal{H}^+f(z_2,z_3)&=&\int_0^1 d\alpha \int_0^{\bar \alpha}d\beta \,\bar \alpha f(z_{23}^\alpha,z_{32}^\beta),
\\\nn
\mathcal{H}^-f(z_2,z_3)&=&\int_0^1 d\alpha \int_{\bar \alpha}^1d\beta \,\bar \alpha f(z_{23}^\alpha,z_{32}^\beta),
\\\nn
P_{12}\mathcal{H}^{e,(1)}f(z_2,z_3)&=&\int_0^1 d\alpha \,\bar \alpha f(z_{32}^\alpha,z_2),
\end{eqnarray}
where $z_{ij}^\alpha=z_i(1-\alpha)+z_j\alpha$. These kernels obey different properties, each one representing various components of conformal transformation. The action of the kernel on $\bar q(z_3)F_{\mu+}(z_2)$ is analogous, but with inverted order of gamma-matrices. The momentum representation can be found in refs.~\cite{Vladimirov:2021hdn,Ji:2014eta}.

\subsection{Evolution kernels for zeroth moments}
\label{app:0moment}

The zeroth moment is defined as
\begin{eqnarray}
f^{(0)}_\mu(z_3)&=&-i\int_{L}^{z_3}dz_2 f(z_2,z_3) e^{-s\delta z_2},
\end{eqnarray}
where $\delta>0$ is a regulator required for determination of phases in the following integrals. The evolution kernels for the zeroth moment can be computed using that
\begin{eqnarray}
\mathcal{H}f^{(0)}(z_3)&=&-i\int_{L}^{z_3}dz_2 \mathcal{H} f(z_2,z_3),
\end{eqnarray}
where $\mathcal{H}$ is elementary kernel defined in eqns.~(\ref{app:Hs}). This computation is easier to perform in terms of the generation functions. Using the shift operator we present
\begin{eqnarray}
f(z_2,z_3)=e^{z_2\partial_{z_2}+z_3\partial_{z_3}}f(0,0),
\end{eqnarray}
where $\partial_{z}$ are derivatives acting on the argument $z$. Then the integrals can be carried out formally. We find
\begin{eqnarray}
\widehat{\mathcal{H}}f^{(0)}(z_3)&=&f^{(0)}(z_3)-i\int_L^{z_3}dz_2 \frac{\partial_{z_2}+\partial_{z_3}}{\partial_3}\ln\(\frac{-s(\partial_2+\partial_3)}{-s\partial_2}\)
f(z_2,z_3),
\\
\mathcal{H}^{+}f^{(0)}(z_3)&=&-\frac{i}{2}\int_L^{z_3}dz_2 \frac{\partial_{z_2}}{\partial_{z_3}}\ln\(\frac{-s(\partial_{z_2}+\partial_{z_3})}{-s\partial_{z_2}}\)
f(z_2,z_3),
\\
\mathcal{H}^{-}f^{(0)}(z_3)&=&\frac{1}{2}f^{(0)}(z_3)+\frac{i}{2}\int_L^{z_3}dz_2 \frac{\partial_{z_3}}{\partial_{z_2}}\ln\(\frac{-s(\partial_{z_2}+\partial_{z_3})}{-s\partial_{z_3}}\)
f(z_2,z_3),
\\
P_{12}\mathcal{H}^{e,(1)}f^{(0)}(z_3)&=&-f^{(0)}(z_3)-i\int_L^{z_3}dz_2 \frac{\partial_{z_2}+\partial_{z_3}}{\partial_{z_2}}\ln\(\frac{-s(\partial_{z_2}+\partial_{z_3})}{-s\partial_{z_3}}\)
f(z_2,z_3).
\end{eqnarray}
In this expression the logarithms and their complex parts are presented and understood in the same way as in eqn.~(\ref{evol:UV:position}) (see footnote \ref{foot:logarithm}). Combining other elements of the evolution equation (\ref{evol:UV:position}) we obtain the LO equation for the zeroth moment of $\Phi_{12}$
\begin{eqnarray}
&&\mu^2 \frac{d}{d\mu^2}\widetilde{\Phi}_{\mu,21}^{(0)[\Gamma]}(z_1,z_3,b)=
a_s(\mu)\Bigg\{
C_F\(1+2\ln\(\frac{\mu^2}{\zeta}\)\)\widetilde{\Phi}_{\mu,21}^{(0)[\Gamma]}(z_1,z_3,b)
\\\nn&&\qquad
-i\int_L^{z_1} dz_2 \Big[
C_A\ln\(\frac{q^+}{+s\partial_{z_3}}\)
+2C_F\ln\(\frac{q^+}{-s\partial_{z_3}}\)
\\\nn &&\qquad\qquad
-2\(C_F-\frac{C_A}{2}\)\(
\frac{\partial_{z_3}}{\partial_{z_2}}\ln\(\frac{q^+}{-s\partial_{z_1}}\)
+\frac{\partial_{z_1}}{\partial_{z_2}}\ln\(\frac{q^+}{-s\partial_{z_3}}\)\)
\Big]\widetilde{\Phi}_{\mu,21}^{[\Gamma]}(z_1,z_2,z_3,b)
\\\nn&&\qquad
-i\int_L^{z_1}dz_2
\Big[
-\frac{C_A}{2}\frac{\partial_{z_2}}{\partial_{z_1}}\ln\(\frac{+s\partial_{z_3}}{-s\partial_{z_2}}\)
+\(C_F-\frac{C_A}{2}\)\ln\(\frac{+s\partial_{z_3}}{-s\partial_{z_1}}\)\Big]\widetilde{\Phi}_{\nu,21}^{[\Gamma\gamma^\nu\gamma^\mu]}(z_1,z_2,z_3,b)\Bigg\},
\end{eqnarray}
where we use that $(\partial_{z_1}+\partial_{z_2}+\partial_{z_3})\Phi=0$, to simplify the expression. Let us emphasize the combination of logarithms in the second line. These logarithms have the same real but distinct imaginary parts. The equation for $\Phi_{12}$ is obtained by replacing $\partial_{z_1}\leftrightarrow\partial_{z_3}$ and reverting the order of gamma-matrices.

In the momentum-fractions space, the kernel reads
\begin{eqnarray}\label{app:P}
&&\mu^2 \frac{d}{d\mu^2}\Phi_{\mu,21}^{(0)[\Gamma]}(x,b)=
a_s(\mu)\Bigg\{
C_F\(1+2\ln\(\frac{\mu^2}{\zeta}\)\)\Phi_{\mu,12}^{(0)[\Gamma]}(x,b)
\\\nn&&\qquad
+\int \frac{[dx]}{x_2-is0}\delta(x-x_3)\Bigg[
2\(C_F-\frac{C_A}{2}\)
\Big[
\frac{x_3}{x_2}\ln\(\frac{|x_3|}{|x_1|}\)
+\frac{i\pi s}{2}\theta^{(0)}_{x_1x_2x_3}\Big]
\Phi_{\mu,21}^{[\Gamma]}(x_1,x_2,x_3,b)
\\\nn&&\qquad
+\Big[
-\frac{C_A}{2}\frac{x_2}{x_1}\ln\(\frac{|x_3|}{|x_2|}\)
+\(C_F-\frac{C_A}{2}\)\ln\(\frac{|x_3|}{|x_1|}\)+i\pi s \theta^{(1)}_{x_1x_2x_3}\Big]\Phi_{\nu,21}^{[\Gamma\gamma^\nu\gamma_\mu]}(x_1,x_2,x_3,b)\Big]\Bigg]\Bigg\}.
\end{eqnarray}
In this expression we set $q^+=|x_3p^+|$ as prescribed by the convention eqn.~(\ref{def:q+}). The expression enclosed by square brackets is regular at $x_2=0$. The complex parts are 
\begin{eqnarray}
\theta^{(0)}_{x_1x_2x_3}=\left\{
\begin{array}{lcc}
-1, &\quad x_{1,2,3}\in & (+,-,-),
\\
1 ,&\quad x_{1,2,3}\in & (+,-,+),
\\
1, &\quad x_{1,2,3}\in & (-,-,+),
\\
1, &\quad x_{1,2,3}\in & (-,+,+),
\\
-1, &\quad x_{1,2,3}\in & (-,+,-),
\\
-1, &\quad x_{1,2,3}\in & (+,+,-),
\end{array}\right.%}
\quad
\theta^{(1)}_{x_1x_2x_3}=\left\{
\begin{array}{lcc}
-\frac{C_A}{2}\frac{x_2}{x_1}, &x_{1,2,3}\in & (+,-,-),
\\
C_F-\frac{C_A}{2} ,&x_{1,2,3}\in & (+,-,+),
\\
0, & x_{1,2,3}\in & (-,-,+),
\\
\frac{C_A}{2}\frac{x_2}{x_1}, &x_{1,2,3}\in & (-,+,+),
\\
-\(C_F-\frac{C_A}{2}\), &x_{1,2,3}\in & (-,+,-),
\\
0, & x_{1,2,3}\in & (+,+,-).
\end{array}\right.%}
\end{eqnarray}
This expression can be compared with the logarithmic part of the hard coefficient function in the factorization theorem \cite{Vladimirov:2021hdn}, and agrees with it. We emphasize the non-trivial cancellation between complex parts of the logarithms.

In the large-$N_c$ limit the expression essentially simplifies
\begin{eqnarray}\label{app:P-largeN}
&&\mu^2 \frac{d}{d\mu^2}\Phi_{\mu,21}^{(0)[\Gamma]}(x,b)=
a_s(\mu)\Bigg\{
C_F\(1+2\ln\(\frac{\mu^2}{\zeta}\)\)\Phi_{\mu,12}^{(0)[\Gamma]}(x,b)
\\\nn&&\qquad
+\int \frac{[dx]}{x_2-is0}\delta(x-x_3)
\Big[
-\frac{C_A}{2}\frac{x_2}{x_1}\ln\(\frac{|x_3|}{|x_2|}\)
+i\pi s \theta^{(1)}_{x_1x_2x_3}\Big]\Phi_{\nu,21}^{[\Gamma\gamma^\nu\gamma_\mu]}(x_1,x_2,x_3,b)\Big]\Bigg\}
\\\nn && \qquad\qquad\qquad+\mathcal{O}\(\frac{a_s}{N_c}\).
\end{eqnarray}
The part of the kernel with $\Phi^{[\Gamma]}$ expresses via the zeroth moment only, whereas the part involving $\Phi^{[\Gamma\gamma^\nu\gamma^\mu]}$ does not. Therefore, the action of the kernel $\mathbb{P}_A$ (\ref{def:PA}) to the zeroth moment expresses via the zeroth moment again in the large-$N_c$ limit, whereas the action of the kernel $\mathbb{P}_B$ (\ref{def:PB}) has a more general structure. 

The definition $q^+$ (\ref{def:q+}) plays the crucial role in the large-$N_c$ simplification. For a general $q^+$ one gets an additional term in the square brackets of  eqn.~(\ref{app:P}) proportional to $\ln(q^+/|x_3p^+|)$. This term survives in the large-$N_c$ limit, and produces the term $2N_c\ln(q^+/|x_3p^+|)\Phi^{[\Gamma]}_{21}$ in the square brackets in eqn.~(\ref{app:P-largeN}). It spoils the simplification of evolution equations for bi-quark TMD distributions. 

\section{On leading behavior of TMD distributions at small-b}
\label{sec:smallb}

At the small values of $b$, the TMD correlators are expressed via the spatially compact light-cone operators. The knowledge of asymptotic small-b expansion is important for phenomenology since it relates collinear and TMD distributions, increasing their mutual predictive power \cite{Bury:2022czx}. In this appendix, we discuss the leading terms of small-$b$ expansion for some of sub-leading TMD distributions, with a particular emphasis on the singular $\sim b^{-1}$ behavior. The list of relations presented here  is incomplete. It lacks some of  the 32 twist-three distributions which have leading matching at higher twist order. The main purpose of this appendix is to present examples of singular small-b behavior, as the phenomenon of the lowering of collinear twist. The complete evaluation and detailed study is a topic left for a separate investigation.

\subsection{Computation of the small-b matching}

The small-b expansion for the twist-two TMD distributions is studied in details. The coefficient functions for leading terms are known at NLO for all TMD distributions \cite{Gutierrez-Reyes:2017glx,Buffing:2017mqm,Scimemi:2019gge,Rein:2022inprep} (except pretzelosity $h_{1T}^\perp$ that has leading term at twist-four, see ref.\cite{Moos:2020wvd}). In some cases the coefficient functions are known at NNLO \cite{Echevarria:2016scs,Gutierrez-Reyes:2018iod,Gutierrez-Reyes:2019rug} and at N$^3$LO \cite{Luo:2020epw,Ebert:2020qef}. 

The tree order of the small-b expansion corresponds to a Taylor expansion of the TMD operator. Clearly, this operation can only increase the geometrical twist of the light-cone operator. Therefore, for an operator with the TMD-twist-(N,M) the leading small-b term has twist-(N+M). Therefore, for the present case the smallest-twist distributions are of twist-three. The leading term is
\begin{eqnarray}\label{smallb:LO-LP}
\widetilde{\Phi}_{\mu,21}^{[\Gamma]}(z_1,z_2,z_3,b)=g \langle p,s|T\{\bar q(z_1n)[z_1n,z_2n]F_{\mu+}[z_2n,z_3n]\frac{\Gamma}{2}q(z_3n)\}|p,s\rangle+\mathcal{O}(b,a_s b^{-1}),
\\\nn
\widetilde{\Phi}_{\mu,12}^{[\Gamma]}(z_1,z_2,z_3,b)=g \langle p,s|T\{\bar q(z_1n)[z_1n,z_2n]F_{\mu+}[z_2n,z_3n]\frac{\Gamma}{2}q(z_3n)\}|p,s\rangle+\mathcal{O}(b,a_s b^{-1}),
\end{eqnarray}
where the infinite parts of Wilson lines are cancelled. The LO/LP expressions for $\Phi_{\mu,12}^{[\Gamma]}$ and $\Phi_{\mu,21}^{[\Gamma]}$ are identical. The corrections $\mathcal{O}(b)$ incorporate higher-twist operators. The corrections $\mathcal{O}(a_s/b)$ incorporate twist-two distributions accompanied by the inverse power of $b$, as it is demonstrated below.

\begin{figure}[t]
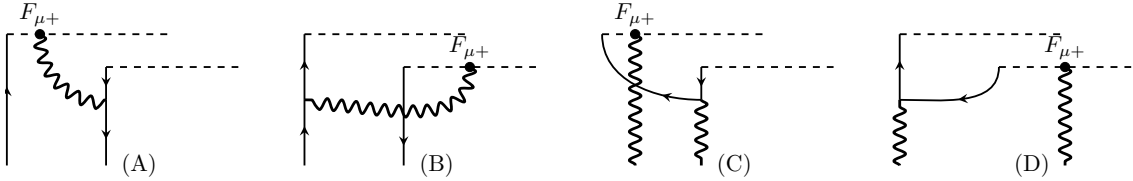

\begin{center}
\includestandalone[width=0.98\textwidth]{Figures/diagrams1}%     without .tex extension
\caption{\label{fig:singular-diag} The diagrams that are singular at $b\to0$, and which provide leading contribution to the matching $\sim a_s/b$ for $\Phi^{[\Gamma]}_{\mu,12}$ (diagrams A and C)  and $\Phi^{[\Gamma]}_{\mu,21}$ (diagrams B and D).}
\end{center}
\end{figure}

The complete computation of the one-loop correction to (\ref{smallb:LO-LP}) is complicated and we do not cover it, for the time being. However, there exists a particular contribution which gives a singular behavior at small-b. This contribution arises from the $2\to1$ diagrams\footnote{
Such diagrams are also present in the renormalization of twist-three collinear operators and twist-two semi-compact operators (see e.g. diagram u2 in fig.8 of ref.~\cite{Vladimirov:2021hdn}). They are power divergent.  However, their renormalization does not mix twists in the $\overline{\text{MS}}$-scheme, where power divergences are removed by definition.}
with two quantum fields turning to a single classical field shown in fig.\ref{fig:singular-diag}. The leading small-b expression for these diagrams is straightforward to compute using the background-field approach, see ref.\cite{Scimemi:2019gge}. The result is UV finite. The quark contribution is given by diagrams (A) and (B)
\begin{eqnarray}\label{smallb:SLO-coord}
\widetilde{\Phi}^{[\Gamma](A)}_{\mu,12}(z_1,z_2,z_3,b)&=&
2ia_s C_F\frac{b^\nu}{b^2}\int_0^1 d\alpha
\langle p,s| \bar q(z_{21}^\alpha n)
\overleftarrow{\partial_+}(\alpha \gamma_\nu \gamma_\mu+\gamma_\mu \gamma_\nu)\frac{\Gamma}{2}q(z_3n)|p,s\rangle,
\\\nn
\widetilde{\Phi}^{[\Gamma](B)}_{\mu,21}(z_1,z_2,z_3,b)&=&
2ia_s C_F\frac{b^\nu}{b^2}\int_0^1 d\alpha 
\langle p,s|\bar q(z_1n)\frac{\Gamma}{2}(\gamma_\nu \gamma_\mu+\alpha\gamma_\mu \gamma_\nu)\partial_+q(z_{23}^\alpha n)|p,s\rangle,
\end{eqnarray}
where we drop the higher power terms (including $\sim b^0$), and $z_{ij}^\alpha$ is defined in eqn.~(\ref{def:zij}). The pure gluon contribution is given by diagrams (C) and (D). At twist-two level their expressions are identical,
\begin{eqnarray}\label{smallb:gluon1}
&&\widetilde{\Phi}^{[\Gamma](C)}_{\mu,12}(z_1,z_2,z_3,b)=
\widetilde{\Phi}^{[\Gamma](D)}_{\mu,21}(z_1,z_2,z_3,b)
\\\nn && \qquad=
2a_sT_F\frac{b_\rho}{b^2}\int_0^1 d\alpha 
\frac{\alpha \Tr(\gamma^\rho\gamma^\nu \gamma^-\Gamma)-\bar\alpha \Tr(\gamma^\nu\gamma^\rho \gamma^-\Gamma)}{2}\langle p,s|F_{\mu+}(z_2n)F_{\nu+}(z_{31}^\alpha n)|p,s\rangle,
\end{eqnarray}
where $T_F=1/2$ is the normalization of $SU(N_c)$ generators. These expressions are $\sim b^\mu/b^2\sim b^{-1}$. Thus, they present the true leading behavior of twist-three TMD distributions at small-b. Note, that we compute only the part proportional to twist-two distributions, neglecting higher twist terms, which could be also singular at $b\to0$.

In the momentum-fraction space the relations (\ref{smallb:SLO-coord}) read
\begin{eqnarray}\label{smallb:SLO-pos1}
&&\Phi_{\mu,12}^{[\Gamma](A)}(x_1,x_2,x_3,b)=
\\\nn &&\qquad
-2a_s C_F\frac{b^\nu}{b^2}\int_0^1 d\alpha \int_{-1}^1 dy
\,y \delta(x_1+\alpha y)\delta(x_2+\bar \alpha y)\(\alpha \Phi^{[\gamma_\nu\gamma_\mu \Gamma]}(y)+\Phi^{[\gamma_\mu\gamma_\nu \Gamma]}(y)\),
\\\nn
&&\Phi_{\mu,21}^{[\Gamma](B)}(x_1,x_2,x_3,b)=
\\\nn &&\qquad
+2a_s C_F \frac{b^\nu}{b^2}\int_0^1 d\alpha \int_{-1}^1 dy
\,y \delta(x_3-\alpha y)\delta(x_2-\bar \alpha y)\(\alpha \Phi^{[\Gamma\gamma_\mu\gamma_\nu ]}(y)+\Phi^{[ \Gamma\gamma_\nu\gamma_\mu]}(y)\),
\end{eqnarray}
where $\Phi^{[\Gamma]}(x)$ is the collinear distributions defined by correlators (\ref{def:z->x}) at $b=0$. The integral over the delta-functions produces 
\begin{eqnarray}\label{smallb:SLO-pos}
&&\Phi_{\mu,12}^{[\Gamma](A)}(x_1,x_2,x_3,b)=
\\\nn&&\qquad
+2a_s C_F \frac{b^\nu}{b^2}
\(
\Phi^{[ \gamma_\mu\gamma_\nu\Gamma]}(x_3)
-\frac{x_1}{x_3} \Phi^{[\gamma_\nu\gamma_\mu\Gamma ]}(x_3)
\)(\theta(x_1,x_2)-\theta(-x_1,-x_2)),\qquad x_1,x_2\neq 0,
\\\nn
&&\Phi_{\mu,21}^{[\Gamma](B)}(x_1,x_2,x_3,b)=
\\\nn&&\qquad
+2a_s C_F\frac{b^\nu}{b^2}
\(
\Phi^{[ \Gamma\gamma_\nu\gamma_\mu]}(-x_1)
-\frac{x_3}{x_1} \Phi^{[ \Gamma \gamma_\mu\gamma_\nu]}(-x_1)
\)
(\theta(x_2,x_3)-\theta(-x_2,-x_3)),\qquad x_2,x_3\neq 0,
\end{eqnarray}
where the relation $x_1+x_2+x_3=0$ is used for simplification. 

In the momentum-fraction space the relations (\ref{smallb:gluon1}) read
\begin{eqnarray}\label{smallb:gluon2}
&&\Phi^{[\Gamma](C)}_{\mu,12}(x_1,x_2,x_3,b)=
\Phi^{[\Gamma](D)}_{\mu,21}(x_1,x_2,x_3,b)
\\\nn && \qquad=
a_sT_F\frac{b_\rho}{b^2}\int_0^1 d\alpha \int_{-1}^1 dy\,y\delta(x_1-\alpha y)\delta(x_3-\bar \alpha y)
\frac{\alpha \Tr(\gamma^\rho\gamma^\nu \gamma^-\Gamma)-\bar\alpha \Tr(\gamma^\nu\gamma^\rho \gamma^-\Gamma)}{2}\Phi_{\mu\nu}(y),
\end{eqnarray}
where $\Phi_{\mu\nu}$ is defined in eqn.(\ref{app:gluon-def}). Integrating over $\delta$-function we obtain
\begin{eqnarray}\label{smallb:gluon3}
&&\Phi^{[\Gamma](C)}_{\mu,12}(x_1,x_2,x_3,b)=
\Phi^{[\Gamma](D)}_{\mu,21}(x_1,x_2,x_3,b)
\\\nn && 
=-a_sT_F\frac{b_\rho}{b^2}
\frac{x_1 \Tr(\gamma^\rho\gamma^\nu \gamma^-\Gamma)-x_3 \Tr(\gamma^\nu\gamma^\rho \gamma^-\Gamma)}{ 2x_2}\Phi_{\mu\nu}(-x_2)(\theta(x_1,x_3)-\theta(-x_1,-x_3)),\quad x_1,x_3\neq 0,
\end{eqnarray}
where we used $x_1+x_2+x_3=0$ for simplification.

At $x_1=0$, or $x_2=0$, or $x_3=0$ these expressions are not defined, but have finite integral over these points. Let us for concreteness inspect the diagram B. Once $x_2$ or $x_3$ turns to zero, the product of $\delta$ function became indefinite, in the sense that its result depends on the order of integration over $\alpha$ and $y$. However, performing the integration over $x_2$ or $x_3$, one specifies the order of integration and the result of integration is definite. Therefore, the functions $\Phi_{\mu,12}^{[\Gamma]}$ and $\Phi_{\mu,21}^{[\Gamma]}$ should be understood as \textit{generalized functions}.

 It is important to mention that the formula (\ref{smallb:gluon3}) has finite zeroth moment, despite an extra factor $1/x_2$. Indeed, computing the zeroth moment of the diagram (D) we find for $x>0$
\begin{eqnarray}
&&\int [dx]\frac{\Phi^{[\Gamma](D)}_{\mu,21}(x_1,x_2,x_3,b)}{x_2}\delta(x_3-x)
\\\nn &&\qquad\qquad=
a_sT_F\frac{b_\rho}{b^2}\int_x^1 \frac{d\alpha}{\alpha}\frac{\bar \alpha \Tr(\gamma^\rho\gamma^\nu \gamma^-\Gamma)-\alpha \Tr(\gamma^\nu\gamma^\rho \gamma^-\Gamma)}{2}\Phi_{\mu\nu}\(\frac{x}{\alpha}\),
\end{eqnarray}
and similar expression for the diagram (C). Therefore, the gluon diagrams do not contribute to the special rapidity divergence discussed in sec.\ref{sec:x2=0}.

\subsection{Leading small-b expressions for TMD distributions}

The matching expression for TMD distributions are obtained by comparing the parameterizations for TMD and collinear matrix elements. The standard parameterization for collinear matrix elements is summarized in appendix \ref{app:collinear} for convenience.  The TMD distributions with non-zero regular matching are
\begin{eqnarray}\nn
f_{\oplus T}(x_1,x_2,x_3,b)&=&T(x_1,x_2,x_3)+\mathcal{O}(a_s),
\\
g_{\ominus T}(x_1,x_2,x_3,b)&=&-\Delta T(x_1,x_2,x_3)+\mathcal{O}(a_s),
\\\nn
h_{\oplus}(x_1,x_2,x_3,b)&=&\delta T_\epsilon(x_1,x_2,x_3)+\mathcal{O}(a_s),
\\\nn
h_{\ominus L}(x_1,x_2,x_3,b)&=&-\delta T_g(x_1,x_2,x_3)+\mathcal{O}(a_s).
\end{eqnarray}
The distributions which have singular matching are
\begin{eqnarray}\nn
f^\perp_{\oplus L}(x_1,x_2,x_3,b)&=&\frac{2a_s}{M^2 b^2}\Big[
-C_F\frac{x_2}{x_1}g_1(-x_1)\(\theta(x_2,x_3)-\theta(-x_2,-x_3)\)
\\\nn &&\qquad +
T_F\frac{x_1-x_3}{x_2}f_g(-x_2)\(\theta(x_1,x_3)-\theta(-x_1,-x_3)\)\Big]
+\mathcal{O}\(b^2,\frac{a_s^2}{b^2}\),
\\\nn
f^\perp_{\ominus}(x_1,x_2,x_3,b)&=&\frac{2a_s}{M^2 b^2}
\Big[
-C_F\frac{x_1-x_3}{x_1}f_1(-x_1)\(\theta(x_2,x_3)-\theta(-x_2,-x_3)\)
\\\nn &&\qquad +
T_F\frac{x_1-x_3}{x_2}f_g(-x_2)\(\theta(x_1,x_3)-\theta(-x_1,-x_3)\)\Big]
+\mathcal{O}\(b^2,\frac{a_s^2}{b^2}\),
\\\nn
g^\perp_{\oplus}(x_1,x_2,x_3,b)&=&\frac{2a_s}{M^2 b^2}\Big[
C_F\frac{x_2}{x_1}f_1(-x_1)\(\theta(x_2,x_3)-\theta(-x_2,-x_3)\)
\\\nn &&\qquad
-T_Fg_g(-x_2)\(\theta(x_1,x_3)-\theta(-x_1,-x_3)\)\Big]
+\mathcal{O}\(b^2,\frac{a_s^2}{b^2}\),
\\\label{smallb:sigula-matching}
g^\perp_{\ominus L}(x_1,x_2,x_3,b)&=&\frac{2a_s}{M^2 b^2}
\Big[C_F
\frac{x_1-x_3}{x_1}g_1(-x_1)\(\theta(x_2,x_3)-\theta(-x_2,-x_3)\)
\\\nn &&\qquad
+T_Fg_g(-x_2)\(\theta(x_1,x_3)-\theta(-x_1,-x_3)\)\Big]
+\mathcal{O}\(b^2,\frac{a_s^2}{b^2}\),
\\\nn
h^{D\perp}_{\ominus T}(x_1,x_2,x_3,b)&=&\frac{2a_sC_F}{M^2 b^2}\frac{x_3}{x_1}h_1(-x_1)\(\theta(x_2,x_3)-\theta(-x_2,-x_3)\)+\mathcal{O}\(b^2,\frac{a_s^2}{b^2}\),
\\\nn
h^{A\perp}_{\ominus T}(x_1,x_2,x_3,b)&=&\frac{2a_sC_F}{M^2 b^2}\frac{x_3}{x_1}h_1(-x_1)\(\theta(x_2,x_3)-\theta(-x_2,-x_3)\)+\mathcal{O}\(b^2,\frac{a_s^2}{b^2}\),
\\\nn
h^{S\perp}_{\ominus T}(x_1,x_2,x_3,b)&=&-\frac{4a_sC_F}{M^2 b^2}h_1(-x_1)\(\theta(x_2,x_3)-\theta(-x_2,-x_3)\)+\mathcal{O}\(b^2,\frac{a_s^2}{b^2}\),
\\\nn
h^{T\perp}_{\ominus T}(x_1,x_2,x_3,b)&=&\frac{4a_sC_F}{M^2 b^2}h_1(-x_1)\(\theta(x_2,x_3)-\theta(-x_2,-x_3)\)+\mathcal{O}\(b^2,\frac{a_s^2}{b^2}\).
\end{eqnarray}
Let us mention, that the T-odd counter-partner distributions to (\ref{smallb:sigula-matching}) are zero at this order. Nonetheless, they receives the singular matching at the two-loop level, at least due to the evolution equations (\ref{evolEQN:even}, \ref{evolEQN:oddA}, \ref{evolEQN:oddB}). Therefore, we have
\begin{eqnarray}
\{ f^\perp_{\ominus L}, 
f^\perp_{\oplus},
g^\perp_{\ominus},
g^\perp_{\oplus L},
h^{D\perp}_{\oplus T},
h^{A\perp}_{\oplus T},
h^{S\perp}_{\oplus T},
h^{T\perp}_{\oplus T}\} = s \times \mathcal{O}\(\frac{a_s^2}{b^2}\).
\end{eqnarray}

These are only a part of small-$b$ relations. To obtain the leading term for the left 12 distributions, one needs to compute a one power higher. Without this computation we cannot strictly identify is the power of the leading small-b term.

The presented here relations violate the na\"ive expectations about power counting of TMD distributions in the large-$p_T$ asymptotic. It explains the problems with ``mismatching'' between the fixed order computations and the leading twist TMD factorization observed in ref.\cite{Bacchetta:2008xw}. In ref. \cite{Bacchetta:2019qkv} authors discuss a possible resolution of this problem by applying ``violated'' counting. Comparing the known asymptotics for the SIDIS structure function $F_{UU}^{\cos\phi}$, and computing the deficit, they derive the large-$k_T$ expression for the TMD bi-quark function $f^\perp$ (\ref{def:biq:V}). Their result agrees with our expressions for $f^\perp_\ominus$ and $g^\perp_{\oplus}$ in the power-scaling, but has the different coefficient. It is due to the fact the authors of ref.\cite{Bacchetta:2008xw} ignore the contribution of twist-four distributions to the large-$p_T$ asymptotic. 

\subsection{Parameterization for collinear distributions}
\label{app:small-b}

The standard parameterization for the collinear distributions of the twist-two \cite{Jaffe:1996zw} reads
\begin{eqnarray}
\langle p,s|\bar q(zn)[zn,0]\frac{\gamma^+}{2}q(0)|p,s\rangle&=&
p^+\int_{-1}^1 dx e^{ix zp^+} f_1(x),
\\
\langle p,s|\bar q(zn)[zn,0]\frac{\gamma^+\gamma^5}{2}q(0)|p,s\rangle&=&
\lambda p^+\int_{-1}^1 dx e^{ix zp^+} g_1(x),
\\
\langle p,s|\bar q(zn)[zn,0]\frac{i\sigma^{\alpha+}\gamma^5}{2}q(0)|p,s\rangle&=&
s_T^\alpha p^+\int_{-1}^1 dx e^{ix zp^+} h_1(x),
\end{eqnarray}
where $f_1$, $g_1$ and $h_1$ are unpolarized, helicity and transversity PDFs (also denoted as $q$, $\Delta q$ and $\delta q$). The $s_T^\mu$ is the transverse part of the spin vector, and $\lambda=M s^+/p^+$ is its longitudinal part. The twist-two PDFs are defined at $-1<x<1$ and for the negative values of $x$ are associated with PDF for anti-quarks
\begin{eqnarray}
f_{1}(x)=\theta(x)f_{1,q\ot h}(x)-\theta(-x)f_{1,\bar q\ot h}(-x),
\end{eqnarray}
and similar for $g_1$ and $h_1$.

The gluon distribution is defined as
\begin{eqnarray}\label{app:gluon-def}
\langle p,s|F_{\mu+}(zn)[zn,0]F_{\nu+}|p,s\rangle&=&\frac{p_+^2}{2}\int_{-1}^1 dx e^{ix zp^+}\,x\,\Phi_{\mu\nu}(x).
\end{eqnarray}
The components of $\Phi_{\mu\nu}$ are
\begin{eqnarray}
\Phi_{\mu\nu}(x)&=&-g_T^{\mu\nu}f_g(x)-i\lambda \epsilon^{\mu\nu}g_{g}(x),
\end{eqnarray}
where $f_g$ and $g_g$ are the unpolarized and helicity gluon distributions.

The twist-three distributions are parametrized as \cite{Scimemi:2018mmi}
\begin{eqnarray}
&&\langle p,s|\bar q(z_1n)F^{\mu+}(z_2n)\frac{\gamma^+}{2}q(z_3n)|p,s\rangle
=
\\\nn&& \qquad\qquad\qquad\qquad p^2_+\epsilon^{\mu\nu}_Ts_{T\nu}M\int [dx]e^{-i(x_1z_1+x_2z_2+x_3z_3)p^+}
T(x_1,x_2,x_3),
\\
&&\langle p,s|\bar q(z_1n)F^{\mu+}(z_2n)\frac{\gamma^+\gamma^5}{2}q(z_3n)|p,s\rangle
=
\\&&\nn\qquad\qquad\qquad\qquad
ip^2_+s_T^\mu M\int [dx]e^{-i(x_1z_1+x_2z_2+x_3z_3)p^+}
\Delta T(x_1,x_2,x_3),
\\
&&\langle p,s|\bar q(z_1n)F^{\mu+}(z_2n)\frac{i\sigma^{\alpha+}\gamma^5}{2}q(z_3n)|p,s\rangle
=
\\\nn &&\qquad\qquad\qquad\qquad
p^2_+\epsilon_T^{\mu\alpha}M\int [dx]e^{-i(x_1z_1+x_2z_2+x_3z_3)p^+}
\delta T_\epsilon(x_1,x_2,x_3)
\\\nn &&\qquad\qquad\qquad\qquad
+ip^2_+\lambda g_T^{\mu\alpha}M\int [dx]e^{-i(x_1z_1+x_2z_2+x_3z_3)p^+}
\delta T_g(x_1,x_2,x_3),
\end{eqnarray}
where we omit Wilson lines in the operator for brevity. The tensors $\epsilon_T^{\mu\nu}$ and $g_T^{\mu\nu}$ are defined in eqn.~(\ref{def:gT,eT}), and $\int [dx]$ is defined in eqn.~(\ref{def:support}). For relation of this notation to other notations see \cite{Scimemi:2018mmi,Braun:2009mi,Braun:2021gvv}. 

% \bibliography{bibFILE}
\input{TMD-evol-sub-leadingP_v2.bbl}
\end{document}

%% file: Figures/diagrams3.tex
\begin{tikzpicture}
%\draw[step=.5cm] (0,0) to[grid with coordinates] (5,3);

%%%%%%%%%%%%%%%%%%%%%%%%%%%%%%%%%%%%%%%%%%%%%%%%%%%%%%%%%%%%%% diag-1
\coordinate (A) at (1, 2);

\draw[thick,postaction={decorate, decoration={
    markings,
    mark=at position 0.6 with {\arrow{stealth}}}}] ($(A)+(0,-2)$)--($(A)+(0,-1)$);

\draw[thick,postaction={decorate, decoration={
    markings,
    mark=at position 0.6 with {\arrow{stealth}}}}] ($(A)+(0,-1)$)--($(A)+(0,0)$);

\draw[very thick,style={decorate, decoration={snake,segment length=6.2}}] 
($(A)+(0.5,0)$) -- ($(A)+(0.5,-2)$);

\draw[very thick,style={decorate, decoration={snake,segment length=6.2}}] 
($(A)+(0.0,-1)$) to[out=0,in=-90] ($(A)+(2.,0.)$);

\draw[blue,very thick,-latex,rounded corners=2pt,dotted] ($(A)+(2.,-0.)$)--($(A)+(2.2,0.2)$)--($(A)+(3.,0.2)$);   

\draw[thick,dashed] ($(A)$) -- ($(A)+(2.5,0)$);

\draw[black,fill=black] ($(A)+(0.5,0)$) circle (2pt);
\node[above] at ($(A)+(0.5,0)$) {$F_{\mu+}$};

\node[] at ($(A)+(2,-2)$) {(c1)};

%%%%%%%%%%%%%%%%%%%%%%%%%%%%%%%%%%%%%%%%%%%%%%%%%%%%%%%%%%%%%% diag-1
\coordinate (A) at (5.5, 2);

\draw[thick,postaction={decorate, decoration={
    markings,
    mark=at position 0.6 with {\arrow{stealth}}}}] ($(A)+(0,-2)$)--($(A)+(0,0)$);

\draw[very thick,style={decorate, decoration={snake,segment length=6.2}}] 
($(A)+(0.5,0)$) -- ($(A)+(0.5,-2)$);

\draw[very thick,style={decorate, decoration={snake,segment length=6.2}}] 
($(A)+(0.5,-1)$) to[out=0,in=-90] ($(A)+(2.,-0.)$);

\draw[blue,very thick,-latex,rounded corners=2pt,dotted] ($(A)+(2.,-0.)$)--($(A)+(2.2,0.2)$)--($(A)+(3.,0.2)$);  

\draw[thick,dashed] ($(A)$) -- ($(A)+(2.5,0)$);

\draw[black,fill=black] ($(A)+(0.5,0)$) circle (2pt);
\node[above] at ($(A)+(0.5,0)$) {$F_{\mu+}$};

\node[] at ($(A)+(2,-2)$) {(c2)};

%%%%%%%%%%%%%%%%%%%%%%%%%%%%%%%%%%%%%%%%%%%%%%%%%%%%%%%%%%%%%% diag-1
\coordinate (A) at (10, 2);

\draw[very thick,style={decorate, decoration={snake,segment length=6.2}}] 
($(A)+(0.0,-1)$) to[out=0,in=-90] ($(A)+(2.,0.)$);

\draw[blue,very thick,-latex,rounded corners=2pt,dotted] ($(A)+(2.,-0.)$)--($(A)+(2.2,0.2)$)--($(A)+(3.,0.2)$);   

\draw[thick,dashed] ($(A)$) -- ($(A)+(2.5,0)$);
\draw[thick,dashed] ($(A)$) -- ($(A)+(0,-2)$);

\node[] at ($(A)+(0.3,-2)$) {$\bar n$};

\node[] at ($(A)+(2,-2)$) {(c3)};

\end{tikzpicture}

%% file: Figures/hexagon21.tex
\begin{tikzpicture}
%\draw[step=.5cm] (-3,-3) to[grid with coordinates] (3,3);
\tikzmath{\s = 1.5;}

%%%%%%%%%%%%%%%%%%%%%%%%%%%%%%%%%%%%%%%%%%%%%%%%%%%%%%%%%%%%%% diag-1
\coordinate (e1) at (0.866025*\s, -0.5*\s);
\coordinate (e2) at (0.*\s, 1.*\s);
\coordinate (e3) at (-0.866025*\s, -0.5*\s);

\draw[-latex,thick,red] (0,0) -- ($3*(e1)$);
\draw[-latex,thick,red] (0,0) -- ($2.6*(e2)$);
\draw[-latex,thick,red] (0,0) -- ($3*(e3)$);

\node[right] at ($3*(e1)$) {$x_1$};
\node[right] at ($2.6*(e2)$) {$x_2$};
\node[left] at ($3*(e3)$) {$x_3$};

\node[above,draw=black!50] at ($0*(e1)+1.4*(e2)-1.4*(e3)$) {\Large $\Phi_{21}$};
\node[right,draw=black!50] at ($1.1*(e1)-1.4*(e2)-0*(e3)$) {$x_1+x_2+x_3=0$};

\draw[] ($-1*(e1)+1*(e2)+0*(e3)$) -- 
		($-1*(e1)+0*(e2)+1*(e3)$) --
		($0*(e1)-1*(e2)+1*(e3)$) --
		($1*(e1)-1*(e2)-0*(e3)$) --
		($1*(e1)+0*(e2)-1*(e3)$) --
		($0*(e1)+1*(e2)-1*(e3)$) --
		($-1*(e1)+1*(e2)+0*(e3)$);

\draw[] (0,0)--($-1*(e1)+1*(e2)+0*(e3)$);
\draw[] (0,0)--($-1*(e1)+0*(e2)+1*(e3)$);
\draw[] (0,0)--($0*(e1)-1*(e2)+1*(e3)$);
\draw[] (0,0)--($1*(e1)-1*(e2)-0*(e3)$);
\draw[] (0,0)--($1*(e1)+0*(e2)-1*(e3)$);
\draw[] (0,0)--($0*(e1)+1*(e2)-1*(e3)$);

\draw[white,fill=white] ($0.6*(e1)-0.3*(e2)-0.3*(e3)$) circle (5pt);
\node[blue!80!green,fill=white,opacity=0.,text opacity=1] at ($0.6*(e1)-0.3*(e2)-0.3*(e3)$) {$\mathbf{(+--)}$};
\node[blue!80!green,fill=white,opacity=0.,text opacity=1] at ($0.3*(e1)-0.6*(e2)+0.3*(e3)$) {$\mathbf{(+-+)}$};
\draw[white,fill=white] ($-0.3*(e1)-0.3*(e2)+0.6*(e3)$) circle (5pt);
\node[blue!80!green,fill=white,opacity=0.,text opacity=1] at ($-0.3*(e1)-0.3*(e2)+0.6*(e3)$) {$\mathbf{(--+)}$};
\node[blue!80!green,fill=white,opacity=0.,text opacity=1] at ($-0.6*(e1)+0.3*(e2)+0.3*(e3)$) {$\mathbf{(-++)}$};
\draw[white,fill=white] ($-0.3*(e1)+0.6*(e2)-0.3*(e3)$) circle (5pt);
\node[blue!80!green,fill=white,opacity=0.,text opacity=1] at ($-0.3*(e1)+0.6*(e2)-0.3*(e3)$) {$\mathbf{(-+-)}$};
\node[blue!80!green,fill=white,opacity=0.,text opacity=1] at ($0.3*(e1)+0.3*(e2)-0.6*(e3)$) {$\mathbf{(++-)}$};

\node[black] at ($0*(e1)+1.05*(e2)-1.05*(e3)$) {\tiny \textbf{(0,1,-1)}};
\node[black] at ($1.16*(e1)+0*(e2)-1.16*(e3)$) {\tiny \textbf{(1,0,-1)}};
\node[black] at ($1.05*(e1)-1.05*(e2)-0*(e3)$) {\tiny \textbf{(1,-1,0)}};
\node[black] at ($0*(e1)-1.05*(e2)+1.05*(e3)$) {\tiny \textbf{(0,-1,1)}};
\node[black] at ($-1.16*(e1)+0*(e2)+1.16*(e3)$) {\tiny \textbf{(-1,0,1)}};
\node[black] at ($-1.05*(e1)+1.05*(e2)-0*(e3)$) {\tiny \textbf{(-1,1,0)}};

%%%%%%%%%%%%%%%%%%%%% -++
\coordinate (C) at ($-2.2*(e1)$);

\draw[fill=black] ($(C)-(0.6,0)$) circle (5pt);
\draw[fill=black] ($(C)+(0.6,0)$) circle (5pt);
\draw[ultra thick] ($(C)+(0.6,0.08)$) -- ($(C)-(0.6,-0.08)$);
\draw[ultra thick] ($(C)+(0.6,0)$) -- ($(C)-(0.6,0)$);
\draw[ultra thick] ($(C)+(0.6,-0.08)$) -- ($(C)-(0.6,0.08)$);

\draw[ thick,postaction={decorate, decoration={markings,
   mark=at position 0.8 with {\arrow{stealth}}}}] ($(C)-(0.7,0)$)--($(C)+(-0.3,0.5)$);

\draw[thick, style={decorate, decoration={snake,amplitude=0.6,segment length=2.2}}] 
($(C)-(0.7,0)$)--($(C)+(-0.3,-0.5)$);

\draw[ thick,postaction={decorate, decoration={markings,
   mark=at position 0.5 with {\arrow{stealth}}}}] ($(C)+(0.3,0.5)$)--($(C)+(0.7,0.)$);

%%%%%%%%%%%%%%%%%%%%% -+-
\coordinate (C) at ($2*(e2)$);
\draw[white,fill=white] (C) circle (6pt);

\draw[fill=black] ($(C)-(0.6,0)$) circle (5pt);
\draw[fill=black] ($(C)+(0.6,0)$) circle (5pt);
\draw[ultra thick] ($(C)+(0.6,0.08)$) -- ($(C)-(0.6,-0.08)$);
\draw[ultra thick] ($(C)+(0.6,0)$) -- ($(C)-(0.6,0)$);
\draw[ultra thick] ($(C)+(0.6,-0.08)$) -- ($(C)-(0.6,0.08)$);

\draw[ thick,postaction={decorate, decoration={markings,
   mark=at position 0.8 with {\arrow{stealth}}}}] ($(C)+(0.7,0.)$)--($(C)+(0.3,-0.5)$);

\draw[thick, style={decorate, decoration={snake,amplitude=0.6,segment length=2.2}}] 
($(C)-(0.7,0)$)--($(C)+(-0.3,0.5)$);

\draw[ thick,postaction={decorate, decoration={markings,
   mark=at position 0.5 with {\arrow{stealth}}}}] ($(C)+(0.3,0.5)$)--($(C)+(0.7,0.)$);
   
%%%%%%%%%%%%%%%%%%%%% ++-
\coordinate (C) at ($-2.2*(e3)$);

\draw[fill=black] ($(C)-(0.6,0)$) circle (5pt);
\draw[fill=black] ($(C)+(0.6,0)$) circle (5pt);
\draw[ultra thick] ($(C)+(0.6,0.08)$) -- ($(C)-(0.6,-0.08)$);
\draw[ultra thick] ($(C)+(0.6,0)$) -- ($(C)-(0.6,0)$);
\draw[ultra thick] ($(C)+(0.6,-0.08)$) -- ($(C)-(0.6,0.08)$);

\draw[ thick,postaction={decorate, decoration={markings,
   mark=at position 0.5 with {\arrow{stealth}}}}] ($(C)+(-0.3,0.5)$)--($(C)-(0.7,0)$);

\draw[thick, style={decorate, decoration={snake,amplitude=0.6,segment length=2.2}}] 
($(C)-(0.5,0)$)--($(C)+(-0.1,0.5)$);

\draw[ thick,postaction={decorate, decoration={markings,
   mark=at position 0.8 with {\arrow{stealth}}}}] ($(C)+(0.7,0.)$)--($(C)+(0.3,0.5)$);
   
\draw[-latex,black!50] ($(C)+(0.2,-0.3)$) -- ($(C)+(1.2,-0.3)$);
\draw[-latex,black!50] ($(C)+(.9,-0.5)$) -- ($(C)+(.9,0.5)$);
\node[right,black!50] at ($(C)+(1.2,-0.3)$) {$x$};
\node[above,black!50] at ($(C)+(.9,0.5)$) {$k_T$};
   
%%%%%%%%%%%%%%%%%%%%% +--
\coordinate (C) at ($2.2*(e1)$);
\draw[white,fill=white] (C) circle (8pt);

\draw[fill=black] ($(C)-(0.6,0)$) circle (5pt);
\draw[fill=black] ($(C)+(0.6,0)$) circle (5pt);
\draw[ultra thick] ($(C)+(0.6,0.08)$) -- ($(C)-(0.6,-0.08)$);
\draw[ultra thick] ($(C)+(0.6,0)$) -- ($(C)-(0.6,0)$);
\draw[ultra thick] ($(C)+(0.6,-0.08)$) -- ($(C)-(0.6,0.08)$);

\draw[ thick,postaction={decorate, decoration={markings,
   mark=at position 0.8 with {\arrow{stealth}}}}] ($(C)+(0.7,0)$)--($(C)-(-0.3,-0.5)$);

\draw[thick, style={decorate, decoration={snake,amplitude=0.6,segment length=2.2}}] 
($(C)+(0.7,0)$)--($(C)+(0.3,-0.5)$);

\draw[ thick,postaction={decorate, decoration={markings,
   mark=at position 0.5 with {\arrow{stealth}}}}] ($(C)-(0.3,-0.5)$)--($(C)-(0.7,0.)$);

%%%%%%%%%%%%%%%%%%%%% +-+
\coordinate (C) at ($-2*(e2)$);

\draw[fill=black] ($(C)-(0.6,0)$) circle (5pt);
\draw[fill=black] ($(C)+(0.6,0)$) circle (5pt);
\draw[ultra thick] ($(C)+(0.6,0.08)$) -- ($(C)-(0.6,-0.08)$);
\draw[ultra thick] ($(C)+(0.6,0)$) -- ($(C)-(0.6,0)$);
\draw[ultra thick] ($(C)+(0.6,-0.08)$) -- ($(C)-(0.6,0.08)$);

\draw[ thick,postaction={decorate, decoration={markings,
   mark=at position 0.8 with {\arrow{stealth}}}}] ($(C)-(0.7,0.)$)--($(C)-(0.3,-0.5)$);

\draw[thick, style={decorate, decoration={snake,amplitude=0.6,segment length=2.2}}] 
($(C)+(0.7,0)$)--($(C)+(0.3,0.5)$);

\draw[ thick,postaction={decorate, decoration={markings,
   mark=at position 0.5 with {\arrow{stealth}}}}] ($(C)-(0.3,0.5)$)--($(C)-(0.7,0.)$);
   
%%%%%%%%%%%%%%%%%%%%% --+
\coordinate (C) at ($2.2*(e3)$);
\draw[white,fill=white] (C) circle (8pt);

\draw[fill=black] ($(C)-(0.6,0)$) circle (5pt);
\draw[fill=black] ($(C)+(0.6,0)$) circle (5pt);
\draw[ultra thick] ($(C)+(0.6,0.08)$) -- ($(C)-(0.6,-0.08)$);
\draw[ultra thick] ($(C)+(0.6,0)$) -- ($(C)-(0.6,0)$);
\draw[ultra thick] ($(C)+(0.6,-0.08)$) -- ($(C)-(0.6,0.08)$);

\draw[ thick,postaction={decorate, decoration={markings,
   mark=at position 0.5 with {\arrow{stealth}}}}] ($(C)+(0.3,0.5)$)--($(C)+(0.7,0)$);

\draw[thick, style={decorate, decoration={snake,amplitude=0.6,segment length=2.2}}] 
($(C)+(0.5,0)$)--($(C)+(0.1,0.5)$);

\draw[ thick,postaction={decorate, decoration={markings,
   mark=at position 0.8 with {\arrow{stealth}}}}] ($(C)-(0.7,0.)$)--($(C)-(0.3,-0.5)$);

\end{tikzpicture}

%% file: Figures/theta_kernel.tex
\begin{tikzpicture}
%\draw[step=.5cm] (-3,-3) to[grid with coordinates] (3,3);
\tikzmath{\s = 1.5;}

%%%%%%%%%%%%%%%%%%%%%%%%%%%%%%%%%%%%%%%%%%%%%%%%%%%%%%%%%%%%%% diag-1
\coordinate (C) at (0,0);

\coordinate (e1) at (0.866025*\s, -0.5*\s);
\coordinate (e2) at (0.*\s, 1.*\s);
\coordinate (e3) at (-0.866025*\s, -0.5*\s);

\node[above,draw=black!50] at ($0.8*(e1)+1.1*(e2)-1.*(e3)+(C)$) {$\Theta_{x_1x_2x_3}$};

\draw[] ($-1*(e1)+1*(e2)+0*(e3)+(C)$) -- 
		($-1*(e1)+0*(e2)+1*(e3)+(C)$) --
		($0*(e1)-1*(e2)+1*(e3)+(C)$) --
		($1*(e1)-1*(e2)-0*(e3)+(C)$) --
		($1*(e1)+0*(e2)-1*(e3)+(C)$) --
		($0*(e1)+1*(e2)-1*(e3)+(C)$) --
		($-1*(e1)+1*(e2)+0*(e3)+(C)$);

\draw[] (C)--($-1*(e1)+1*(e2)+0*(e3)+(C)$);
\draw[] (C)--($-1*(e1)+0*(e2)+1*(e3)+(C)$);
\draw[] (C)--($0*(e1)-1*(e2)+1*(e3)+(C)$);
\draw[] (C)--($1*(e1)-1*(e2)-0*(e3)+(C)$);
\draw[] (C)--($1*(e1)+0*(e2)-1*(e3)+(C)$);
\draw[] (C)--($0*(e1)+1*(e2)-1*(e3)+(C)$);

\draw[fill=red!40,postaction={pattern=north west lines}] (C)--($-1*(e1)+1*(e2)+0*(e3)+(C)$)--($0*(e1)+1*(e2)-1*(e3)+(C)$)--(C);
\draw[fill=blue!40,postaction={pattern=vertical lines}] (C)--($-1*(e1)+1*(e2)+0*(e3)+(C)$)--($-1*(e1)+0*(e2)+1*(e3)+(C)$)--(C);
\draw[fill=red!40,postaction={pattern=north east lines,dashed}] (C)--($1*(e1)-1*(e2)+0*(e3)+(C)$)--($0*(e1)-1*(e2)+1*(e3)+(C)$)--(C);
\draw[fill=blue!40,postaction={pattern=horizontal lines,dashed}] (C)--($1*(e1)-1*(e2)+0*(e3)+(C)$)--($+1*(e1)+0*(e2)-1*(e3)+(C)$)--(C);

\node[fill=red!40,rounded corners=10pt] at ($-0.4*(e1)+0.8*(e2)-0.4*(e3)+(C)$) {\scalebox{1}{$C_F-\frac{C_A}{2}$}};
\node[fill=blue!40,rounded corners=10pt] at ($-0.7*(e1)+0.35*(e2)+0.35*(e3)+(C)$) {\scalebox{1}{$-\frac{C_A}{2}$}};
\node[fill=red!40,rounded corners=10pt] at ($0.4*(e1)-0.8*(e2)+0.4*(e3)+(C)$) {\scalebox{0.8}{$-\Big(C_F-\frac{C_A}{2}\Big)$}};
\node[fill=blue!40,rounded corners=10pt] at ($0.7*(e1)-0.35*(e2)-0.35*(e3)+(C)$) {\scalebox{1}{$\frac{C_A}{2}$}};

%%%%%%%%%%%%%%%%%%%%%%%%%%%%%%%%%%%%%%%%%%%%%%%%%%%%%%%%%%%%%% diag-1
\coordinate (C) at (6.5,0);

\coordinate (e1) at (0.866025*\s, -0.5*\s);
\coordinate (e2) at (0.*\s, 1.*\s);
\coordinate (e3) at (-0.866025*\s, -0.5*\s);

\node[above,draw=black!50] at ($0.8*(e1)+1.1*(e2)-1.*(e3)+(C)$) {$\mathbb{P}_{x_2x_1}$};

\draw[] ($-1*(e1)+1*(e2)+0*(e3)+(C)$) -- 
		($-1*(e1)+0*(e2)+1*(e3)+(C)$) --
		($0*(e1)-1*(e2)+1*(e3)+(C)$) --
		($1*(e1)-1*(e2)-0*(e3)+(C)$) --
		($1*(e1)+0*(e2)-1*(e3)+(C)$) --
		($0*(e1)+1*(e2)-1*(e3)+(C)$) --
		($-1*(e1)+1*(e2)+0*(e3)+(C)$);

\draw[red,thick] (C)--($-1*(e1)+1*(e2)+0*(e3)+(C)$);
\draw[] (C)--($-1*(e1)+0*(e2)+1*(e3)+(C)$);
\draw[] (C)--($0*(e1)-1*(e2)+1*(e3)+(C)$);
\draw[red,thick] (C)--($1*(e1)-1*(e2)-0*(e3)+(C)$);
\draw[] (C)--($1*(e1)+0*(e2)-1*(e3)+(C)$);
\draw[] (C)--($0*(e1)+1*(e2)-1*(e3)+(C)$);

\draw[very thick,dashed,blue] ($0.5*(e1)-1*(e2)+0.5*(e3)+(C)$)--($-1*(e1)+0.5*(e2)+0.5*(e3)+(C)$);

\draw[blue,fill=black] ($-0.4*(e1)-0.1*(e2)+0.5*(e3)+(C)$) circle (2pt);
\draw[blue,fill=white] ($-0.1*(e1)-0.4*(e2)+0.5*(e3)+(C)$) circle (2pt);
%\draw[white,fill=white] ($-0.5*(e1)+0.5*(e2)+0*(e3)+(C)$) circle (3.5pt);
\node[left] at ($-0.4*(e1)-0.1*(e2)+0.5*(e3)+(C)$) {\scalebox{0.6}{$(x_1,x_2,x_3)$}};
\node[left] at ($-0.1*(e1)-0.4*(e2)+0.5*(e3)+(C)$) {\scalebox{0.6}{$(x_2,x_1,x_3)$}};
\node[right] at ($-1*(e1)+0.5*(e2)+0.5*(e3)+(C)$) {$x_3>0$};

\end{tikzpicture}

%% file: Figures/diagrams2.tex
\begin{tikzpicture}
%\draw[step=.5cm] (0,0) to[grid with coordinates] (5,3);

%%%%%%%%%%%%%%%%%%%%%%%%%%%%%%%%%%%%%%%%%%%%%%%%%%%%%%%%%%%%%% diag-1
\coordinate (A) at (1, 2);

\draw[thick,postaction={decorate, decoration={
    markings,
    mark=at position 0.6 with {\arrow{stealth}}}}] ($(A)+(0,-2)$)--($(A)+(0,-1)$);

\draw[thick,postaction={decorate, decoration={
    markings,
    mark=at position 0.6 with {\arrow{stealth}}}}] ($(A)+(0,-1)$)--($(A)+(0,0)$);
\draw[thick,postaction={decorate, decoration={
    markings,
    mark=at position 0.8 with {\arrow{stealth}}}}] ($(A)+(1.5,-0.5)$)--($(A)+(1.5,-2)$);    

\draw[very thick,style={decorate, decoration={snake,segment length=6.2}}] 
($(A)+(0.5,0)$) -- ($(A)+(0.5,-2)$);

\draw[very thick,style={decorate, decoration={snake,segment length=6.2}}] 
($(A)+(0.0,-1)$) to[out=0,in=-90] ($(A)+(2.,-0.5)$);

\draw[blue,very thick,-latex,rounded corners=2pt,dotted] ($(A)+(2.,-0.5)$)--($(A)+(2.2,-0.3)$)--($(A)+(3.5,-0.3)$);   

\draw[thick,dashed] ($(A)$) -- ($(A)+(2.5,0)$);
\draw[thick,dashed] ($(A)+(1.5,-0.5)$) -- ($(A)+(3.5,-.5)$);

\draw[black,fill=black] ($(A)+(0.5,0)$) circle (2pt);
\node[above] at ($(A)+(0.5,0)$) {$F_{\mu+}$};

\node[] at ($(A)+(2,-2)$) {(r1)};

%%%%%%%%%%%%%%%%%%%%%%%%%%%%%%%%%%%%%%%%%%%%%%%%%%%%%%%%%%%%%% diag-1
\coordinate (A) at (5.5, 2);

\draw[thick,postaction={decorate, decoration={
    markings,
    mark=at position 0.6 with {\arrow{stealth}}}}] ($(A)+(0,-2)$)--($(A)+(0,0)$);

\draw[thick,postaction={decorate, decoration={
    markings,
    mark=at position 0.8 with {\arrow{stealth}}}}] ($(A)+(1.5,-0.5)$)--($(A)+(1.5,-2)$);    

\draw[very thick,style={decorate, decoration={snake,segment length=6.2}}] 
($(A)+(0.5,0)$) -- ($(A)+(0.5,-2)$);

\draw[very thick,style={decorate, decoration={snake,segment length=6.2}}] 
($(A)+(0.5,-1)$) to[out=0,in=-90] ($(A)+(2.,-0.5)$);

\draw[blue,very thick,-latex,rounded corners=2pt,dotted] ($(A)+(2.,-0.5)$)--($(A)+(2.2,-0.3)$)--($(A)+(3.5,-0.3)$);  

\draw[thick,dashed] ($(A)$) -- ($(A)+(2.5,0)$);
\draw[thick,dashed] ($(A)+(1.5,-0.5)$) -- ($(A)+(3.5,-.5)$);

\draw[black,fill=black] ($(A)+(0.5,0)$) circle (2pt);
\node[above] at ($(A)+(0.5,0)$) {$F_{\mu+}$};

\node[] at ($(A)+(2,-2)$) {(r2)};

%%%%%%%%%%%%%%%%%%%%%%%%%%%%%%%%%%%%%%%%%%%%%%%%%%%%%%%%%%%%%% diag-1
\coordinate (A) at (10, 2);

\draw[thick,postaction={decorate, decoration={
    markings,
    mark=at position 0.6 with {\arrow{stealth}}}}] ($(A)+(0,-2)$)--($(A)+(0,0)$);

\draw[thick,postaction={decorate, decoration={
    markings,
    mark=at position 0.6 with {\arrow{stealth}}}}] ($(A)+(1.5,-1)$)--($(A)+(1.5,-2)$);    
\draw[thick,postaction={decorate, decoration={
    markings,
    mark=at position 0.6 with {\arrow{stealth}}}}] ($(A)+(1.5,-0.5)$)--($(A)+(1.5,-1)$);    
    
\draw[very thick,style={decorate, decoration={snake,segment length=6.2}}] 
($(A)+(0.5,0)$)-- ($(A)+(0.5,-2)$);

\draw[very thick,style={decorate, decoration={snake,segment length=6.2}}] 
($(A)+(2,0)$) to [out=-90,in=0] ($(A)+(1.5,-1)$);

\draw[blue,very thick,-latex,rounded corners=2pt,dotted] ($(A)+(2,0)$)--($(A)+(2.2,-0.2)$)--($(A)+(3,-0.2)$);    

\draw[thick,dashed] ($(A)$) -- ($(A)+(2.5,0)$);
\draw[thick,dashed] ($(A)+(1.5,-0.5)$) -- ($(A)+(3.5,-.5)$);

\draw[black,fill=black] ($(A)+(0.5,0)$) circle (2pt);
\node[above] at ($(A)+(0.5,0)$) {$F_{\mu+}$};

\node[] at ($(A)+(2,-2)$) {(r3)};

%%%%%%%%%%%%%%%%%%%%%%%%%%%%%%%%%%%%%%%%%%%%%%%%%%%%%%%%%%%%%% diag-1
\coordinate (A) at (14.5, 2);

\draw[thick,postaction={decorate, decoration={
    markings,
    mark=at position 0.6 with {\arrow{stealth}}}}] ($(A)+(0,-2)$)--($(A)+(0,0)$);

\draw[thick,postaction={decorate, decoration={
    markings,
    mark=at position 0.6 with {\arrow{stealth}}}}] ($(A)+(1.5,-1)$)--($(A)+(1.5,-2)$);    
\draw[thick,postaction={decorate, decoration={
    markings,
    mark=at position 0.6 with {\arrow{stealth}}}}] ($(A)+(1.5,-0.5)$)--($(A)+(1.5,-1)$);    
    
\draw[very thick,style={decorate, decoration={snake,segment length=6.2}}] 
($(A)+(0.5,0)$) to[out=-90,in=-180] ($(A)+(1.5,-1)$);

\draw[blue,very thick,-latex,rounded corners=2pt,dotted] ($(A)+(0.5,0)$)--($(A)+(0.7,-0.2)$)--($(A)+(2.5,-0.2)$);    

\draw[thick,dashed] ($(A)$) -- ($(A)+(2.5,0)$);
\draw[thick,dashed] ($(A)+(1.5,-0.5)$) -- ($(A)+(3.5,-.5)$);

\draw[black,fill=black] ($(A)+(0.5,0)$) circle (2pt);
\node[above] at ($(A)+(0.5,0)$) {$F_{\mu+}$};

\node[] at ($(A)+(2,-2)$) {(r4)};

\end{tikzpicture}

%% file: Figures/diagrams1.tex
\begin{tikzpicture}
%\draw[step=.5cm] (0,0) to[grid with coordinates] (5,3);

%%%%%%%%%%%%%%%%%%%%%%%%%%%%%%%%%%%%%%%%%%%%%%%%%%%%%%%%%%%%%% diag-1
\coordinate (A) at (1, 2);

\draw[thick,postaction={decorate, decoration={
    markings,
    mark=at position 0.6 with {\arrow{stealth}}}}] ($(A)+(0,-2)$)--($(A)+(0,0)$);

\draw[thick,postaction={decorate, decoration={
    markings,
    mark=at position 0.6 with {\arrow{stealth}}}}] ($(A)+(1.5,-1)$)--($(A)+(1.5,-2)$);    
\draw[thick,postaction={decorate, decoration={
    markings,
    mark=at position 0.6 with {\arrow{stealth}}}}] ($(A)+(1.5,-0.5)$)--($(A)+(1.5,-1)$);    
    
\draw[very thick,style={decorate, decoration={snake,segment length=6.2}}] 
($(A)+(0.5,0)$) to[out=-90,in=-180] ($(A)+(1.5,-1)$);

\draw[thick,dashed] ($(A)$) -- ($(A)+(2.5,0)$);
\draw[thick,dashed] ($(A)+(1.5,-0.5)$) -- ($(A)+(3.5,-.5)$);

\draw[black,fill=black] ($(A)+(0.5,0)$) circle (2pt);
\node[above] at ($(A)+(0.5,0)$) {$F_{\mu+}$};

\node[] at ($(A)+(2,-2)$) {(A)};

%%%%%%%%%%%%%%%%%%%%%%%%%%%%%%%%%%%%%%%%%%%%%%%%%%%%%%%%%%%%%% diag-1
\coordinate (A) at (5.5, 2);

\draw[thick,postaction={decorate, decoration={
    markings,
    mark=at position 0.6 with {\arrow{stealth}}}}] ($(A)+(0,-2)$)--($(A)+(0,-1)$);

\draw[thick,postaction={decorate, decoration={
    markings,
    mark=at position 0.6 with {\arrow{stealth}}}}] ($(A)+(0,-1)$)--($(A)+(0,0)$);
\draw[thick,postaction={decorate, decoration={
    markings,
    mark=at position 0.8 with {\arrow{stealth}}}}] ($(A)+(1.5,-0.5)$)--($(A)+(1.5,-2)$);    
    
\draw[very thick,style={decorate, decoration={snake,segment length=6.2}}] 
($(A)+(2.5,-0.5)$) to[out=-90,in=0] ($(A)+(0,-1)$);

\draw[thick,dashed] ($(A)$) -- ($(A)+(2.5,0)$);
\draw[thick,dashed] ($(A)+(1.5,-0.5)$) -- ($(A)+(3.5,-.5)$);

\draw[black,fill=black] ($(A)+(2.5,-0.5)$) circle (2pt);
\node[above] at ($(A)+(2.5,-0.5)$) {$F_{\mu+}$};

\node[] at ($(A)+(2,-2)$) {(B)};

%%%%%%%%%%%%%%%%%%%%%%%%%%%%%%%%%%%%%%%%%%%%%%%%%%%%%%%%%%%%%% diag-1
\coordinate (A) at (10, 2);

\draw[thick,postaction={decorate, decoration={
    markings,
    mark=at position 0.3 with {\arrow{stealth}}}}] ($(A)+(1.5,-1)$)to[out=-180,in=-90]($(A)+(0,0)$);

\draw[very thick,style={decorate, decoration={snake,segment length=6.2}}] ($(A)+(1.5,-1)$)--($(A)+(1.5,-2)$);    
\draw[thick,postaction={decorate, decoration={
    markings,
    mark=at position 0.6 with {\arrow{stealth}}}}] ($(A)+(1.5,-0.5)$)--($(A)+(1.5,-1)$);    
    
\draw[very thick,style={decorate, decoration={snake,segment length=6.2}}] 
($(A)+(0.5,0)$) -- ($(A)+(0.5,-2)$);

\draw[thick,dashed] ($(A)$) -- ($(A)+(2.5,0)$);
\draw[thick,dashed] ($(A)+(1.5,-0.5)$) -- ($(A)+(3.5,-.5)$);

\draw[black,fill=black] ($(A)+(0.5,0)$) circle (2pt);
\node[above] at ($(A)+(0.5,0)$) {$F_{\mu+}$};

\node[] at ($(A)+(2,-2)$) {(C)};

%%%%%%%%%%%%%%%%%%%%%%%%%%%%%%%%%%%%%%%%%%%%%%%%%%%%%%%%%%%%%% diag-1
\coordinate (A) at (14.5, 2);

\draw[very thick,style={decorate, decoration={snake,segment length=6.2}}] ($(A)+(0,-2)$)--($(A)+(0,-1)$);

\draw[thick,postaction={decorate, decoration={
    markings,
    mark=at position 0.6 with {\arrow{stealth}}}}] ($(A)+(0,-1)$)--($(A)+(0,0)$);
\draw[thick,postaction={decorate, decoration={
    markings,
    mark=at position 0.5 with {\arrow{stealth}}}}] ($(A)+(1.5,-0.5)$)to[out=-90,in=0]($(A)+(0,-1)$);    
    
\draw[very thick,style={decorate, decoration={snake,segment length=6.2}}] 
($(A)+(2.5,-0.5)$) --($(A)+(2.5,-2)$);

\draw[thick,dashed] ($(A)$) -- ($(A)+(2.5,0)$);
\draw[thick,dashed] ($(A)+(1.5,-0.5)$) -- ($(A)+(3.5,-.5)$);

\draw[black,fill=black] ($(A)+(2.5,-0.5)$) circle (2pt);
\node[above] at ($(A)+(2.5,-0.5)$) {$F_{\mu+}$};

\node[] at ($(A)+(2,-2)$) {(D)};

\end{tikzpicture}

%% file: TMD-evol-sub-leadingP_v2.bbl
\providecommand{\href}[2]{#2}\begingroup\raggedright\endgroup